\documentclass{jfm}
\usepackage{graphicx}
\usepackage{accents}
\usepackage{epstopdf, epsfig}
\usepackage{color}

\title{Viscoelastic thin film lubrication in finite width channels}

\author{Humayun Ahmed\aff{1}
  \corresp{\email{jfm@damtp.cam.ac.uk}},
 \and Luca Biancofiore\aff{1,2}}

\affiliation
{
\aff{1}Department of Mechanical Engineering, Bilkent University, Ankara, Turkey
\aff{2}Department of Industrial Engineering Information and Economics, University of L’Aquila, Piazzale Ernesto Pontieri Monteluco di Roio, L’Aquila 67100, Italy
}

\begin{document}

\maketitle

\begin{abstract}
Lubricant viscoelasticity arises due to a finite polymer relaxation time ($\lambda$) and can provide beneficial effects. In applications, such as bearings, gears, biological joints, etc., where the height-to-length ratio is small ($H_0 / \ell_x$) and the shear due to the wall velocity ($U_0$) is high, a simplified two-dimensional computational analysis across the channel length and height reveals a finite increase in the load carrying capacity of the film purely due to polymer elasticity. In channels with a finite length-to-width ratio, $a$, the spanwise effects can be significant, but the resulting mathematical model is computationally intensive. In this work, we propose simpler reduced-order models, namely via a (i) first-order perturbation in the Deborah number ($\lambda U_0 / H_0$), and the (ii) viscoelastic Reynolds approach extended from \textit{Ahmed, H., \& Biancofiore, L. (2021). A new approach for modeling viscoelastic thin film lubrication. Journal of Non-Newtonian Fluid Mechanics, 292, 104524}. We predict the variation in the net vertical force exerted on the channel walls (for a fixed film height) versus increasing viscoelasticity, and the channel aspect ratio. The models predict an increase in the net force, which is zero for the Newtonian case, versus both the Deborah number and the channel aspect ratio. Interestingly, for a fixed $De$, this force varies strongly between the two limiting cases (i) $a << 1$; an infinitely wide, and (ii) $a >> 1$; an infinitely short channel channel, implying a change in the polymers response. Furthermore, we observe a different trend (i) for a spanwise varying channel, in which a peak is observed between the two limits, and (ii) for a spanwise uniform channel, where the largest load value is for $a << 1$. When $a$ is O($1$), the viscoelastic response varies strongly and spanwise effects cannot be ignored. 
\end{abstract}


\section{Introduction}
\par Fluid mixtures comprising of elastic material and subjected to fast flow processes exhibit strongly non-Newtonian traits.  Examples of such fluid flow are commonly found in both biological and mechanical systems, such as capillary blood flow \citep{pandey2016lubrication}, human synovial joints consisting of a lubricating fluid and rapid motion of the joint \citep{tichy_said_spherical,yousfi2013analytical} and tear films subjected to rapid movement of the eyelids \citep{dunn2013lubrication}, and mechanical systems. The lubrication of these components has been around for several centuries and serves an important function in maintaining the performance of the machine elements. In thin film hydrodynamic lubrication, a thin layer of fluid (liquid or gas) persists between two moving surfaces (bearings, gears, artificial joints etc.). 
\par The lubricant, when engineered, is often a mixture comprising a base Newtonian solvent which is then enhanced via the addition of polymeric additives \citep{mortier2010chemistry}. These additives are often necessary in counteracting the unavoidable effects due to high shear stresses and high pressure such as, (i) shear thinning \citep{ahmed2022modified} in which the shear stress causes a decline in viscosity, (ii) viscoplasticity; an irreversible decrease in lubricant viscosity versus the applied shear stress, (iii) piezoviscosity \citep{jeng1986piezoviscous}; an increase in the lubricant viscosity versus the film pressure, (iv) thermal degradation due to uncoiling of polymeric chains as the film temperature increases, and also (v) cavitation of the film; generation of a liquid-vapor mixture region due to a drop in film pressure below cavitation pressure \citep{gamaniel2021effect,sari2024effect}. These effects are studied extensively in tribology to optimize the performance of machine elements \citep{ahmed2022modified}. The emergence of viscoelasticity, arising purely due to a finitely elastic polymer additive, is somewhat overlooked owing to the non-linear interaction of these phenomena which makes it difficult to isolate any single effect, at least experimentally. 

\par In classical thin-film hydrodynamic lubrication (where the bounding surfaces are separated and assumed rigid) the film thickness, $H_0$, is some orders of magnitude less than the channel length, $\ell_x$, (characterized by the ratio $\epsilon = H_0 / \ell_x << 1$). There are several operating modes depending on the motion of the surfaces, e.g., sliding, rolling, squeezing and a combination of thereof \citep{szeri2010fluid}. The lubrication mode and the channel surface configuration unsurprisingly have a strong influence on the film pressure distribution and magnitude. For channel configurations where the bounding surfaces are in relative motion, i.e., a sliding configuration in which at least one surface has a fixed speed, $U_0$, (such as, e.g., bearings, meshing of gear teeth), the film pressure is generated only if some surface gradient is present. The strength of these pressure gradients are measured via the bearing number $\Lambda = \eta_0 U_0 \ell_x / p_a H_0^2$, i.e., the ratio between the viscous and pressure forces \citep{kundu2015fluid}, which tends to be large leading to high pressure gradients. A combination of the two (large $\Lambda$ and surface sliding) exposes the elastic polymeric additives to a high applied shear rate that persists along the entire channel preventing their otherwise relaxed state (when the polymeric stress decays to zero or reaches its Newtonian value). We focus primarily on the impact of viscoelasticity on the film pressure (and its ability to bear the applied load, the so-called load carrying capacity) in the sliding case which has received comparatively little attention when compared with other non-linear phenomena mentioned previously.
\par For the thin sliding channel, we have (i) a high shear strain rate, $\dot{\gamma} \approx U_0 / H_0$, and (ii) a residence time on the order of $\ell_x / U_0$, where, $U_0$ is some relative surface speed. When elastic additives are exposed to these two distinct time scales they tend to exhibit characteristics strongly differing from the classical Newtonian behavior such as the onset of shear thinning. However, when the Deborah number $De = \lambda U_0 / \ell_x$, i.e., the ratio of the polymer relaxation time, $\lambda \approx \eta_p / G$, ($\eta_p$ is the polymer's viscosity and $G$ is the shear modulus) to the residence time scale $\ell_x / U_0$, is not negligible then viscoelastic effects also emerge. In addition, the Weissenberg number, i.e., the ratio of the polymer relaxation time to the shear time scale $Wi = \lambda \dot{\gamma}$, tends to acquire large values since, the shear strain rate can be large in practice. These two parameters when plotted against one another span the Pipkin space which characterizes the problem \citep{tanner2000engineering}, e.g., linearly viscoelastic ($De << 1$, $Wi << 1$) or  viscometric ($De << 1$, $Wi >> 1$), etc. Furthermore, for thin film lubrication they are related via the thin film ratio $De = \epsilon Wi$ \citep{ahmed2021new}. 
\par Experimental evidence suggests that increasing the fluid elasticity has an observable impact, beyond the typical viscous effects. For example, in the case of sliding hydrodynamic lubrication (such as those in bearings and gears), it increases the lubricating film's load carrying capacity and thereby enhances the performance of the machine components. Tests conducted via a cone-and-plate rheometer have demonstrated a finite increase in the load carrying capacity when viscoelasticity is present \citep{schuh2017design}. In fact, when elastic compounds in the form of amino acid ionic liquids are added to water based lubricants (which remains Newtonian even under extreme flow conditions), an increase in the first normal stress difference is found \citep{feng2024rheological}, corresponding to a positive net force. Similarly, a thin-film squeezed between two parallel plates predicts an increased first normal stress difference and the load carrying capacity \citep{tichy_winer_squeeze}.  However, owing to the small dimensions of the contact region, point-wise variation of the pressure is difficult to obtain via experimental approaches.
\par Numerical studies on the flow of polymeric material through a thin film enable visualization of the pressure, velocity and stress distributions within the channel necessary for calculating the load carrying capacity, flow rate and friction, respectively. 
A full-scale mathematical model comprises the mass and momentum conservation laws coupled with a valid constitutive relation for the polymer extra stress tensor, $\boldsymbol{\tau}^*$, for example, the  Oldroyd-B model approximating the elastic material as an ideal Hookean spring element, being able to stretch several order of magnitudes in comparison to its equilibrium length. When the equations are is solved via direct numerical simulations (DNS), it was discovered that a critical numerical instability prevents convergence for large values of the $De$ or $Wi$, termed as the high Weissenberg number problem (HWNP) \citep{keunings1986high,owens2002computational}. Its manifestation was traced to a loss of symmetric positive definite property of the conformation tensor, $\mathsfbi{C}$, related to the polymeric stress via the relation $\boldsymbol{\tau}^* = \eta_p / \lambda (\mathsfbi{I} - \mathsfbi{C})$, as $De$ increases beyond a certain value. This problem improves, only temporarily, when the maximum possible extension of the polymer chains is restricted as in FENE-CR, and further by shear thinning e.g., the FENE-P model.

Extensive efforts in devising novel discretization methods did not yield significant improvements as the degeneracy of the eigenvalues of $\mathsfbi{C}$ was traced to the use of polynomial based approximations for the stress or conformation tensor. Specifically, beyond a certain critical $De$ (or $Wi$), negative eigenvalues tend to emerge which are unphysical \citep{chakraborty2010viscoelastic}. This was  resolved to a great extent in the seminal work of \cite{fattal2004constitutive} which employed a logarithmic transformation of $\mathsfbi{C}$ (the so-called logarithmic conformation representation or simply LCR) and its success was immediately demonstrated on some classical problems \citep{fattal2005time}, later extended to more complicated and numerically challenging cases, such as the 4:1 contraction problem \citep{alves2021numerical}. Still, for some critical value of $De$, simulation accuracy becomes questionable and eventually fails. The HWNP only having been delayed prompted further research and recent efforts have shown that it can be triggered also due to (i) a change in type of the numerical system \citep{renardy2021mathematician} or (ii) improper interpolation of the eigenvalues of $\mathsfbi{C}$ \citep{zhang2023role}, but dampened via artificial diffusion methods \citep{fernandes2017improved}. Interestingly, for thin-film lubrication problems, the HWNP is not alleviated as the grid is refined by including more nodes/elements, contrary to earlier findings that suggest the opposite. This was detected very early on for simulations of journal bearings \citep{beris1986finite} in which a saturation of the stability versus mesh refinements was observed. 

\par For thin film boundary driven lubrication, the use of DNS for predicting the pressure profile along the channel via a computationally efficient implementation of the LCR in RheoTool \citep{alves2021numerical} was limited to small values of the $De$ \citep{ahmed2021new, ahmed2023modeling}. When combined with other non-linear effects such as, film cavitation, the critical $De$ value was further reduced \citep{gamaniel2021effect}. These difficulties are alleviated via  simplifications resulting from the application of lubrication theory \citep{szeri2010fluid} on the governing set of equations. This effort yields a set of reduced equations that can be solved to give the classical Reynolds equation for the Newtonian case. However, it does not simplify the constitutive relations. Efforts have therefore been directed at methods which attempt to further reduce the non-linear coupling between pressure, velocity and polymer extra stress components. In this regard, the first major attempt was the generalized Reynolds equation \citep{johnson1977shear} celebrated for its capacity in capturing cross-film variation of the lubricants properties (shear thinning, thermal effects, liquid compressibility, etc.) was modified to include the effects of elasticity. However, the problem was approached in a Newtonian like manner only considering the effects of the non-Newtonian shear stress and ignoring the normal stress (since for the Newtonian case, the latter is on the order of $\epsilon^2$) \citep{wolf_kubo}. Furthermore, only a portion substantive derivative appearing in the upper-convected rate operator was kept for simplicity leading to an over-simplified prediction of the polymeric stress. Owing to these limitations, the elasticity of the lubricant film was not considered significant. Nonetheless, these early efforts, driven by a practical need to resolve the role of viscoelasticity in mechanical components, highlight the complexity of the problem.
\par The problem in obtaining a simplified analytical model lies in the non-linear coupling of the velocity field (and its gradients) with the polymer extra stress tensor. Noting that (i) the Deborah number appears naturally in the constitutive relations and (ii) can be argued to be small for some cases, a perturbation approach in the $De$ was applied which yielded an equation for the $De$-order pressure depending only on the channel surface function, $h$, for the upper-convected Maxwell fluid \citep{tichy_lin_1996}, and, also for the second-order fluid \citep{tichy_sawer_SOF_1998}. A more robust approach based on the reciprocal theorem that allows expansions up to arbitrary powers of $De$ yielded a third-order expansion in for purely pressure driven flows for the Oldroyd-B model \citep{boyko2021reciprocal}, comparing favorably with DNS over a wide range. Similarly, we find other studies for pressure driven flows that leverage the perturbation analysis and demonstrate its effectiveness in predicting the relevant quantities, such as, e.g. flow-rate, pressure drop across the channel etc. \citep{housiadas2024pressure, housiadas2024viscoelastic}. These models generally express the desired quantity (usually pressure, or flow rate) in function of the surface height but a recent notable work by \cite{boyko2024flow} achieves near-analytical expressions for the pressure and stress by (i) leveraging a curvilinear coordinate transformation and (ii) noting the velocity field is Newtonian when the polymer concentration is small. These assumptions allow for the elimination of the cross-film velocity component (in the curvilinear space). In a related work, a numerical treatment of the thin-film reduced system reveals the stress relaxation mechanism and how it contributes to a pressure drop for contracting channels \citep{hinch2024fast}. These simplified numerical models demonstrate the strongly contrasting behaviour of a viscoelastic fluid mainly owing to the generation of streamwise normal stresses (depending quadratically on the applied shear strain rate for small $De$) which are O($\epsilon^2$) for a Newtonian fluid, and, are therefore, negligible in a thin film. Furthermore, they reveal how the surface profile and its gradients (converging or diverging gaps) determine whether the extra stress will enhance or inhibit the pressure for problems neglecting spanwise effects.  
\par The same measure of sensitivity in increasing viscoelasticity is not observed for the velocity field which is only weakly perturbed \citep{zheng2023squeeze}. In certain special cases, the viscoelastic influence on the kinematics vanishes entirely and the flow field remains Newtonian \citep{phan1984lubrication,phan1985squeeze,phan1983viscoelastic} (see the Tanner and Pipkin theorem for a two-dimensional Cartesian configuration).  In fact, for the two-dimensional channels (representing the cross-section of an infinitely wide real channel) the elastic component of the velocity field vanishes when the channel inlet and outlet height are equal and, also, when the polymeric viscosity, $\eta_p$, is small in comparison to the solvent $\eta_s$ (the ultra-dilute limit). Exploiting the weak dependence of $\textbf{u}$ on the Deborah number, it is possible to decouple the momentum equation from the constitutive relation, and obtain a Reynolds type equation for the pressure, the so-called the viscoelastic Reynolds approach or VR \citep{ahmed2021new}. For sliding lubricated contacts, a straight-forward numerical treatment shows a better prediction to the classical first-order perturbation versus the Deborah number and compares favorably with the DNS. It predicts the emergence of a non-linear trend in the load carrying capacity versus the $De$, demonstrating not only the positive contribution of the polymer normal stress but a mechanism that differs significantly from the Newtonian case in which the primary contribution is only due to viscous effects \citep{tichy_said_spherical}. 
\par The studies mentioned thus far focus on a two-dimensional approximation of real lubricated channels arising from either symmetry arguments or being infinite across the third dimension. However, practical problems involving the thin film lubrication of two sliding surfaces deal with channels that have a finite spanwise width with respect to the length. These channels are arguably three-dimensional when the length scales are on the same order of magnitude, leading to a larger set of scalar-differential equations to be solved for the viscoelastic case. It is evident from the outset that the lack of analytical solutions even for the Newtonian case significantly increases the overall complexity of the problem. This necessitates the use of a numerical procedure and simplified reduced-order models. 
\par
\begin{figure}
\centering
\includegraphics[width=0.99\linewidth]{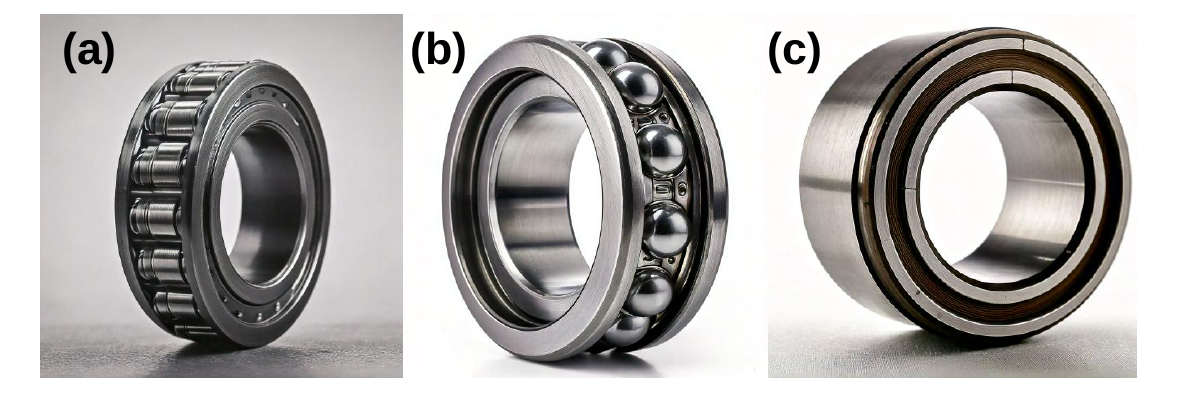}
\caption{A schematic diagram of (a) a cylindrical roller element bearing ($\ell_z > \ell_x$ for the rollers), (b) a ball bearing ($\ell_z \approx \ell_x$), and (c) a journal bearing ($\ell_z < \ell_x$). The images were generated using Adobe Firefly.}
\label{fig_moti}
\end{figure}

 Our aim is to understand the influence of a finite spanwise width on the polymeric response. To do so, we consider different length-to-width ratios to mimic different types of sliding channels, depicted in Fig. \ref{fig_moti}. (i) The contact region in cylindrical roller bearing elements has a constant cross-section and is very wide in comparison to its length (Fig. \ref{fig_moti}(a)), (ii) ball bearings have equal dimensions at the contact point (Fig. \ref{fig_moti}(b)), (iii) while the contact region in non-textured journal bearings that have no surface gradients along the width and the contact region is long in comparison to its width (Fig. \ref{fig_moti}(c)), in contrast to surface textures like pockets or surface roughness. There are, therefore, two geometric considerations when dealing with three dimensional problems, namely the surface gradients and the channel aspect ratio (length-to-width ratio). 
This paper extends the viscoelastic Reynolds approach as a potential reduced-order modeling technique for the flow of a viscoelastic lubricant through a thin finite-width channel examining mainly the load carrying capacity per unit width versus $De$ and the channel aspect ratio . In addition, we compare the results against a model linearised in $De$ to shed light on its usefulness for the three-dimensional case. Furthermore, we attempt to explain the relation between the observed trends in the load and the polymer stress distribution.

\section{Problem formulation}\label{sec:rules_submission}
\par In this section, we present the mathematical problem for the boundary driven flow of a viscoelastic lubricant through a three-dimensional (3D) channel. In section \ref{sec_govEqns}, the governing equations along with the relevant dimensionless parameters characterizing the flow are presented. In section \ref{sec_ReducedOrderModels}, these equations are then simplified in two ways; (i) using classical perturbation technique and (ii) extending the VR method \citep{ahmed2021new}.
\subsection{Governing equations}
\label{sec_govEqns}
\par The flow of a viscoelastic fluid mixture comprising a Newtonian base solvent and a dilute concentration of elastic polymeric additives through a channel with impermeable walls is mathematically described via the system of equations
\begin{subeqnarray}
\label{eqn_VESys1}
& \boldsymbol{\nabla}^{*} \cdot  \boldsymbol{u}^{*} = 0,
\\
\label{eqn_VESys2}
& \rho_0\frac{D \boldsymbol{u}^{*}}{D t^{*}} = -\boldsymbol{\nabla}^{*} p^{*} +  \eta_s\boldsymbol{\nabla}^{*2}\boldsymbol{u}^{*} + \boldsymbol{\nabla}^{*} \cdot \boldsymbol{\tau}^{*},
\\
\label{eqn_VESys3}
& \boldsymbol{\tau}^* + \lambda \accentset{\triangledown}{\boldsymbol{\tau}}^* = \eta_p \mathsfbi{D}^*,
\end{subeqnarray}
where, $\rho^{*}$ is the fluid density, $\eta_s$ is the solvent viscosity, $\eta_p$ is the polymer viscosity, $\lambda$ is the polymer relaxation time, $\textbf{u}^{*}$ is the velocity field, $t^{*}$ is the time, $p^{*}$ is the fluid pressure, $\boldsymbol{\tau}^{*}$ is the polymer extra stress tensor, $\mathsfbi{D}^{*} = \boldsymbol{\nabla}^{*}\boldsymbol{u}^{*} + \boldsymbol{\nabla}^{*}\boldsymbol{u}^{*T}$ is the deformation rate tensor, $\mathsfbi{L} = \boldsymbol{\nabla}^{*}\boldsymbol{u}^{*}$ is the velocity gradient tensor, $\accentset{\triangledown}{\boldsymbol{\tau}}^* = \frac{D \boldsymbol{\tau}^*}{Dt^*} -\mathsfbi{L}^*\boldsymbol{\tau}^* - \boldsymbol{\tau}^* \mathsfbi{L}^{*T}$ is the upper convected rate. Eq. \ref{eqn_VESys3}c is the Oldroyd-B constitutive relation \citep{bird1987dynamics}.

\par Using the length and velocity scaling
\begin{subeqnarray}
\label{eqn_geo_scaling}
& x^* = x \ell_x, y^* = y H_0, z^* = z \ell_z,
\\
\label{eqn_vel_scaling}
& u^* = u U_0, v^* = v \epsilon U_0, w^* = w U_0,
\end{subeqnarray}
where, $\epsilon = \frac{H_0}{\ell_x}$ is the thin film ratio, and a stress scaling \citep{tichy_lin_1996,li2014non,ahmed2021new,boyko2022pressure,housiadas2023lubrication},
\begin{subeqnarray}
\label{eqn_nonDimSys1_stress}
& \tau_{xx} = \tau_{xx}^*\frac{ h_0^2}{\eta_0 U_0 \ell_x},    \tau_{zx} = \tau_{zx}^*\frac{h_0^2}{\eta_0 U_0 \ell_x},    \tau_{xy} = \tau_{xy}^*\frac{h_0}{\eta_0 U_0},
\\
\label{eqn_nonDimSys2_stress}
& \tau_{yy} = \tau_{yy}^*\frac{ \ell_x}{\eta_0 U_0},  \tau_{yz} = \tau_{yz}^*\frac{ h_0}{\eta_0 U_0},  \tau_{zz} = \tau_{zz}^*\frac{ h_0^2}{\eta_0 U_0 \ell_x},
\end{subeqnarray}
in which all the normal stress components are on the order of the film pressure, we arrive at a reduced system of equations, 
\begin{subeqnarray}
\label{eqn_thinfilmSys1_full}
& \frac{\partial u }{\partial x}  + \frac{\partial v}{\partial y} + a\frac{\partial w}{\partial z}= 0,
\\
& \frac{\partial p}{\partial x} = \beta \frac{\partial^2 u}{\partial y^2} + \frac{\partial\tau_{xx}}{\partial x} + \frac{\partial\tau_{yx}}{\partial y} + a \frac{\partial\tau_{zx}}{\partial z},
\\
& \frac{\partial p}{\partial y}  = 0,
\\
& a\frac{\partial p}{\partial z} = \beta \frac{\partial^2 w}{\partial y^2} + \frac{\partial\tau_{xz}}{\partial x} + \frac{\partial\tau_{yz}}{\partial y} + a\frac{\partial\tau_{zz}}{\partial z},
\\
& \tau_{xx} + De \big(\frac{D\tau_{xx}}{Dt} -2\tau_{xy}\frac{\partial u}{\partial y} - 2\tau_{xx}\frac{\partial u}{\partial x} - 2a\frac{\partial u}{\partial z}\tau_{xz} \big) = 0,
\\
\nonumber
& \tau_{xy} + De \big(\frac{D\tau_{xy}}{Dt} -\tau_{yy}\frac{\partial u}{\partial y} - \tau_{xx}\frac{\partial v}{\partial x} - \frac{\partial u}{\partial x}\tau_{xy} - \frac{\partial v}{\partial y}\tau_{xy} - a\frac{\partial u}{\partial z}\tau_{yz} 
\\
& - a\frac{\partial v}{\partial z}\tau_{xz} \big) = (1 - \beta)\frac{\partial u}{\partial y},
\\
\nonumber
& \tau_{xz} + De \big(\frac{D\tau_{xz}}{Dt} -\tau_{xz}\frac{\partial u}{\partial x} - \tau_{xx}\frac{\partial w}{\partial x} -\tau_{yz}\frac{\partial u}{\partial y} -\tau_{xy}\frac{\partial w}{\partial y}
\\
& -a\tau_{zz}\frac{\partial u}{\partial z} -a\tau_{xz}\frac{\partial w}{\partial z} \big) = 0,
\\
& \tau_{yy} + De \big(\frac{D\tau_{yy}}{Dt} -2\tau_{xy}\frac{\partial v}{\partial x} - 2\tau_{yy}\frac{\partial v}{\partial y} - 2a\frac{\partial v}{\partial z}\tau_{yz} \big) = 2(1 - \beta)\frac{\partial v}{\partial y},
\\
\nonumber
& \tau_{yz} + De \big(\frac{D\tau_{yz}}{Dt} -\tau_{xz}\frac{\partial v}{\partial x} - \tau_{xy}\frac{\partial w}{\partial x} - \tau_{yz}\frac{\partial v}{\partial y} - \tau_{yy}\frac{\partial w}{\partial y} - a\tau_{zz}\frac{\partial v}{\partial z} 
\\
& - a\tau_{yz}\frac{\partial w}{\partial z}\big) = (1 - \beta)\frac{\partial w}{\partial y},
\\
& \tau_{zz} + De \big(\frac{D\tau_{yy}}{Dt} - 2\tau_{xz}\frac{\partial w}{\partial x} - 2\tau_{xy}\frac{\partial w}{\partial y} - 2a\frac{\partial w}{\partial z}\tau_{zz} \big) = 0.
\label{eqn_thinfilmSys2}
\end{subeqnarray}
where the Deborah number $De$ is defined as $De=\frac{\lambda U_0}{\ell_x}$, $\beta$ is the solvent concentration in terms of vscosity $\beta=\frac{\eta_s}{\eta_p+\eta_s}$ and  $a = \frac{\ell_x}{\ell_z}$ is the channel aspect ratio (length-to-width ratio), and $\frac{D}{Dt} = u\frac{\partial }{\partial x} + v\frac{\partial }{\partial y} + aw\frac{\partial }{\partial z}$ is the material derivative operator. 
Inherent in the derivation is the assumption of a vanishingly small Reynolds number ($Re=\frac{\rho_0 U_0 H_0}{\eta_0} << 0$). The system of Eqs. \ref{eqn_thinfilmSys1_full} is similar to that obtained in \cite{boyko2023non} but with a different definition of the channel aspect ratio, $a$, that allows retrieving the two-dimensional case when applying the limit ${a\rightarrow 0}$ in Eqs. \ref{eqn_thinfilmSys1_full}. In addition, values of $a$ of order $\epsilon^{-1}$ cannot be used, since terms on the order of $\epsilon^2a^2$ have also been ignored when applying lubrication theory. This is true for both the Newtonian and the viscoelastic case. 
\par The non-linear coupled system yields analytical solutions in certain special cases, such as a constant shear flow \citep{oliveira2002exact}. However, for most practical purposes it must be solved numerically which is challenging owing to the non-linear coupling between the flow field and the polymer stress (intensifying with increasing $De$), and the onset of the high Weissenberg number problem (HWNP) which prevents solutions at large $De$ (or $Wi$). 
\par For this work, the system of Eqs. \ref{eqn_thinfilmSys1_full} and any reduced model obtained thereof, are subject to the boundary conditions  
\begin{subeqnarray}
\label{eqns_boundaryConds1}
& u(x, 0, z) = 1, u(x, h, z) = 0,  
\\
& v(x, 0, z) = v(x, h, z) = 0,
\\
& w(x, 0, z) = w(x, h, z) = 0,
\\
& p(0, z) = p(1,z) = p(x,1)= 0, \mbox{ and } \frac{\partial p}{\partial z}(x,0) = 0,
\\
& \frac{\partial \boldsymbol{\tau}}{\partial x}(0,y,z) = 0, \frac{\partial \boldsymbol{\tau}}{\partial z}(x,y,0) = 0, \frac{\partial \boldsymbol{\tau}}{\partial y}(x,0,z) = 0,
\end{subeqnarray}
that describe a configuration in which the lower channel wall ($y=0$) is sliding at a fixed speed and the upper wall (at $y=h(x,z)$) is stationary. We assume symmetry along the spanwise axis ($z=0$) and therefore, employ a zero-gradient condition. These conditions imply gauge pressure at the open boundaries, see  \cite{tichy_lin_1996,tichy_sawer_SOF_1998, ahmed2021new} for a discussion, and a brief analysis covering the deviation from this condition is presented in section \ref{sec_boundaryCondition}. 
Particularly, in 3D contacts, assuming zero pressure at the lateral boundary (i.e., $z=1$) is common to model standard mechanical compenents, such as bearings, pads, etc. \citep{rastogi1991accounting,bertocchi2013fluid,ccam2023numerical}.
Note that we assume that the environment is pressurized to avoid the presence of cavitation, that is often found in mechanical components at ambient pressure \citep{dowson1979cavitation,gamaniel2021effect,ccam2023numerical}.

\subsection{Reduced-order models}
\label{sec_ReducedOrderModels}
\par The thin-film reduced system of equations is a non-linear and fully-coupled mathematical problem that does not readily yield any analytical solutions. In this section, we introduce two potential reduced-order models that decouple the momentum equations from the scalar constitutive equations, namely the linearized model (LIN) following a first-order perturbation in $De$ and the viscoelastic Reynolds numerical approach (VR) which assumes a weakly perturbed velocity field (extended here to three-dimensional channels \citep{ahmed2021new}).
\subsubsection{$De$-order model}
\par The linearized model is obtained by substituting the expansion
\begin{equation}
    \{p, \boldsymbol{u}, \boldsymbol{\tau}\} = \{p, \boldsymbol{u}, \boldsymbol{\tau}\}^{(0)} + De \{p, \boldsymbol{u}, \boldsymbol{\tau}\}^{(1)} + O(De^2),
\end{equation}
into Eqs. \ref{eqn_thinfilmSys1} and neglecting all terms on the order of $De^2$, where, $\{\cdot \}^{(0)}$, $\{\cdot \}^{(1)}$ denote the leading-order and $De$-order, respectively. Collecting the relevant terms and simplifying, we find the leading-order or Newtonian system of equations
\begin{subeqnarray}
\label{eqn_thinfilmSys1_leadingOrder}
& \frac{\partial u^{(0)} }{\partial x}  + \frac{\partial v^{(0)} }{\partial y} + a\frac{\partial w^{(0)} }{\partial z}= 0,
\\
& \frac{\partial p^{(0)} }{\partial x} = \beta \frac{\partial^2 u^{(0)}}{\partial y^2} + \frac{\partial\tau_{xx}^{(0)} }{\partial x} + \frac{\partial\tau_{yx}^{(0)} }{\partial y} + a \frac{\partial\tau_{zx}^{(0)} }{\partial z},
\\
& \frac{\partial p^{(0)}}{\partial y}  = 0,
\\
& a\frac{\partial p^{(0)}}{\partial z} = \beta \frac{\partial^2 w^{(0)} }{\partial y^2} + \frac{\partial\tau_{xz}^{(0)} }{\partial x} + \frac{\partial\tau_{yz}^{(0)} }{\partial y} + a\frac{\partial\tau_{zz}^{(0)} }{\partial z},
\\
& \tau_{xx}^{(0)} = 0,
\\
& \tau_{xy}^{(0)} = (1 - \beta)\frac{\partial u^{(0)} }{\partial y},
\\
& \tau_{xz}^{(0)} = 0,
\\
& \tau_{yy}^{(0)} = 2(1 - \beta)\frac{\partial v^{(0)} }{\partial y},
\\
& \tau_{yz}^{(0)} = (1 - \beta)\frac{\partial w^{(0)} }{\partial y},
\\
& \tau_{zz}^{(0)} = 0,
\label{eqn_thinfilmSys2}
\end{subeqnarray}
and the $De$-order system of equations
\begin{subeqnarray}
\label{eqn_thinfilmSys1}
& \frac{\partial u^{(1)} }{\partial x}  + \frac{\partial v^{(1)} }{\partial y} + a\frac{\partial w^{(1)}}{\partial z}= 0,
\\
& \frac{\partial p^{(1)}}{\partial x} = \beta \frac{\partial^2 u^{(1)} }{\partial y^2} + \frac{\partial\tau_{xx}^{(1)} }{\partial x} + \frac{\partial\tau_{yx}^{(1)} }{\partial y} + a \frac{\partial\tau_{zx}^{(1)} }{\partial z},
\\
& \frac{\partial p^{(1)} }{\partial y}  = 0,
\\
& a\frac{\partial p^{(1)} }{\partial z} = \beta \frac{\partial^2 w^{(1)} }{\partial y^2} + \frac{\partial\tau_{xz}^{(1)} }{\partial x} + \frac{\partial\tau_{yz}^{(1)} }{\partial y} + a\frac{\partial\tau_{zz}^{(1)} }{\partial z },
\\
& \tau_{xx}^{(1)} = -De \big(\frac{D\tau_{xx}^{(0)}}{Dt} -2\tau_{xy}^{(0)}\frac{\partial u^{(0)} }{\partial y} - 2\tau_{xx}^{(0)}\frac{\partial u^{(0)} }{\partial x} - 2a\frac{\partial u^{(0)} }{\partial z}\tau_{xz}^{(0)} \big) ,
\\
\nonumber
& \tau_{xy}^{(1)} = -De \big(\frac{D\tau_{xy}^{(0)}}{Dt} -\tau_{yy}^{(0)}\frac{\partial u^{(0)} }{\partial y} - \tau_{xx}^{(0)}\frac{\partial v^{(0)}}{\partial x} - \frac{\partial u^{(0)}}{\partial x}\tau_{xy}^{(0)} - \frac{\partial v^{(0)} }{\partial y}\tau_{xy}^{(0)} - a\frac{\partial u^{(0)}}{\partial z}\tau_{yz}^{(0)} 
\\
& - a\frac{\partial v^{(0)}}{\partial z}\tau_{xz}^{(0)} \big) + (1 - \beta)\frac{\partial u^{(1)} }{\partial y},
\\
\nonumber
& \tau_{xz} + De \big(\frac{D\tau_{xz}}{Dt} -\tau_{xz}\frac{\partial u}{\partial x} - \tau_{xx}\frac{\partial w}{\partial x} -\tau_{yz}\frac{\partial u}{\partial y} -\tau_{xy}\frac{\partial w}{\partial y}
\\
& -a\tau_{zz}\frac{\partial u}{\partial z} -a\tau_{xz}\frac{\partial w}{\partial z} \big) = 0,
\\
& \tau_{yy} + De \big(\frac{D\tau_{yy}}{Dt} -2\tau_{xy}\frac{\partial v}{\partial x} - 2\tau_{yy}\frac{\partial v}{\partial y} - 2a\frac{\partial v}{\partial z}\tau_{yz} \big) = 2(1 - \beta)\frac{\partial v}{\partial y},
\\
\nonumber
& \tau_{yz} + De \big(\frac{D\tau_{yz}}{Dt} -\tau_{xz}\frac{\partial v}{\partial x} - \tau_{xy}\frac{\partial w}{\partial x} - \tau_{yz}\frac{\partial v}{\partial y} - \tau_{yy}\frac{\partial w}{\partial y} - a\tau_{zz}\frac{\partial v}{\partial z} 
\\
& - a\tau_{yz}\frac{\partial w}{\partial z}\big) = (1 - \beta)\frac{\partial w}{\partial y},
\\
& \tau_{zz} + De \big(\frac{D\tau_{yy}}{Dt} - 2\tau_{xz}\frac{\partial w}{\partial x} - 2\tau_{xy}\frac{\partial w}{\partial y} - 2a\frac{\partial w}{\partial z}\tau_{zz} \big) = 0.
\label{eqn_thinfilmSys2_leadingOrder}
\end{subeqnarray}
The solution to the leading-order system of Eqs. (subject to the conditions Eqs. \ref{eqns_boundaryConds1}) gives
give the Newtonian pressure, and velocity field \citep{szeri2010fluid},
\begin{subeqnarray}
\label{eqns_Newtonian_sys1}
& \frac{\partial }{ \partial x}\big( \frac{h^3}{12}\frac{\partial p}{\partial x} \big) +a^2 \frac{\partial }{ \partial z}\big( \frac{h^3}{12}\frac{\partial p}{\partial z} \big)
= \frac{1}{2}\frac{\partial h}{\partial x},
\\
& u = \frac{1}{2}\frac{\partial p}{\partial x}(y^2 - yh) + (1 - \frac{y}{h}),
\\
& v = -\frac{y^3}{6}\big( \frac{\partial^2 p}{\partial x^2} + a^2 \frac{\partial^2 p}{\partial z^2} \big) + \frac{y^2}{2}\big( \frac{\partial^2 p}{\partial x^2}\frac{h}{2} + a^2 \frac{\partial^2 p}{\partial z^2}\frac{h}{2} + \frac{1}{2}\frac{\partial p}{\partial x}\frac{\partial h}{\partial x} + \frac{a^2}{2}\frac{\partial p}{\partial z}\frac{\partial h}{\partial z}\big),
\\
\label{eqns_Newtonian_sys2}
& w = \frac{a}{2}\frac{\partial p}{\partial z}(y^2 - yh).
\end{subeqnarray}
It is important to note that Eqs. \ref{eqns_Newtonian_sys1} do not yield an analytical solution for even the simplest surface profiles. The matter is made worse when factoring in discontinuities along the surface, such as a textured pocket \citep{schuh2017design}. There are two limiting cases for Eq. \ref{eqns_Newtonian_sys1}(a); (i) $a << 1$, describing an infinitely wide channel and (ii) $a >> 1$, which models a slender channel.  
\par The solution to the $De$-order system requires additional effort owing to the increase in the number of equations and terms thereof. The final system is a lengthy set of equations for the pressure, and velocity components. For the sake of brevity, we present only the $De$-order pressure equation
\begin{subeqnarray}
\label{chap_PF_eqn_DeborahOrder1}
\nonumber
& \frac{\partial}{\partial x}\big( \frac{h^3}{12}\frac{\partial p^{(1)}}{\partial x}\big) 
+
a^2 \frac{\partial}{\partial z}\big( \frac{h^3}{12}\frac{\partial p^{(1)}}{\partial z}\big) 
=
\\
\nonumber
& - a^4 (1-\beta)
\bigg( \frac{1}{48}  h^{5} \frac{\partial p^{(0)}}{\partial z}  \frac{\partial^{3} {p^{(0)}}}{\partial z^{3}}
+ \frac{1}{48} h^{5} (\frac{\partial^{2} {p^{(0)}} }{\partial z^{2}} )^{2}
+ \frac{7}{48}  h^{4} \frac{\partial h}{\partial z}  \frac{\partial {p^{(0)}}}{\partial z}  \frac{\partial^{2} {p^{(0)}}}{\partial z^{2}} 
\\
\nonumber
& + \frac{1}{48} h^{4} \frac{\partial^{2} h}{\partial z^{2}}  (\frac{\partial p^{(0)}}{\partial z} )^{2}
+ \frac{1}{12} h^{3} (\frac{\partial h}{\partial z} )^{2} (\frac{\partial {p^{(0)}}}{\partial z} )^{2} \bigg)
\\
\nonumber
& -  a^{2} (1-\beta) \bigg(
\frac{h^5}{48} \frac{\partial p^{(0)}}{\partial x} \frac{\partial^3 p^{(0)}}{\partial z^2 \partial x}
+
\frac{h^5}{40} \frac{\partial^2 p^{(0)}}{\partial x^2} \frac{\partial^2 p^{(0)}}{\partial z^2}
+
\frac{h^5}{48} \frac{\partial p^{(0)}}{\partial z} \frac{\partial^3 p^{(0)}}{\partial x^2 \partial z}
+
\\
\nonumber
& \frac{h^5}{60} \big( \frac{\partial^2 p^{(0)}}{\partial x \partial z}\big)^2 
+
\frac{h^4}{12} \frac{\partial h}{\partial x} \frac{\partial p^{(0)}}{\partial x} \frac{\partial^2 p^{(0)}}{\partial z^2}
+
\frac{h^4}{16} \frac{\partial h}{\partial x} \frac{\partial p^{(0)}}{\partial z} \frac{\partial^2 p^{(0)}}{\partial x\partial z } 
-
\frac{h^3}{24} \frac{\partial h}{\partial x} \frac{\partial^3 p^{(0)}}{\partial z^2 \partial x} +  
\\
\nonumber
& +
\frac{h^4}{16} \frac{\partial h}{\partial z} \frac{\partial 
p^{(0)}}{\partial x} \frac{\partial^2 p^{(0)}}{\partial x \partial z} 
 +
\frac{h^4}{12} \frac{\partial h}{\partial z} \frac{\partial^2 
p^{(0)}}{\partial x^2} \frac{\partial p^{(0)}}{\partial z} 
 +
\frac{h^3}{6} \frac{\partial h}{\partial x} \frac{\partial h}{\partial z} \frac{\partial p^{(0)}}{\partial x} \frac{\partial p^{(0)}}{\partial z} 
-
\frac{h^2}{8}\frac{\partial p^{(0)}}{\partial z} \frac{\partial^2 p^{(0)}}{\partial x \partial z} 
\bigg)
\\
\nonumber
& + (1-\beta ) \bigg( \frac{h^5}{48} \big( \frac{\partial p^{(0)}}{\partial x} \frac{\partial^3 p^{(0)}}{\partial x^3} + \frac{\partial^2 p^{(0)}}{\partial x^2} \big) 
+
\frac{7 h^4}{48} \frac{\partial h}{\partial x} \frac{\partial p^{(0)}}{\partial x} \frac{\partial^2 p^{(0)}}{\partial x^2} 
+ 
\frac{h^4}{48} \frac{\partial^2 h}{\partial x^2} \frac{\partial p^{(0)}}{\partial x} \big( \frac{\partial p^{(0)}}{\partial x}\big)^2 
+
\\
& \frac{h^3}{12} \big( \frac{\partial h}{\partial x} \big)^2 \big( \frac{\partial p^{(0)}}{\partial x} \big)^2 
-
\frac{h^3}{24} \frac{\partial^3 p{(0)}}{\partial x^3} 
- 
\frac{h^2}{8} \frac{\partial h}{\partial x} \frac{\partial^2 p^{(0)}}{\partial x^2} 
-
\frac{1}{12} \frac{\partial^2 h}{\partial x^2}\bigg) .
\label{chap_PF_eqn_lin3d}
\end{subeqnarray}
The expressions in both the systems, i.e., the leading-order and the $De$-order, involve higher powers of $a$. Furthermore, third-order, and mixed partial derivatives of the leading-order pressure emerge. If $a \rightarrow 0$, then the linearized system of equations for a two-dimensional problem are easily retrieved. 
\subsubsection{Viscoelastic Reynolds approach}
\par An alternative numerical approach, the VR method, significantly alleviates these issues. In this numerical treatment, we enforce the perturbation only in the velocity field, which implies a weak dependence of the flow kinematics due the presence of the polymers. This has been reported for several cases involving thin film lubrication. Hence, we assume $\boldsymbol{u} = \boldsymbol{u}^{(0)} + De \boldsymbol{u}^{(1)} + O(De^2)$. This assumption implies that the velocity field responds linearly to increasing viscoelasticity (as $De$ increases), whereas the stress components, in particular along the streamwise and spanwise normal directions, can vary strongly and naturally, following their inherent non-linearity. More importantly, with VR we decouple the momentum equation from the stress constitutive relation, since the velocity field is known \textit{apriori} from LIN. The Reynolds type equation including the polymeric stress components is obtained, see Appendix \ref{app_VR3D},
\begin{subeqnarray}
\label{eqn_VR1}
&\frac{\partial}{\partial x} \big( \frac{H^3}{12} \frac{\partial p}{\partial x} \big) + a^2 \frac{\partial}{\partial z} \big( \frac{H^3}{12} \frac{\partial p}{\partial z} \big) =  \frac{\partial \mathcal{F}}{\partial x} + \frac{\partial \mathcal{G}}{\partial z}  ,
\\
& \mathcal{F} = \frac{\beta}{2}H + \frac{H^3}{2}\overline{[(Y-Y^2)\frac{\partial \tau_{xx}}{\partial x}]} + \frac{H^2}{2}\overline{[(2Y-1)\tau_{xy}]} + a\frac{H^3}{2}\overline{[(Y-Y^2)\frac{\partial \tau_{xz}}{\partial z}]},
\\
& \mathcal{G} = a^2\frac{H^3}{2}\overline{[(Y-Y^2)\frac{\partial \tau_{zz}}{\partial z}]} + \frac{H^2}{2}\overline{[(2Y-1)\tau_{zy}]} + a\frac{H^3}{2}\overline{[(Y-Y^2)\frac{\partial \tau_{xz}}{\partial x}]},
\label{eqn_VR2}
\end{subeqnarray}
where, $\overline{[\dots]} = \int_{0}^{1} [\dots] dY$, and $Y = \frac{y}{h(x,z)}$. For an extremely wide channel, $a << 1$, Eqs. \ref{eqn_VR1} reduce to the two-dimensional case. In addition, if $h(0, z) = h(1, z)$, then $\boldsymbol{u}^{(1)} = 0$ and, based on the expressions obtained from the application of the reciprocal theorem \citep{boyko2021reciprocal}), the higher-order expansions for $De$ (for a two-dimensional pressure driven flow problem) also vanish, greatly simplifying the ensuing numerical procedure. 
\par We resort to a numerical treatment via the finite difference method (FDM) of the leading-order (Eqs. \ref{eqn_thinfilmSys1}), the $De$-order (Eqs. \ref{chap_PF_eqn_lin3d}) systems, and the VR equations (Eqs. \ref{eqn_VR1}) subject to the boundary conditions given by Eqs. \ref{eqns_boundaryConds1}. Details regarding the numerical procedure are presented in \cite{ahmed2023modeling}. Briefly, we employ a curvilinear transformation for the velocity given in Appendix \ref{app_curvilinear}, see \citep{boyko2024flow} for a detailed treatment of all physical quantities, allowing a flux conserving implicit treatment of the stress advection, presented in Appendix \ref{app_num_meth}, and finally, the relevant differential operators in the three-dimensional case are summarized in Appendix \ref{app_coordinate_transformation}.

\section{Results and discussion}
\par In this section, we examine the influence of viscoelasticity on the load carrying capacity per unit width $F_\ell^* = 2\frac{1}{\ell_z} \int_{0}^{\ell_x}\int_{0}^{\ell_z} \sigma_{yy}^* dx^* dz^*$ which, upon re-scaling using Eqs. \ref{eqn_geo_scaling} and Eqs. \ref{eqn_nonDimSys1_stress}, reduces to $F_\ell = \int_{0}^{1}\int_{0}^{1} p dx dz$ (since $\tau_{yy}$ is O($\epsilon^2$), while $p$ is O($1$)). The dependency of $F_\ell$ on the polymer elasticity measured via $De$ and the additional influence of the channel's spanwise variation measured via the aspect ratio $a$, are examined using the VR approach, and the $De$-order perturbed model (LIN). The choice of the channel surface geometry is given in section \ref{sec_channelGeo}, and some consideration on the different boundary conditions for the stress are provided in section \ref{sec_boundaryCondition}. The solvent concentration in terms of viscosity is set to $\beta=0.8$ all along the paper.
\subsection{Channel geometry}
\label{sec_channelGeo}
\par The success of the numerical solution of the reduced constitutive relations Eqs. \ref{eqn_thinfilmSys1} and Eqs. \ref{eqn_VR1} is strongly dependent on the channel geometry. For sharply varying surfaces, such as pocketed textures and contractions, convergence of the numerical procedure becomes a challenge \citep{schuh2017design}. However, the main mechanisms involved in a viscoelastic flow can also be studied via a smoothly varying channel surface profile, avoiding several numerical hurdles \citep{hinch2024fast}. 
\par
\begin{figure}
\centering
\includegraphics[scale = 0.15]{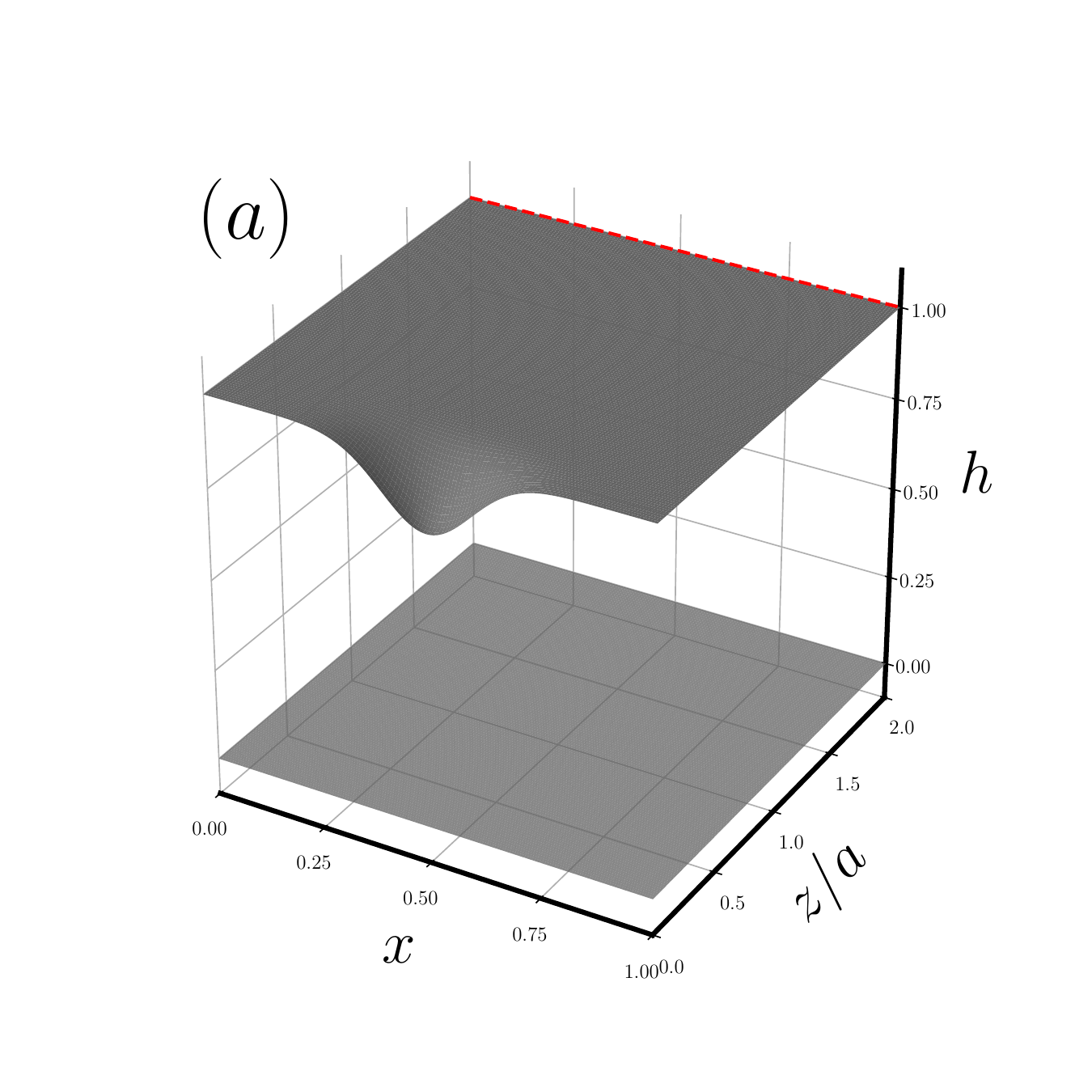}
\includegraphics[scale = 0.15]{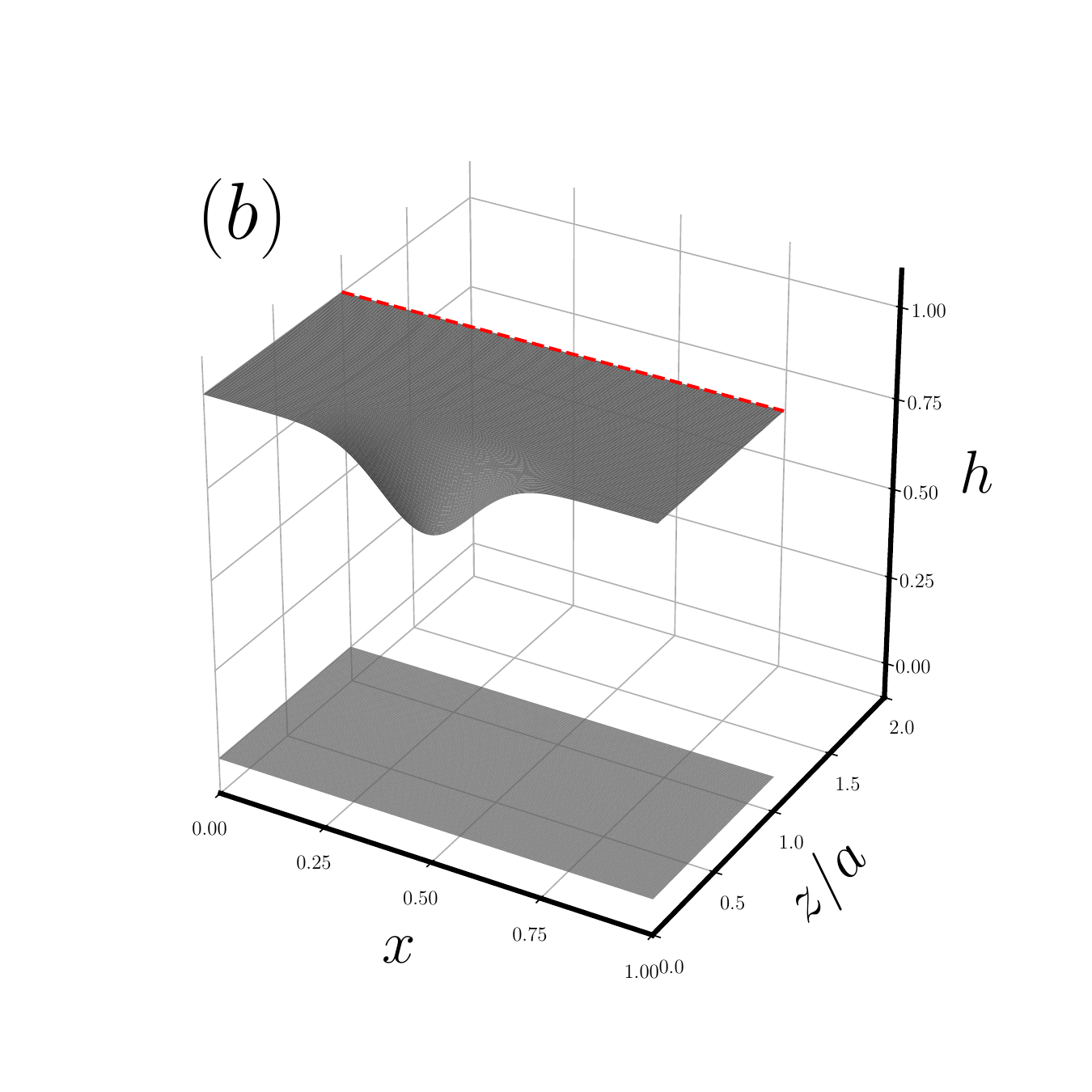}
\includegraphics[scale = 0.15]{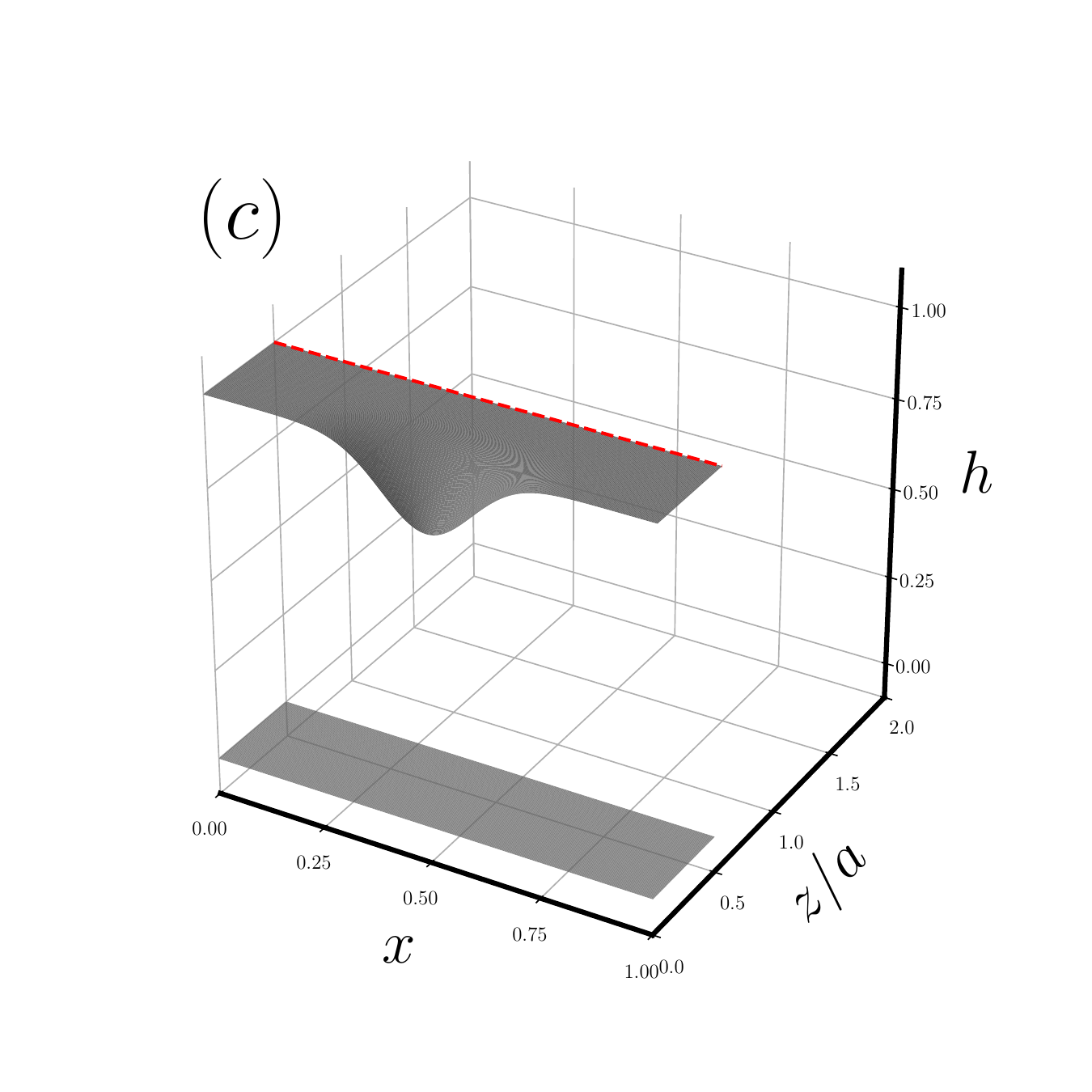}
\includegraphics[scale = 0.15]{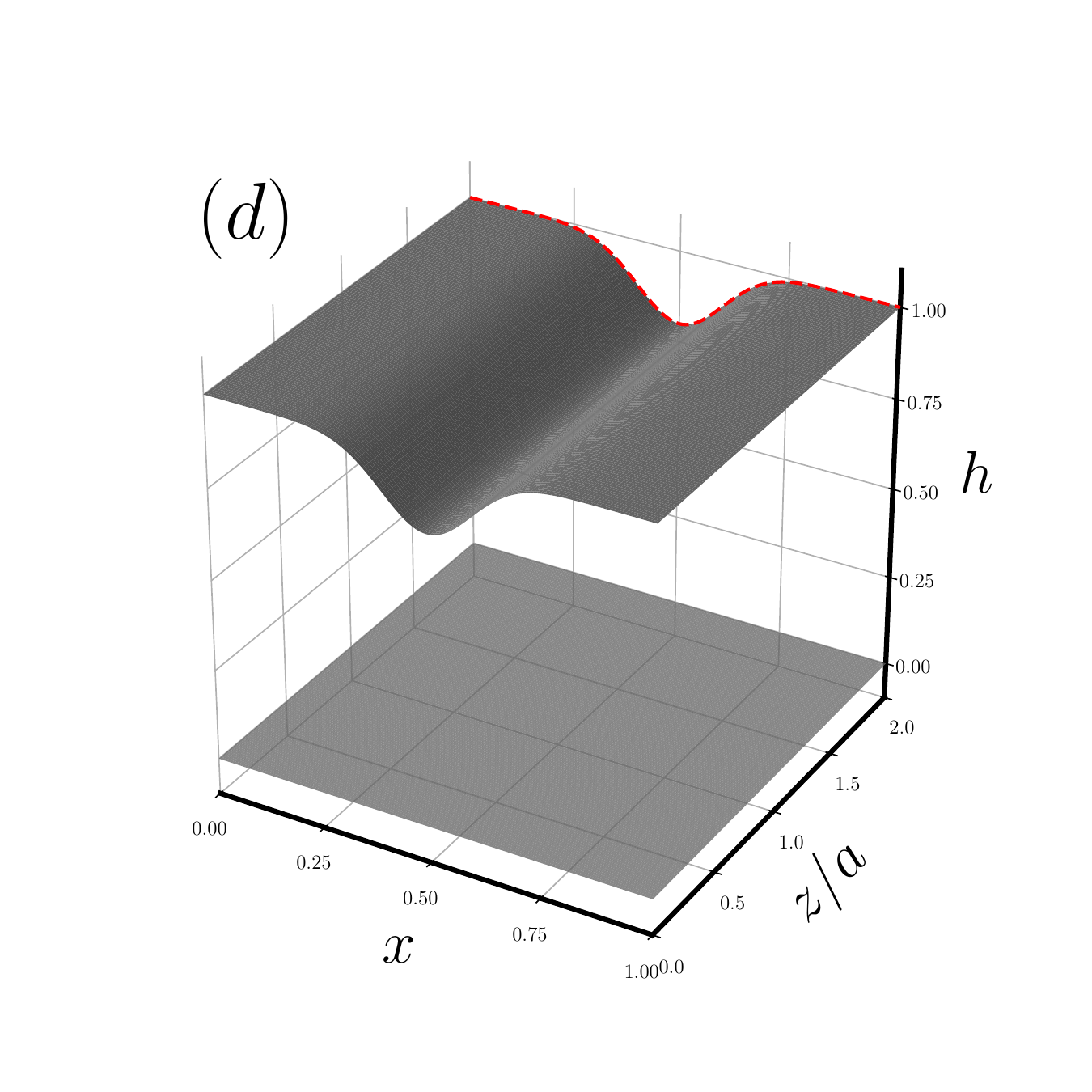}
\includegraphics[scale = 0.15]{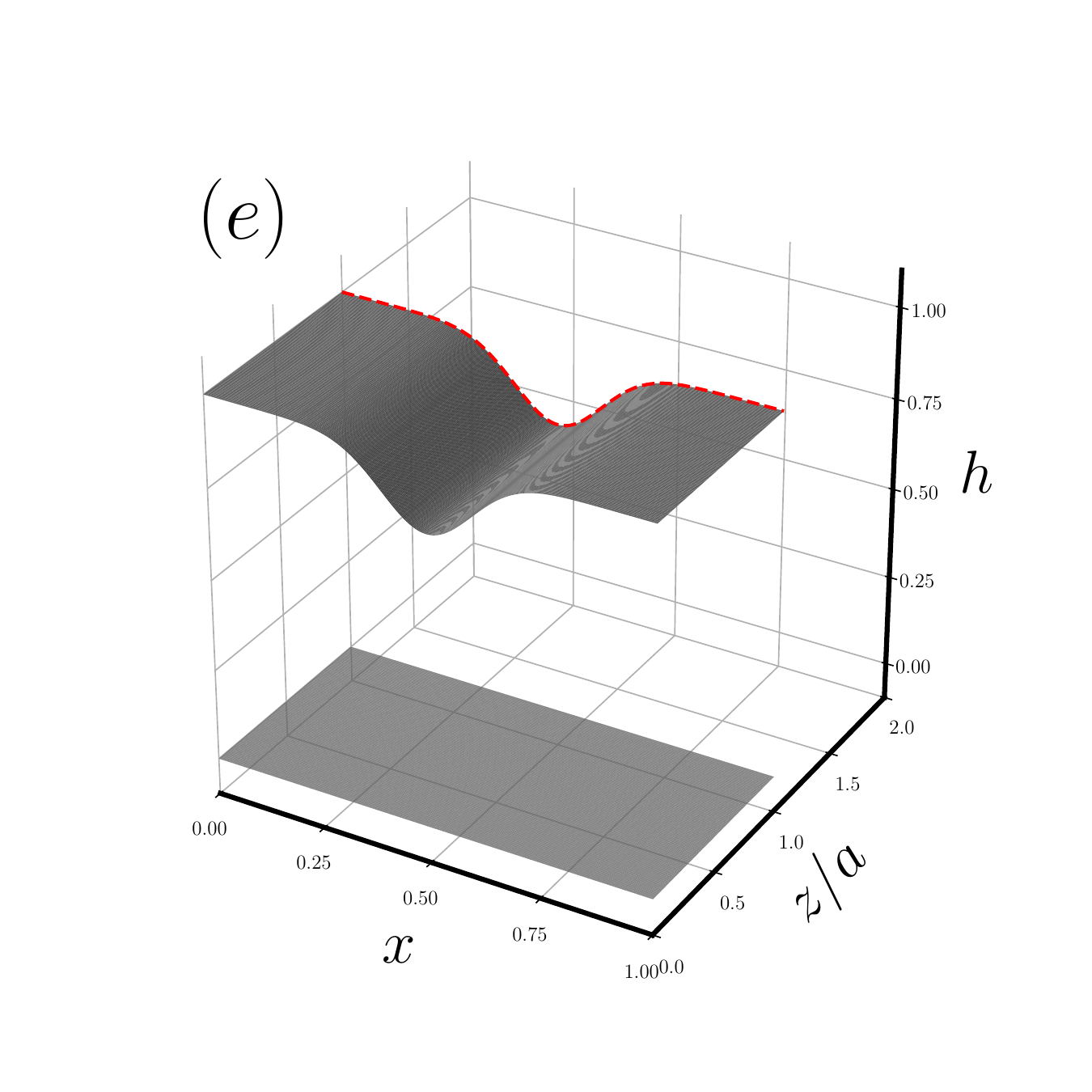}
\includegraphics[scale = 0.15]{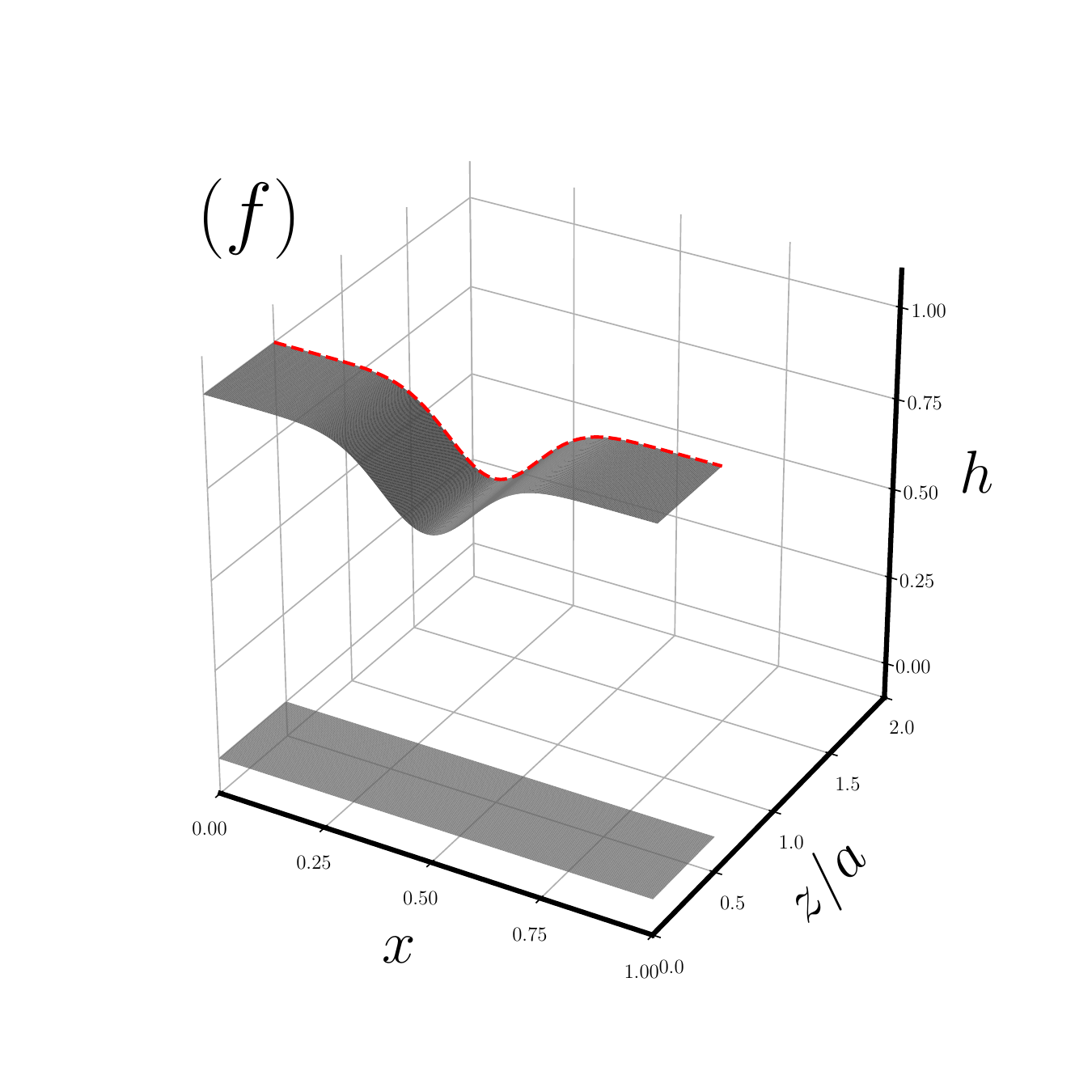}
\caption{The channel surface height variation for three different values of the aspect ratio, $a = 1/2$, $a = 1$ and $a=2$ for (a)-(c) the spanwise-varying (Eq. \ref{eqn_gauss_full}) and (d)-(f) the extruded (Eq. \ref{eqn_gauss_extruded}). The channel depth is $d = 0.2$, while the spread is $s = 0.1$.}
\label{fig_gauss}
\end{figure}

 We consider two surface configurations; a spanwise-varying surface modeled via a symmetric Gaussian distribution that allows controlling the spread of the curved region and its depth,
\begin{equation}
\label{eqn_gauss_full}
h = 1 - d \exp{\bigg( -\frac{( x - X_c)^2}{s^2}-\frac{(z - Z_c)^2}{s^2 a^2} \bigg) },
\end{equation}
where, $d$ is the depth, $s$ is the spread, along the spanwise and streamwise directions, and $X_c$ and $Z_c$ represent the coordinates of the center. Note that in this work $s$, $X_c$ and $Z_c$ are always set to $s = 0.1$, $X_c = 0.5$ and $Z_c = 0$. Eq. \ref{eqn_gauss_full} represents a surface protrusion that is sensitive to the channel width and has gradients along both the streamwise (sliding) and spanwise directions, and is depicted in Figs. \ref{fig_gauss}(a)-(c). 
\par The extruded case approximates channels that are simply an extrusion of the cross-section along the streamwise direction, i.e., the $(x,y)$-plane, making the surface height $h$ independent of $z$. These types of channels are found in  cylindrical roller bearings, slider bearings, line contacts, etc. The surface height for this case is obtained by modifying Eq.
\ref{eqn_gauss_full}
\begin{equation}
\label{eqn_gauss_extruded}
h = 1 - d \exp{\bigg( -\frac{(x - X_c)^2}{s^2} \bigg) },
\end{equation}
and the resulting surface profiles are depicted in Figs. \ref{fig_gauss}(d)-(f). 

\subsection{Comparison between LIN and VR}
\label{sec_LIN_vs_VR}
\par
\begin{figure}
\centering
\includegraphics[scale = 0.2]{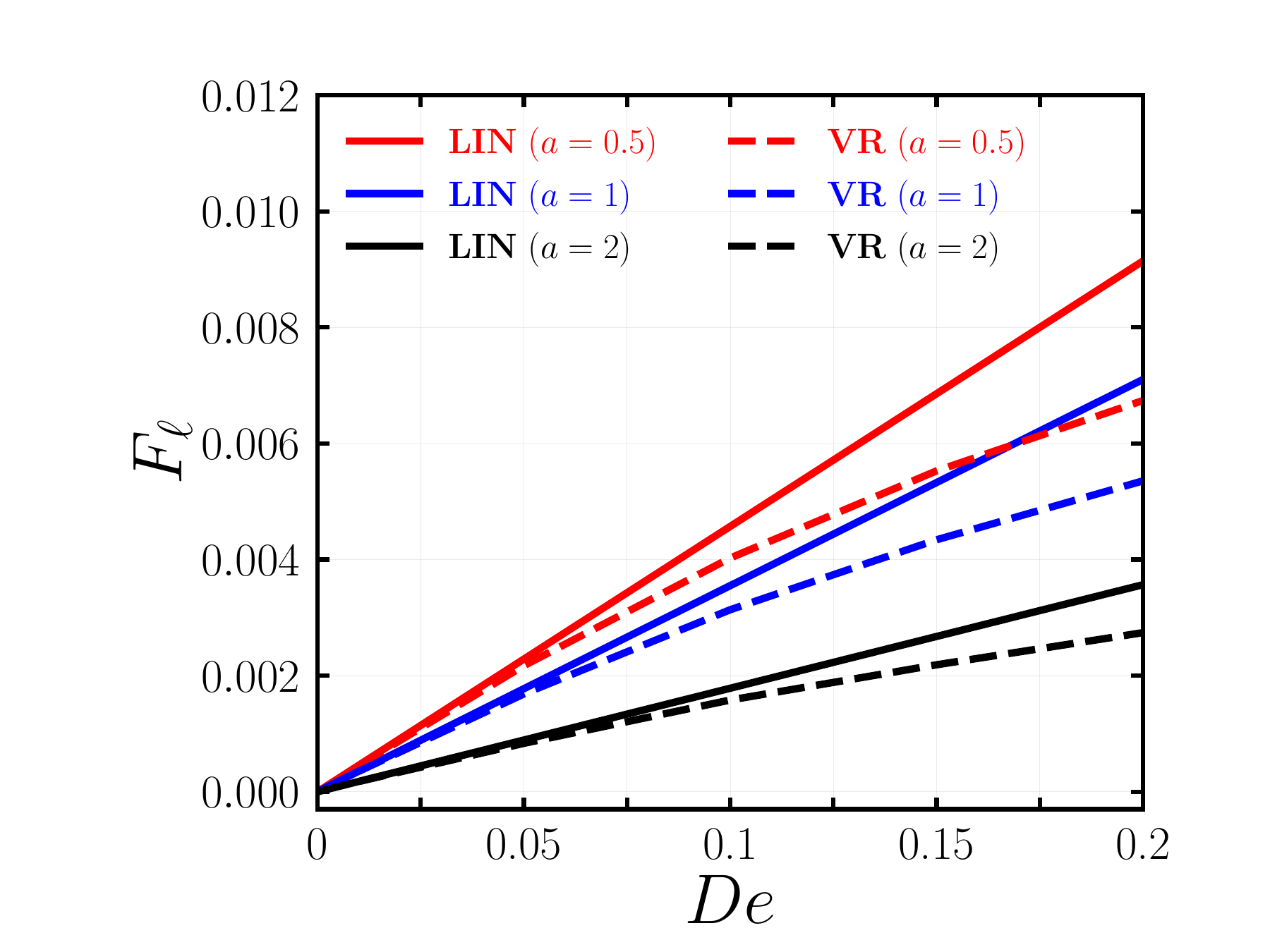}
\caption{The variation in the load carrying capacity per unit width, predicted by the $De$-order linear (LIN) model and the VR approach, for the extruded surface versus the Deborah number for three different aspect ratios, using $ d=0.2$.}
\label{fig_load_LIN_VR}
\end{figure}

 Typically, the pressure will increase when the viscosity of the lubricant is increased, leading to larger shear stresses and, consequently, higher pressures to overcome the increased resistance to flow. As such, a purely Newtonian fluid ($De = 0$) gives an antisymmetric pressure distribution along $x$ such that $\int_{0}^{1}pdx = 0$ with $p > 0$ ($p < 0$) for converging (diverging) sections of the channel. Contrary to this, the viscoelastic mixture (for constant viscosity) delivers a net positive load. In Fig. \ref{fig_load_LIN_VR}, we compare the load carrying capacity versus the Deborah number for different channel aspect ratios using $h=h(x)$, using $d=0.2$, and $\beta = 0.8$.  
The simplified models (LIN and VR) predict an expected increase in the net force as we enhance the polymer elastic contribution \citep{ahmed2021new, ahmed2023modeling}. Solving for larger $De$ did not change the trend given in Fig. \ref{fig_load_LIN_VR}, but increased the numerical complexity, particularly for the limiting cases of large and small $a$. Such limitations were not present for the $De$-order model.  However, despite the computational advantage offered by the linear model, a noticeable deviation from VR is observed at large $De$.
\par
\begin{figure}
\centering
\includegraphics[scale = 0.2]{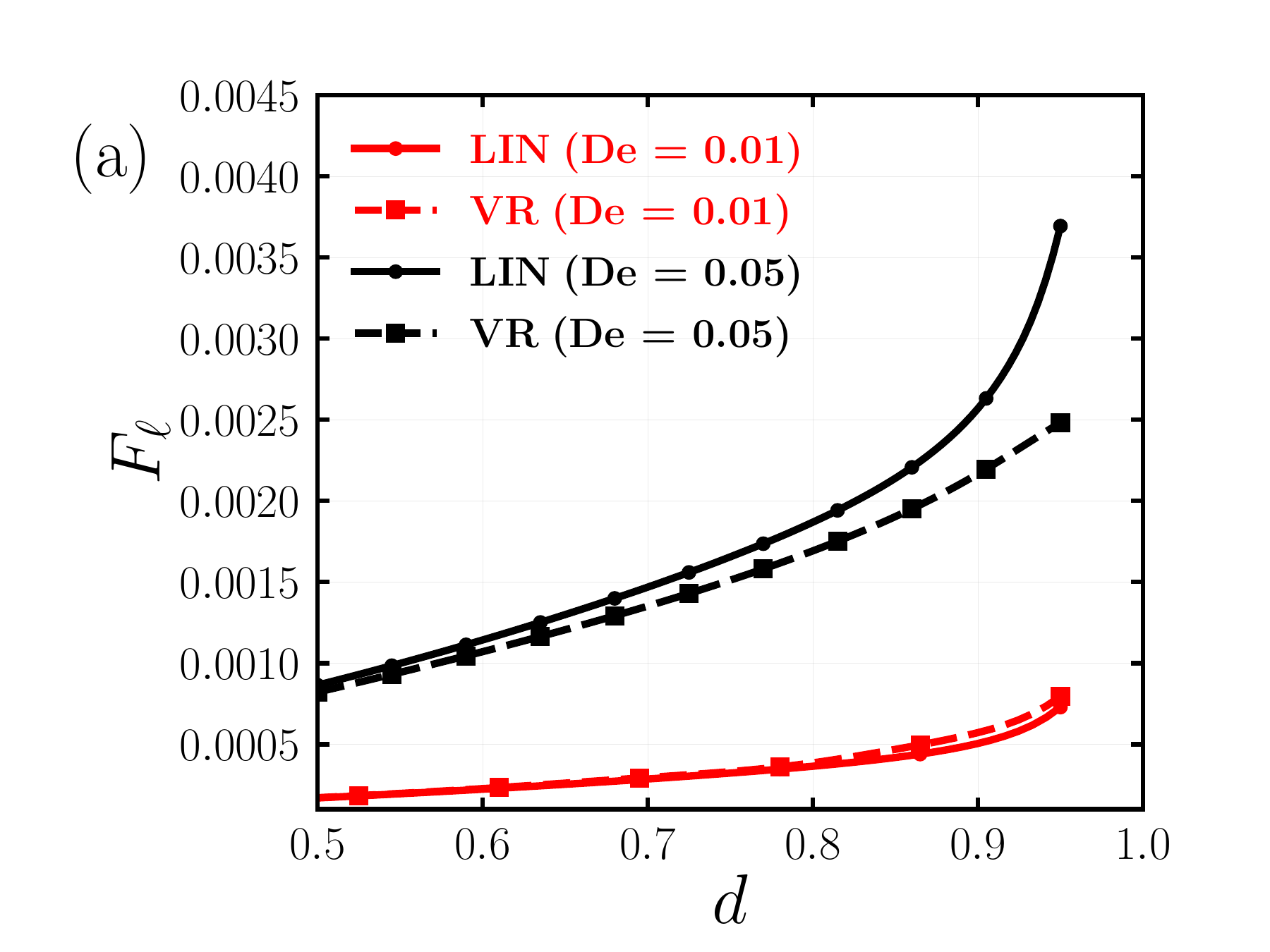}
\includegraphics[scale = 0.2]{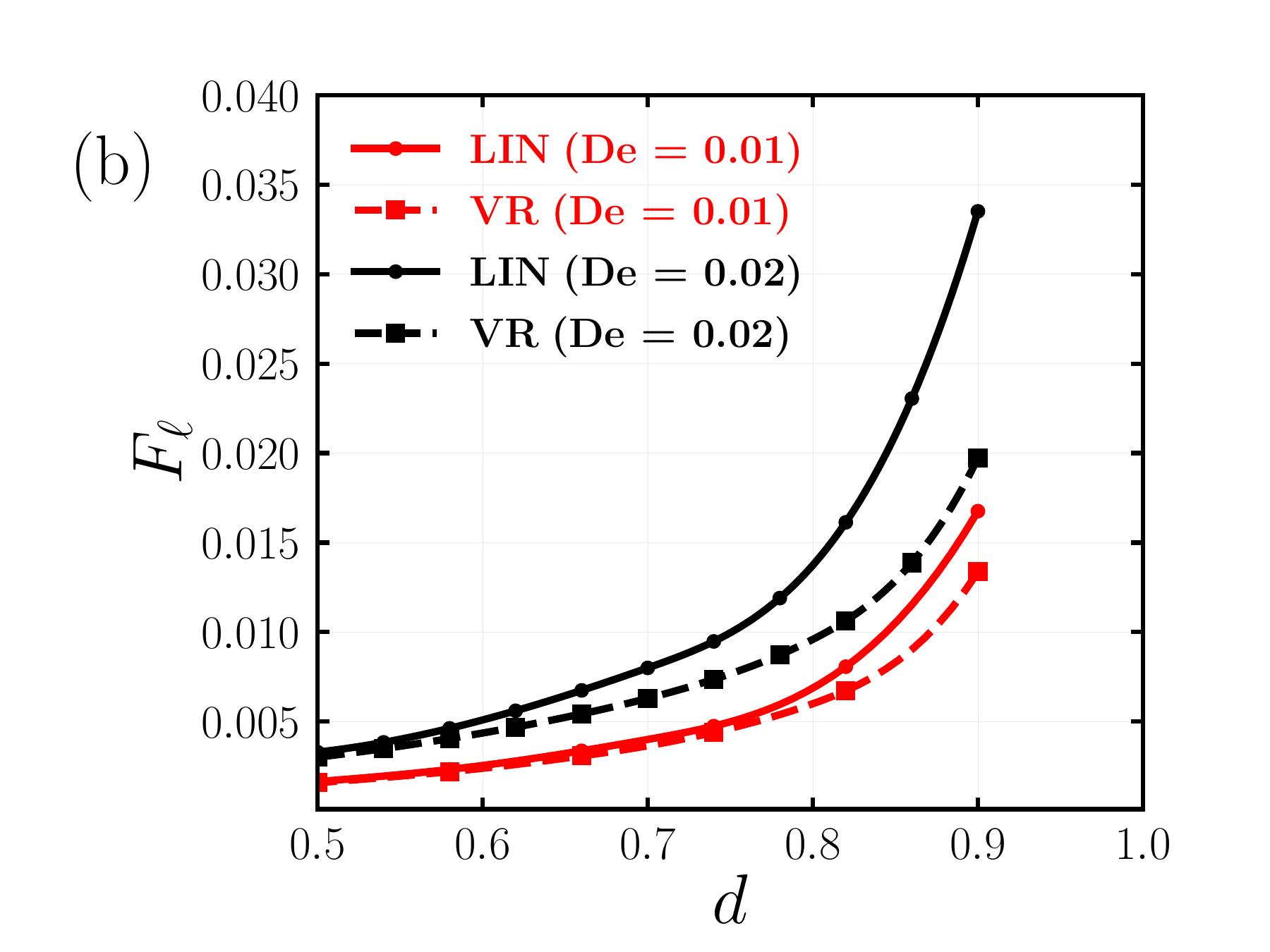}
\caption{The variation in the load carrying capacity per unit width, predicted by the $De$-order linear (LIN) model and the VR approach, for (a) the spanwise-varying and (b) the extruded channel versus the channel depth, using $a=1$.}
\label{fig_LIN_VR_d}
\end{figure}

 The decreased accuracy versus increasing $De$ is not unexpected and has been similarly observed for the equivalent two-dimensional sliding cases. However, in thin-film lubrication, the accuracy of the linear perturbation approach also depends strongly on the channel surface gradients which do not explicitly appear in the Deborah number. 
We examine the variation in the load versus the channel depth, as shown in Fig. \ref{fig_LIN_VR_d}, for the two different channel configurations.
The steepening of the channel implies an increase in the average shear rate distribution which directly contributes to an enhancement in the total stretch of the polymer additives. Regardless of whether the surface varies along the spanwise direction, a deviation in the estimated load is observed as the depth increases. In fact, for the extruded channel profile, the difference manifests strongly even for $De = 0.02$. 
\par
\begin{figure}
\centering
\includegraphics[width=0.5\linewidth]{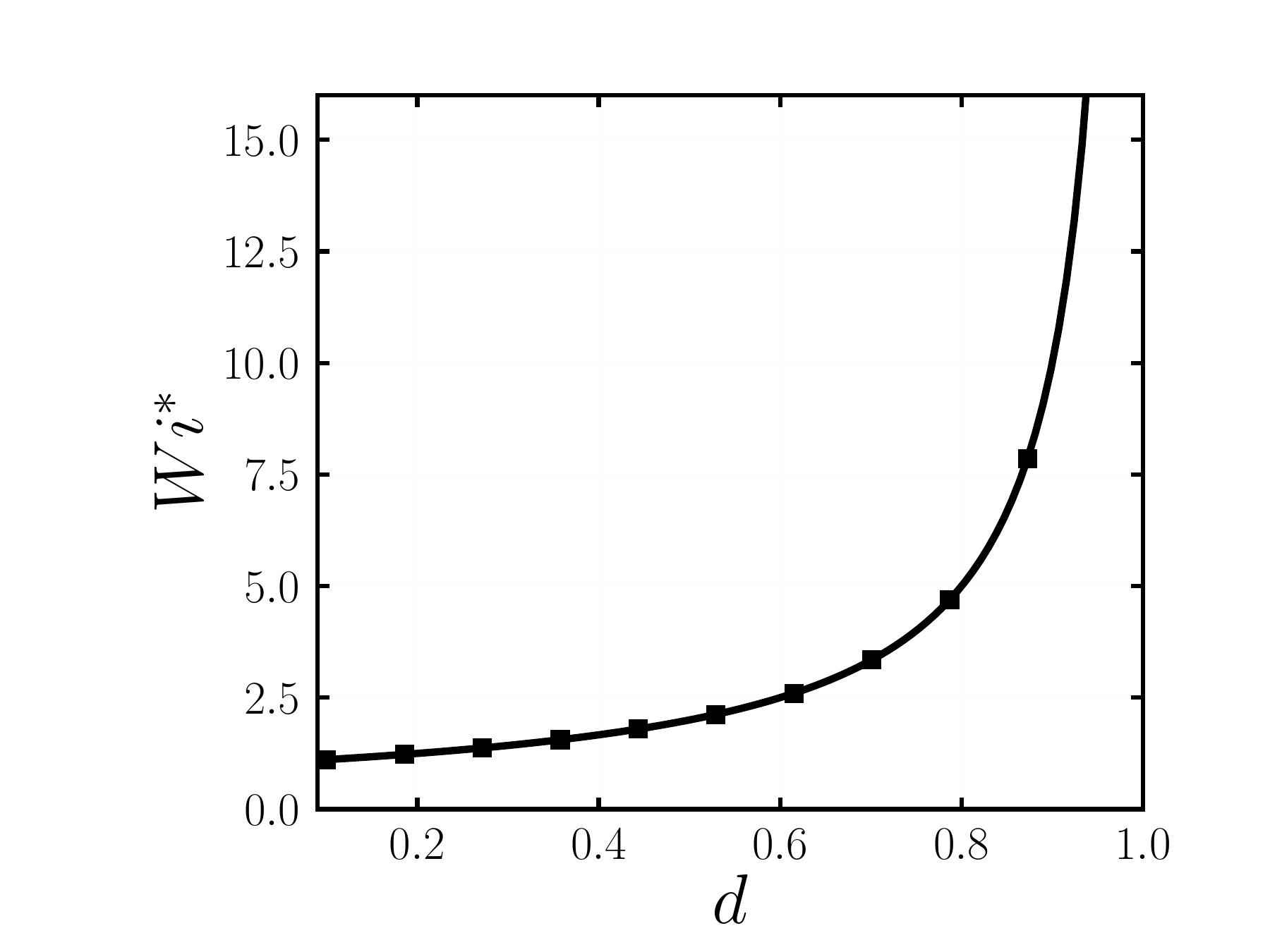}
\caption{The effective Weissenberg number (Eq. \ref{eqn_Wi_eff}) versus the depth $d$.}
\label{fig_Wi_eff}
\end{figure}

 In the most general case, the manifestation of viscoelasticity depends strongly on both the Deborah and Weissenberg numbers which reduce to the relation $De = \epsilon Wi$ in the thin-film limit, and therefore, the Deborah number typically remains less than one (since $\epsilon << 1$).
Interestingly, the first-order perturbation model in $De$ operates in the linearly viscoelastic limit, but as the channel surface steepens (increasing $d$) we observe a discrepancy between the VR and LIN even for small magnitudes of $De$. We attribute this to the increase in the local Weissenberg number
\begin{equation}
\label{eqn_Wi_eff}
Wi^* = \frac{\lambda U_0}{(1 - d)H_0},
\end{equation}
where, the reference film height $(1 - d)H_0$ is the minimum of the channel height. For small values of $d$, the maximum and minimum channel height are on the same order of magnitude and either choice is suitable.
However, as $d$ increases, $Wi^*$ rises by nearly an order of magnitude, as shown in Fig. \ref{fig_Wi_eff} for the spanwise varying and the extruded channels. Notice the exponential increase in $Wi^*$ as $d$ increases.
An increase in $Wi^*$ is a sympton of stronger viscoelastic effects also at low $De$. Therefore, at large $d$  we can expect a significant non-linearity for smaller Deborah numbers with respect to the low $d$ case. 
Alternative scaling approaches do however extract the $Wi$ but at the expense of the Deborah number. For instance, when applying the shear scaling for all the stress components, i.e., $\tau_{ij}^* = \frac{\eta_0 U_0}{H_0} \tau_{ij}$, in the constitutive relation and the momentum equations \citep{akyildiz2004viscoelastic,venkatesh2022peeling,abbaspur2023analytical}.

\subsection{Load carrying capacity}
\textcolor{black}{In this section, we examine first the influence of the channel aspect ratio on the load carrying capacity for both the spanwise varying and the extruded channel types (see section \ref{sec_channelGeo}). Furthermore, we delineate the contributions from the different components of the polymer extra stress tensor, paying particular attention to the dominant streamwise normal stress component. The proceeding analysis is conducted only via the VR approach owing to the inaccuracy of the first-order model for larger $De$, as demonstrated in section \ref{sec_LIN_vs_VR}.}
\subsubsection{Effect of the aspect ratio}
\par
\begin{figure}
\centering
\includegraphics[scale = 0.2]{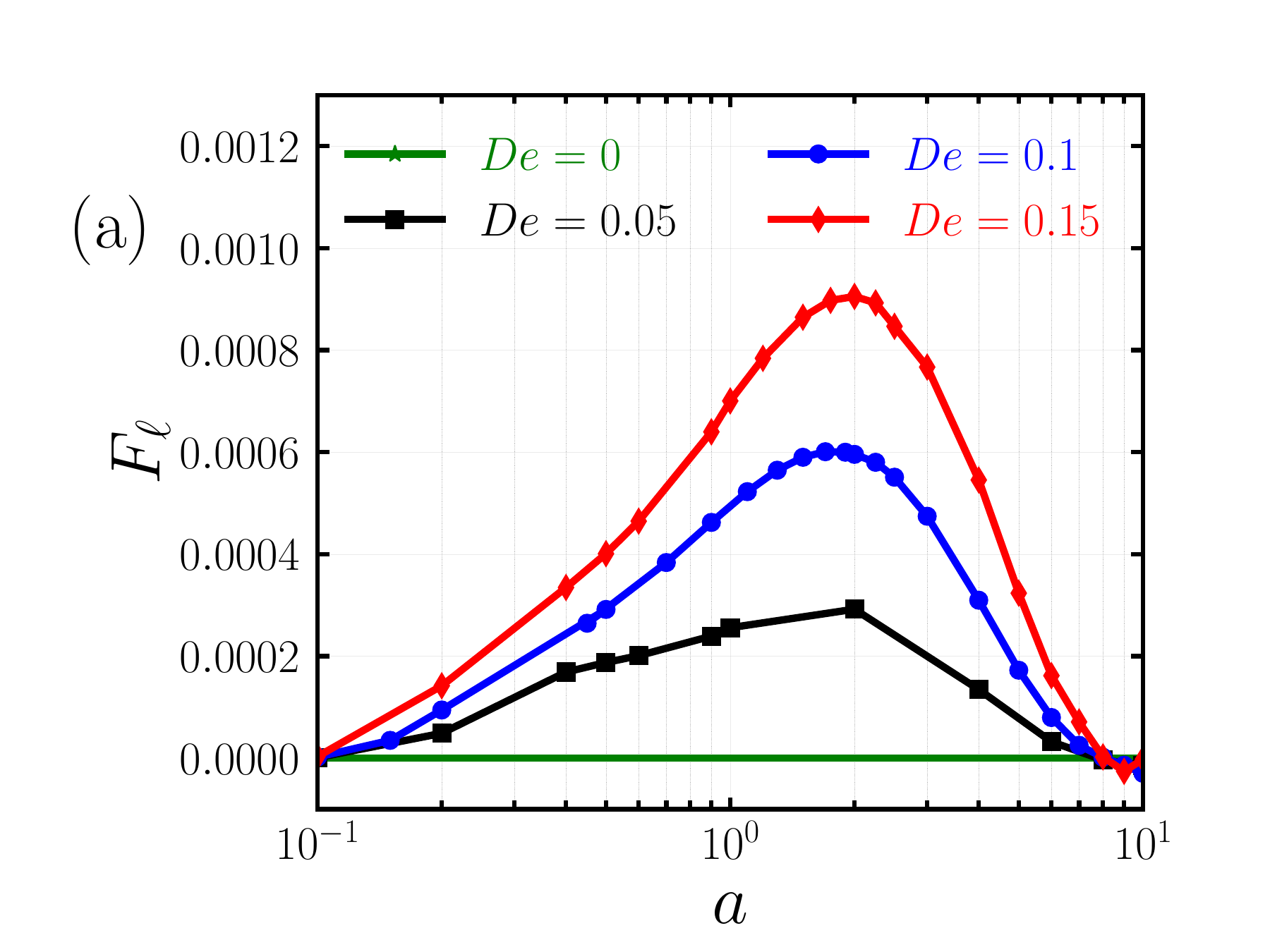}
\includegraphics[scale = 0.2]{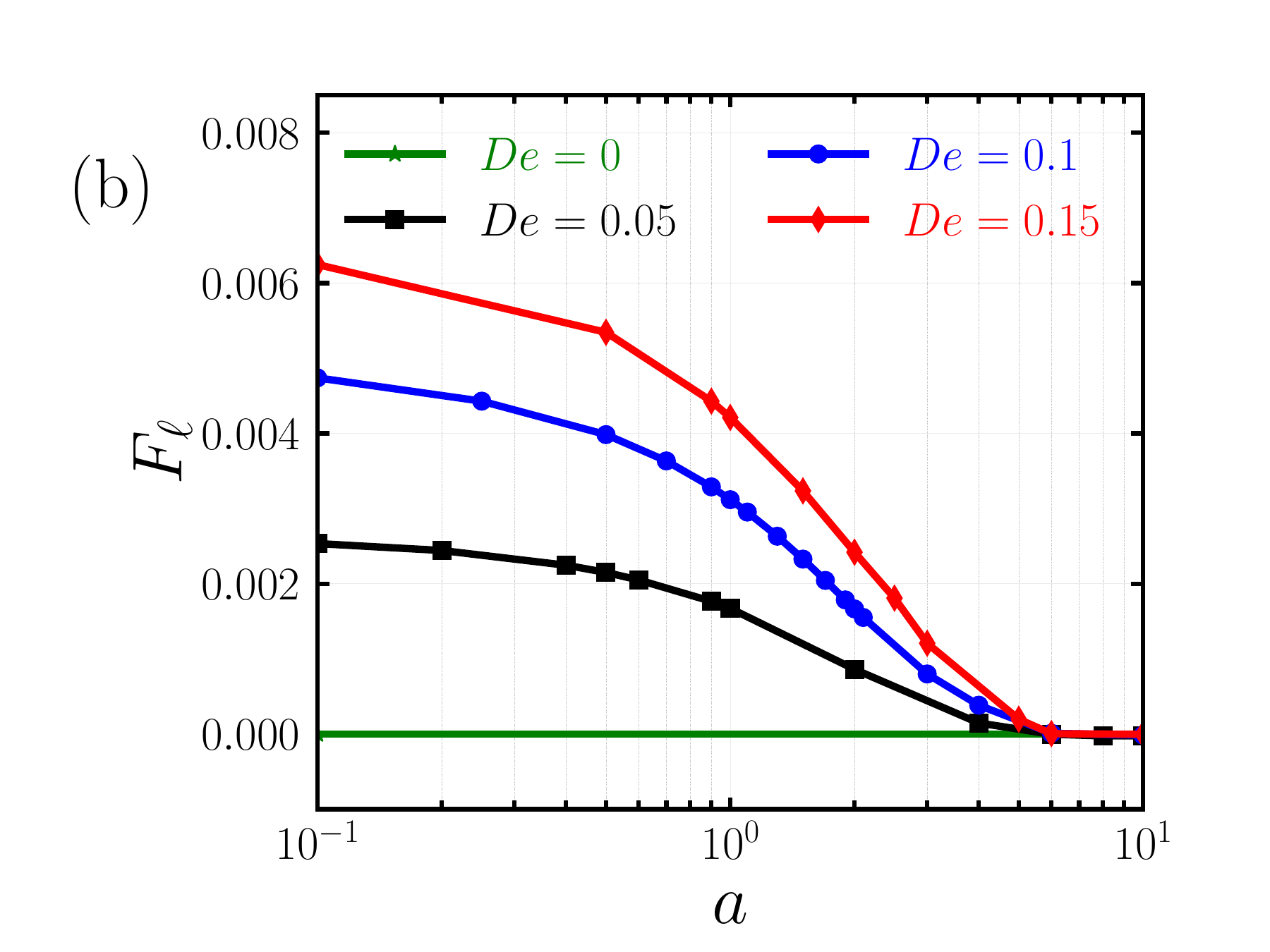}
\caption{Load variation versus the channel aspect ratio, for (a) the spanwise-varying and (b) the extruded channel, for three different values of the Deborah number using $d=0.2$.}
\label{fig_fl_Wi100_150}
\end{figure}

 We vary the channel aspect ratio and examine the influence of the polymeric stress on the load carrying capacity, depicted in Fig. \ref{fig_fl_Wi100_150}, for $De=0.05$, $De = 0.1$ and $De = 0.15$. Increasing $De$ has the desired beneficial effect of increasing the load, however the response to varying channel dimensions is quite different between the two configurations. We further observe that the Newtonian response is identically zero for all $a$, implying that the extra force is due to purely viscoelastic effects.
\par For the spanwise varying case represented in Fig. \ref{fig_fl_Wi100_150}(a), the load varies strongly versus increasing $a$. Close to the two limiting cases; (i) a wide channel ($a < 1$) and (ii) a slender channel ($a > 1$), the load is small and rises as we deviate from these limits, reaching a maximum when the width is on the order of the channel length. The variation is essentially governing by two factors; (i) the fraction of curved region and (ii) the fluid leakage along the spanwise boundary. When the width is large, the curved portion covers only a small portion of the total channel wall, which is insufficient to generate strong pressure gradients. As width and the length starts to be comparable (i.e., around $a=1$), the surface gradients increase proportionally generating strong pressure gradients driving the flow. This leads to an enhanced stretching of the polymer chains. However, upon further decrease in the width $a > 1$, the close proximity of the spanwise boundaries allows the fluid (forcibly drawn in due to the sliding action) to escape, leading to a decline in the pressure. 
\par In the extruded channel we find a starkly contrasting trend of the load versus the channel aspect ratio as can be seen in Fig. \ref{fig_fl_Wi100_150}(b). The maximum value is observed in the lower limit ($a << 1$) which tends to the two-dimensional channel. Unlike the spanwise varying case, surface gradients are absent across the width but are present along the streamwise direction for all $a$. Hence, pressure gradients along $x$ are always present, driving the flow and stretching the polymer additives over a larger portion of the channel, diminishing only as the spanwise boundaries are brought close to the bulk (in the upper limit $a >> 1$). In both cases, the increasing channel aspect ratio promotes side leakage that can lead to a loss of useful polymeric stretch diminishing, thus, the load carrying capacity. 


\begin{figure}
\centering
\includegraphics[scale = 0.2]{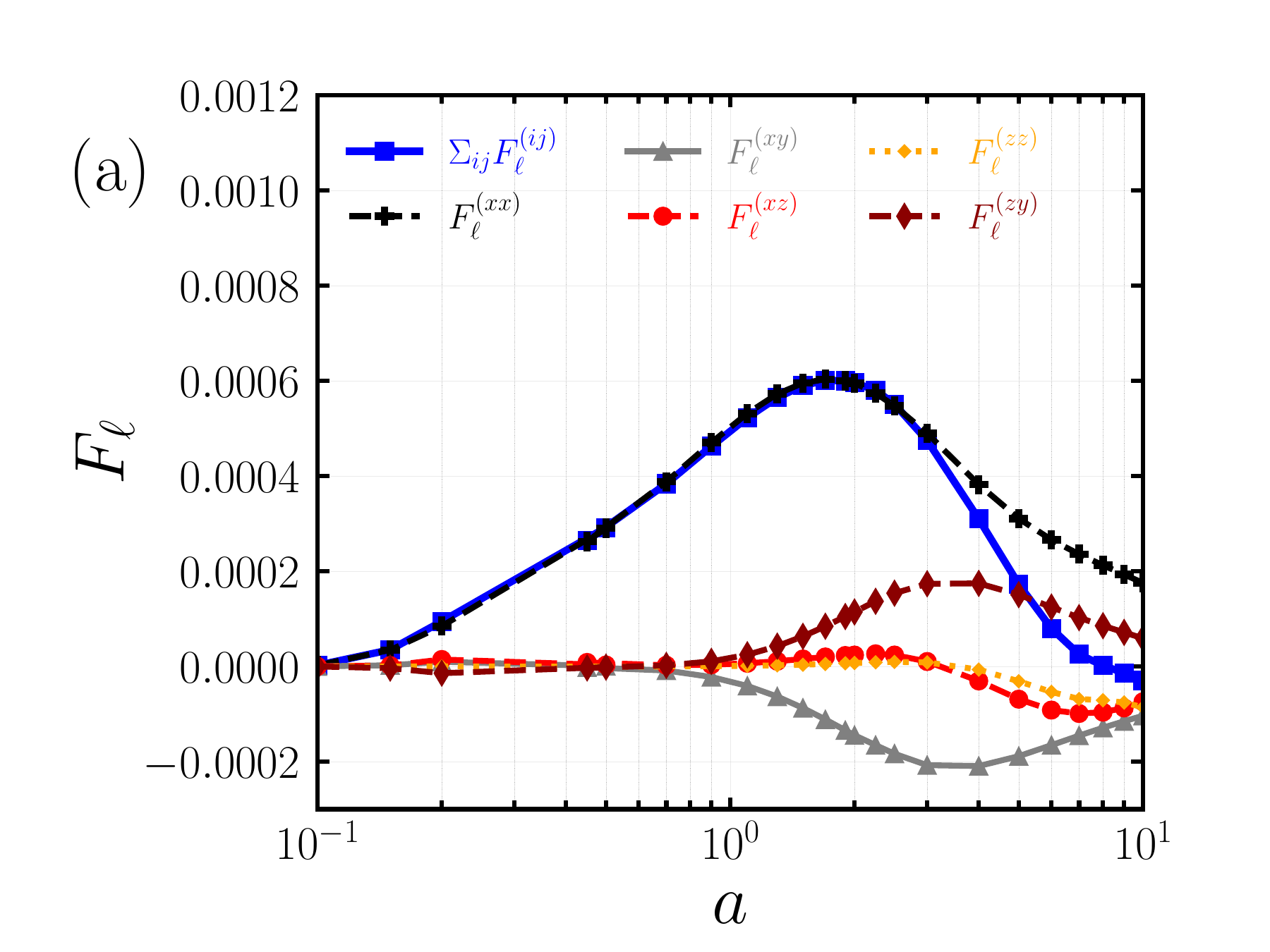}
\includegraphics[scale = 0.2]{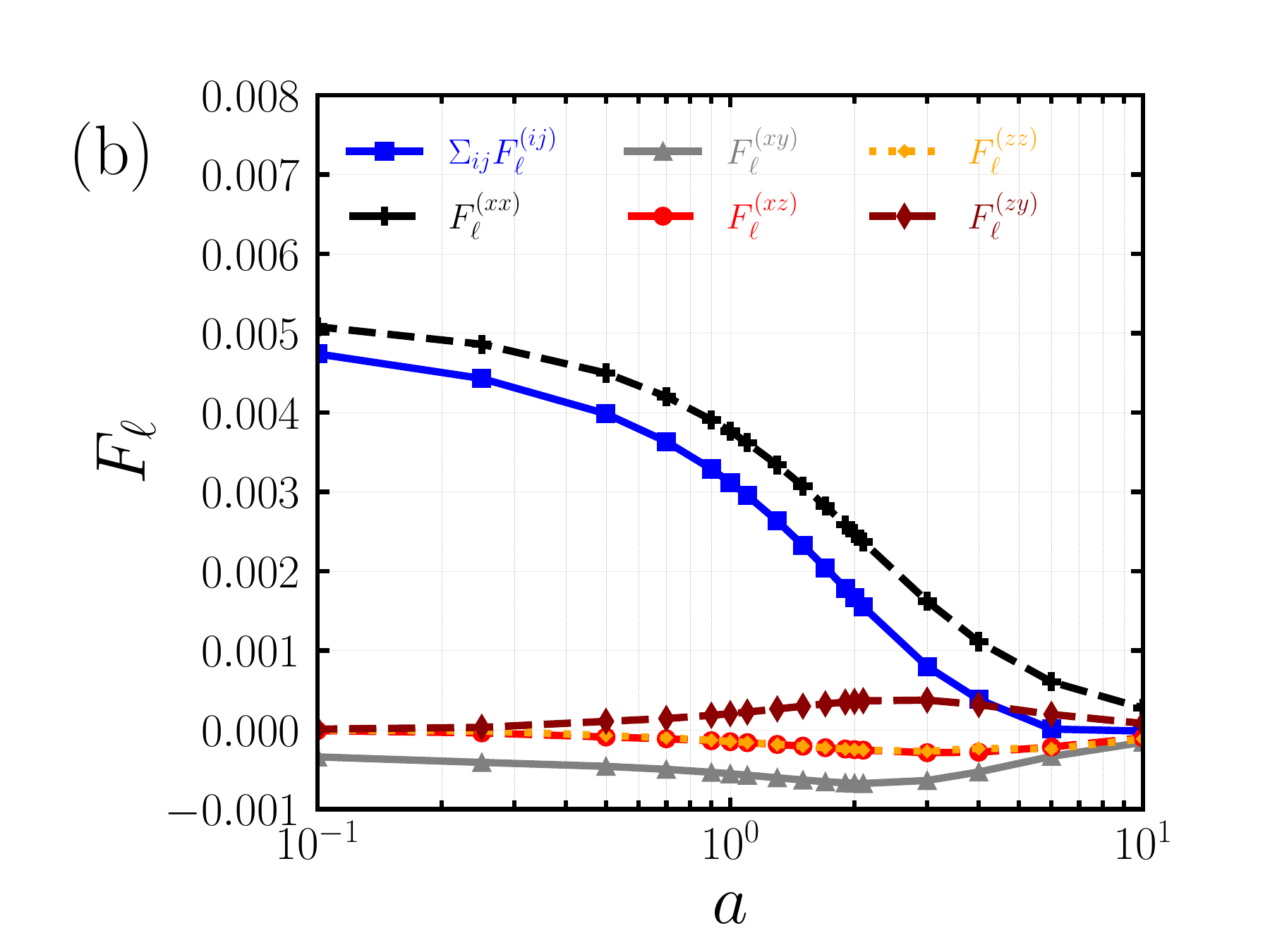}
\caption{The variation in the components of the load carrying capacity per unit width for the (a) spanwise varying and the (b) extruded surfaces versus the channel aspect ratio for $De = 0.1$.}
\label{fig_load_comps}
\end{figure}
\subsubsection{Streamwise normal stress}
\par The load carrying capacity for finite $De$ is the result of a contribution from each component of the polymer stress, $[\tau_{xx}, \tau_{xy}]$ and $[\tau_{xx}, \tau_{xy}, \tau_{xz}, \tau_{yz}, \tau_{zz}]$ for the two-dimensional (2D) and the three-dimensional (3D) cases, respectively. We decompose the individual contributions to the load by solving VR but retaining only the desired stress components on the right hand side of Eq. \ref{eqn_VR1}. The result is shown in Fig. \ref{fig_load_comps}, for $d = 0.2$, $De = 0.1$, and $\beta = 0.8$.
\par The primary contribution comes from the streamwise normal stress ($\tau_{xx}$), similar to what was found by \cite{ahmed2021new} for the 2D case, and a secondary effect from the remaining stress components which, upon summation, appears to cancel out. The streamwise component dominates, not surprisingly, due to the boundary motion along $x$ which pulls and stretches the polymers. Similar to the two-dimensional case, the shear induces a strong stretch along the flow (streamlines), since $\tau_{xx}$ depending quadratically on the shear rate even for very small $De$. During the converging sections of the channel, this causes a build-up of the normal stress (stretching) that relieves the pressure, since the net force required to propel the fluid is now both the pressure and streamwise normal stress. The reverse occurs during the divergent segments where the shear rate is arguably lower causing the polymers to relax. This stretch-relaxation mechanism (also measured via the tension in the streamlines) was first reported by \cite{hinch2024fast} and relates the pressure variation to the tension in the streamlines. A notable difference in this work and that of \cite{hinch2024fast} is the prevailing presence of shear due to the Couette flow which prevents complete relaxation to zero stress, and the smallest value of the normal stress is that of the pure-shear driven state in the absence of pressure gradients. 
\par
\begin{figure}
\centering
\includegraphics[scale = 0.2]{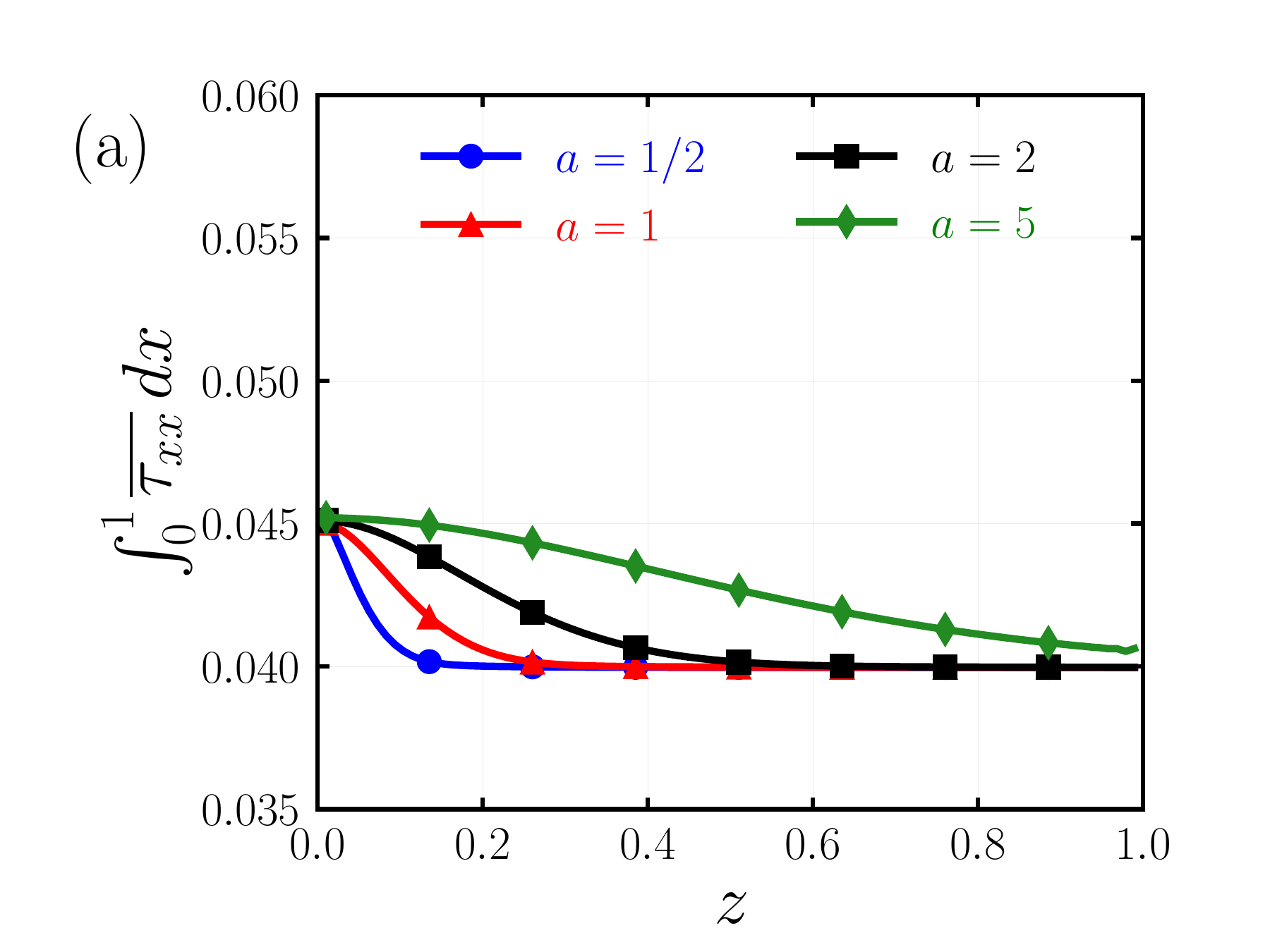}
\includegraphics[scale = 0.2]{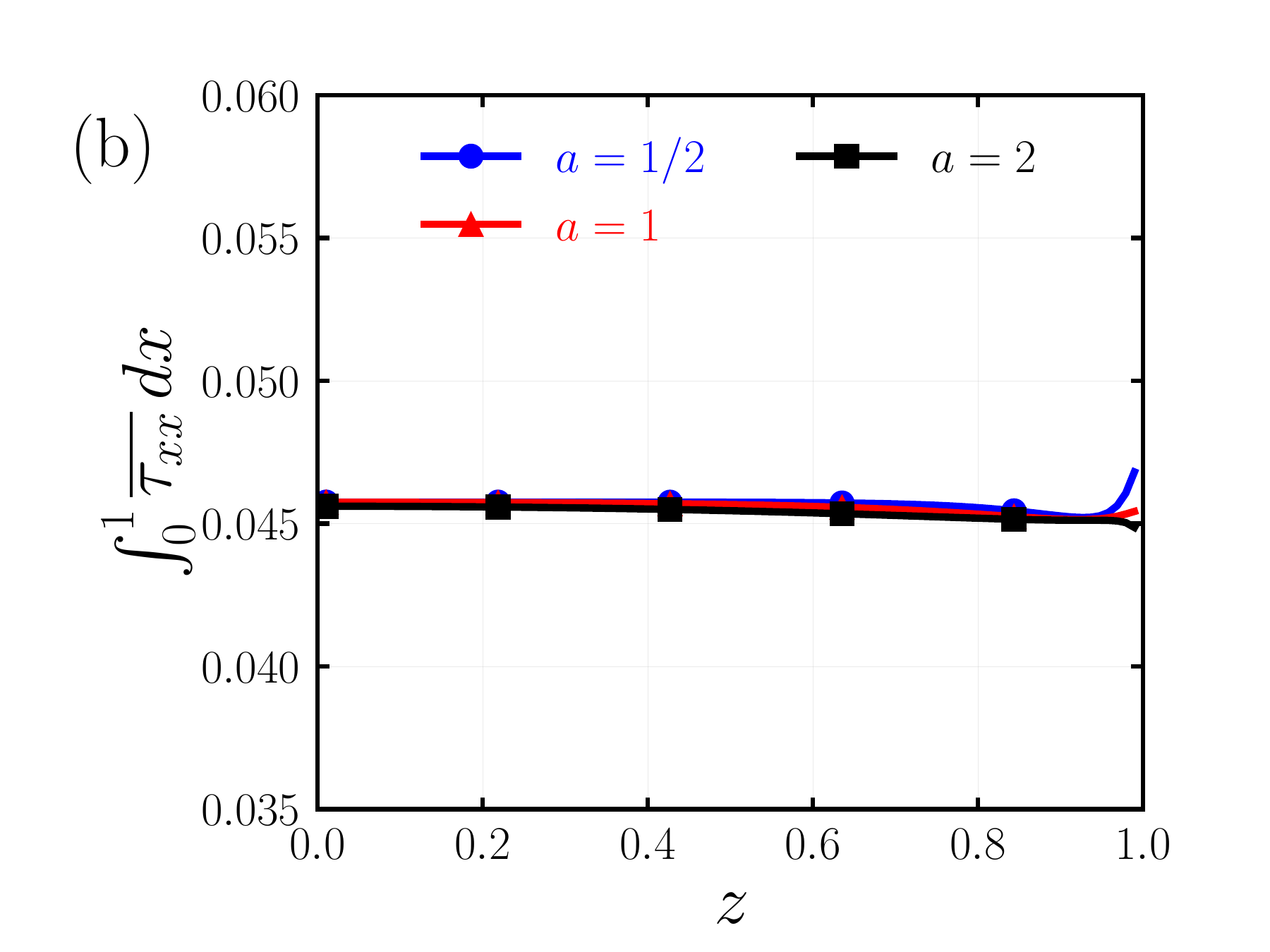}
\caption{The film-averaged normal stress distribution along the spanwise direction (averaged along $x$) for the (a) spanwise varying and the (b) extruded channel, for different aspect ratios, using $De = 0.1$, and $d = 0.2$. }
\label{fig_relax}
\end{figure}

 In this three-dimensional channel, the normal stress distribution along $x$, varies strongly versus $a$, in both the spanwise varying and extruded configurations. The variation is not only attributed to decreasing pressure gradients, hence, lesser shearing in the channel, but also insufficient room for relaxation due to a shortening of the channel width. This is reflected in Fig. \ref{fig_relax}, which shows the distribution of the film averaged streamwise normal stress along the spanwise direction for different values of the aspect ratio ($De = 0.1$, $\beta = 0.8$, and $d = 0.2$). 
textcolor{blue}{When spanwise surface gradients are present, see Fig. \ref{fig_relax}(a), the channel height reaches a maximum beyond $z > s = 0.1$ and the shear rate reduces, causing the polymers to relax (reaching the pure shear value $\tau_{xx} = 2(1-\beta)De \dot{\gamma}_{xy}^2$, where $\dot{\gamma}_{xy} = 1$). Furthermore, we continue to shrink the channel by increasing $a$ (decreasing the channel width), losing them across the spanwise boundary due to side leakage (which is neglected completely in the two-dimensional cases). }
\par However, for the extruded channel, shown in Fig. \ref{fig_relax}(b), the polymers are exposed to a greater shear rate ($\dot{\gamma}_{xy}\approx \frac{1}{\min(h)}=\frac{1}{(1-d)}$), preventing them from relaxing along the width. A sudden jump is observed at the exit due to a strong leakage of the flow along the spanwise boundary. It is evident that varying the aspect ratio impacts (i) the room available for the polymer to stretch/relax and (ii) the rate at which the polymers escape along the spanwise open boundaries dictated by the spanwise velocity $w$. The latter always contributes to a decline in the load carrying capacity.
\par
\begin{figure}
\centering
\includegraphics[width=0.3\linewidth]{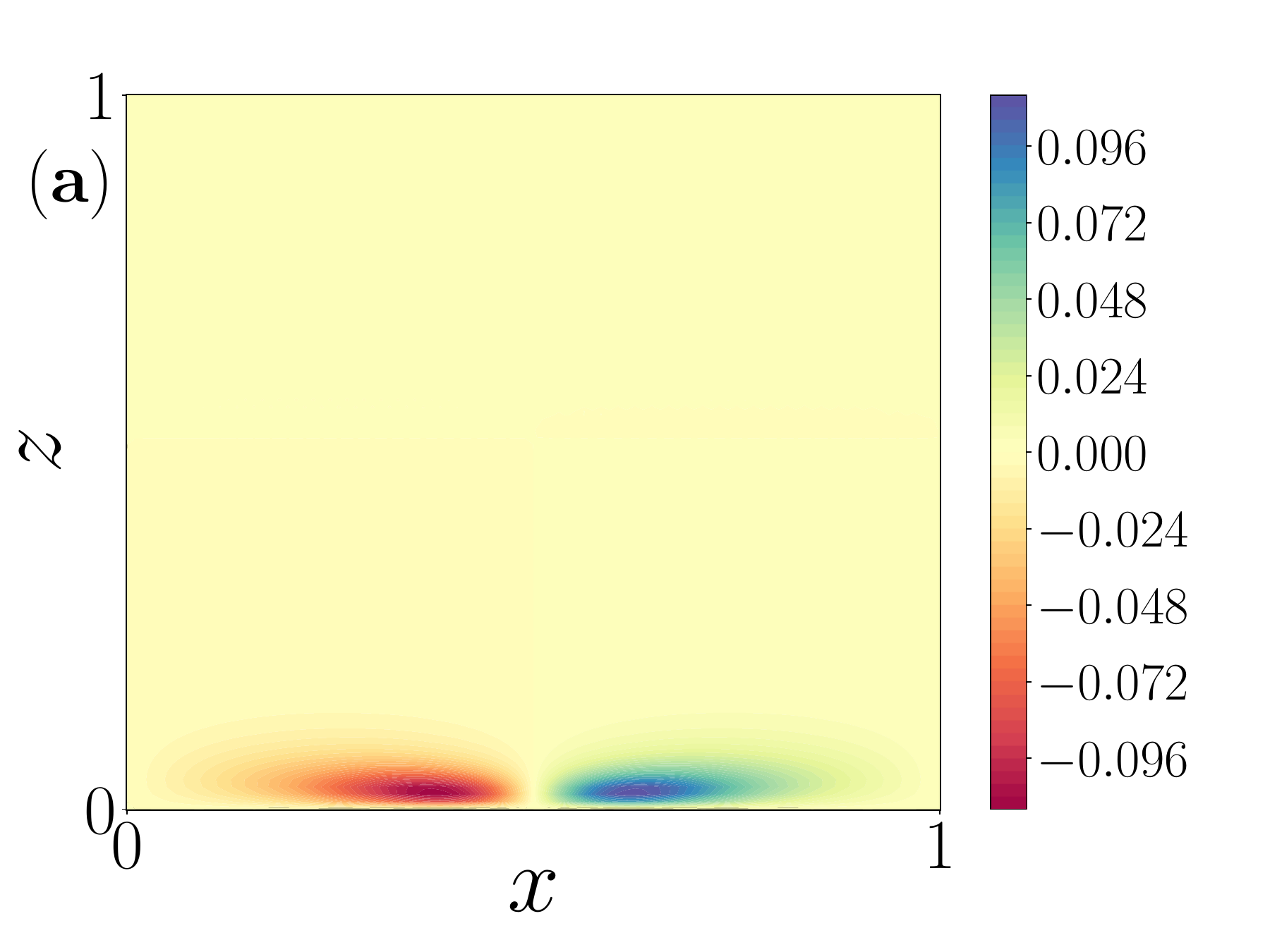}
\includegraphics[width=0.3\linewidth]{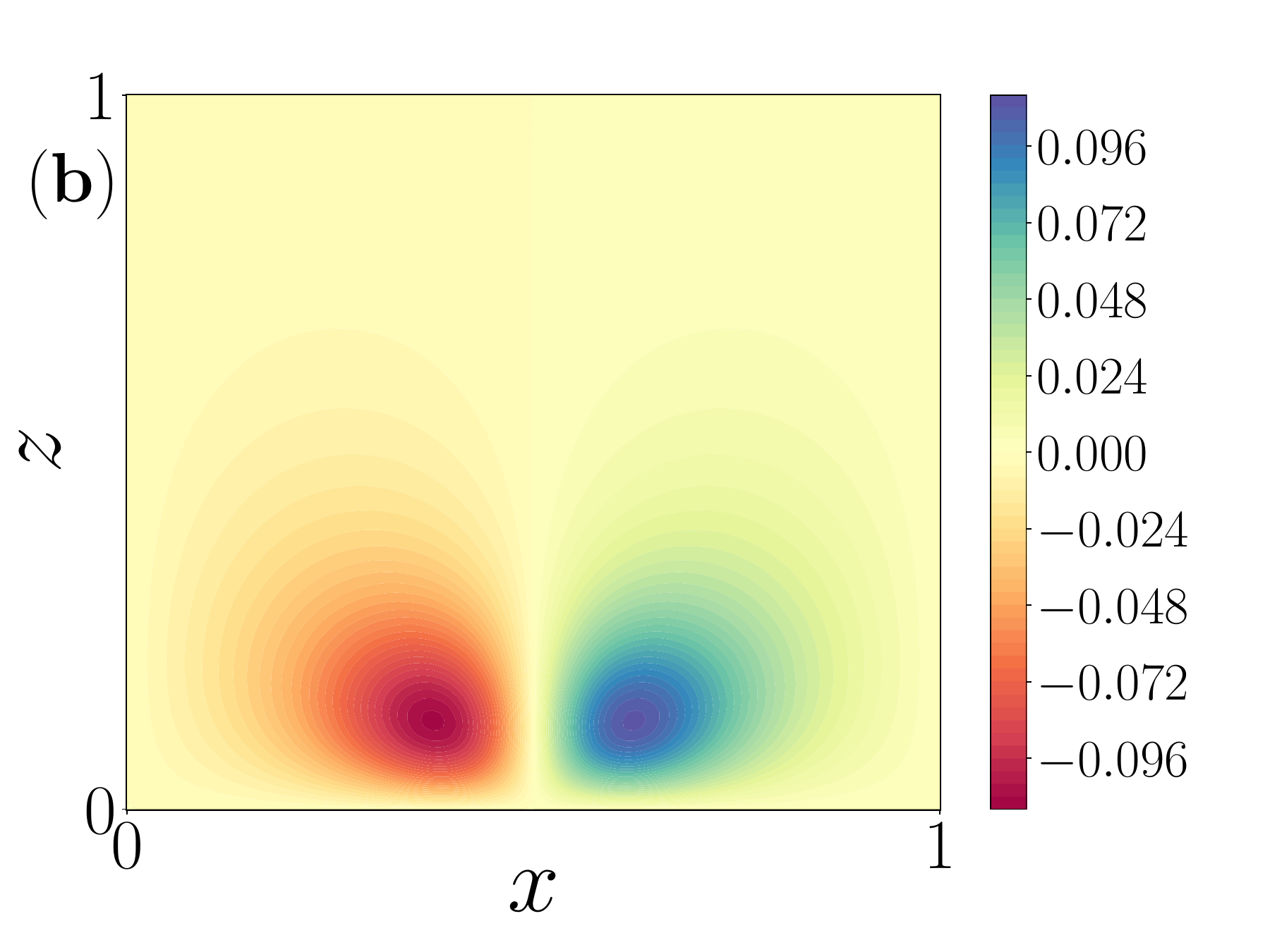}
\includegraphics[width=0.3\linewidth]{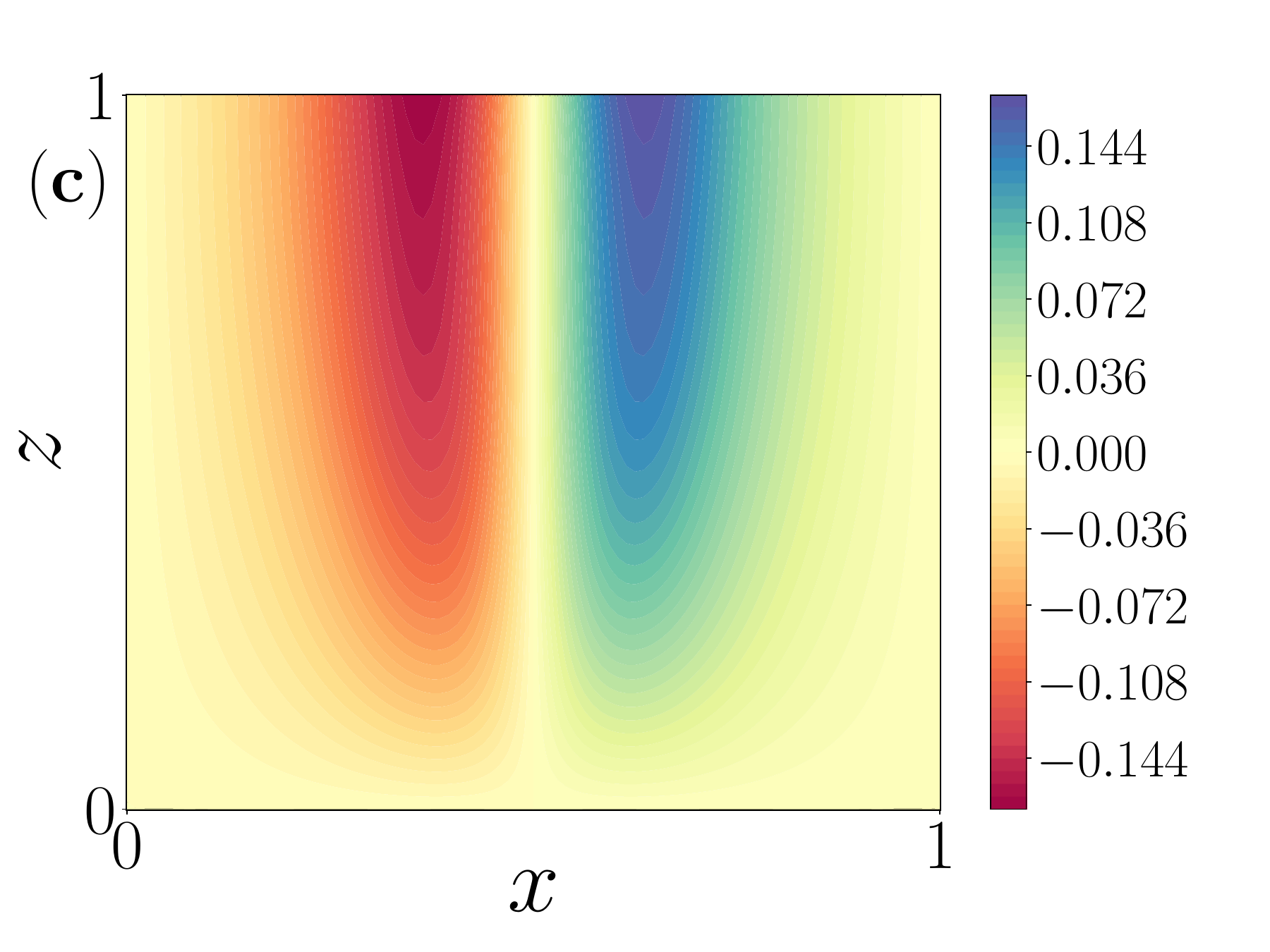}
\includegraphics[width=0.3\linewidth]{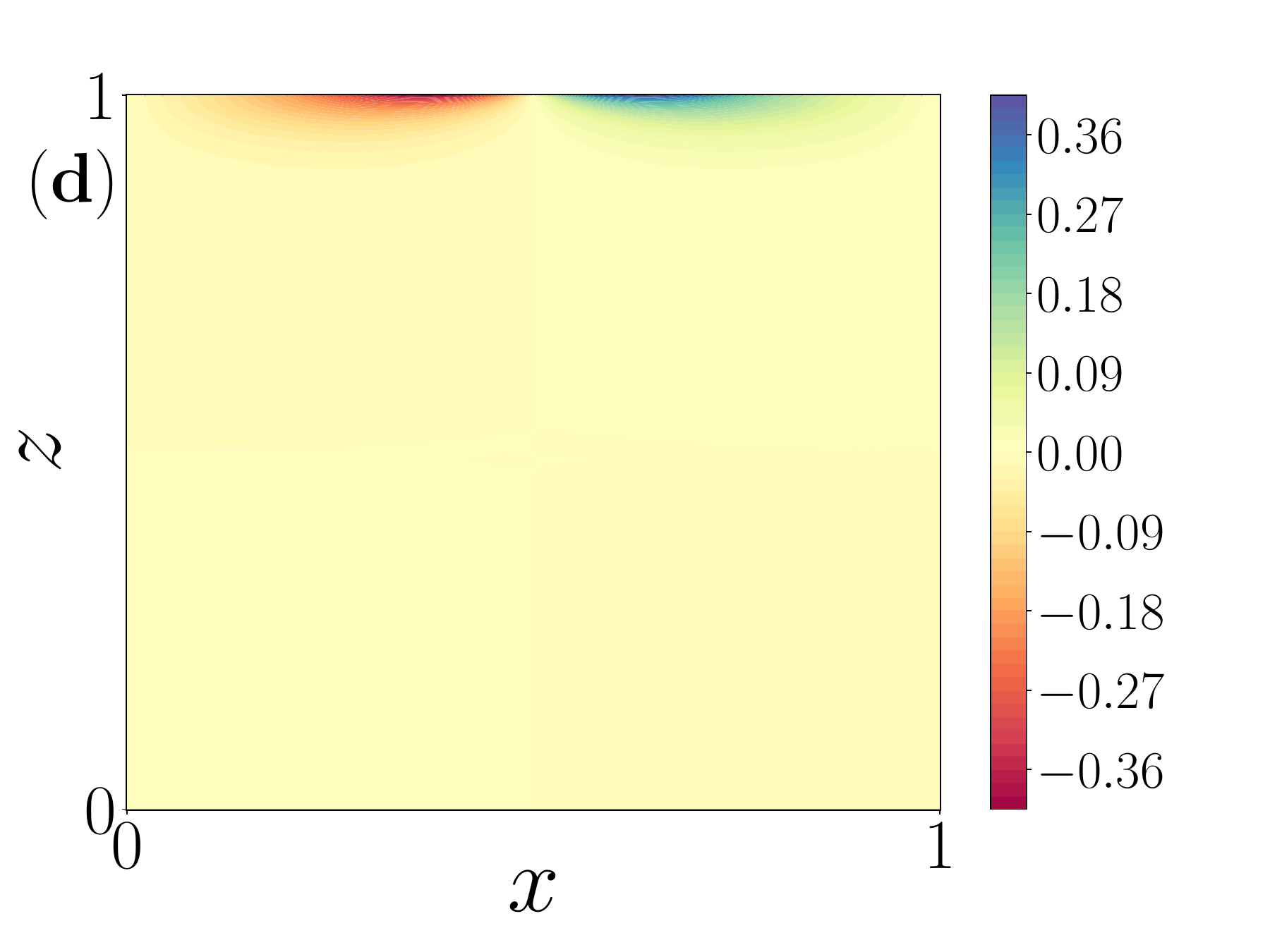}
\includegraphics[width=0.3\linewidth]{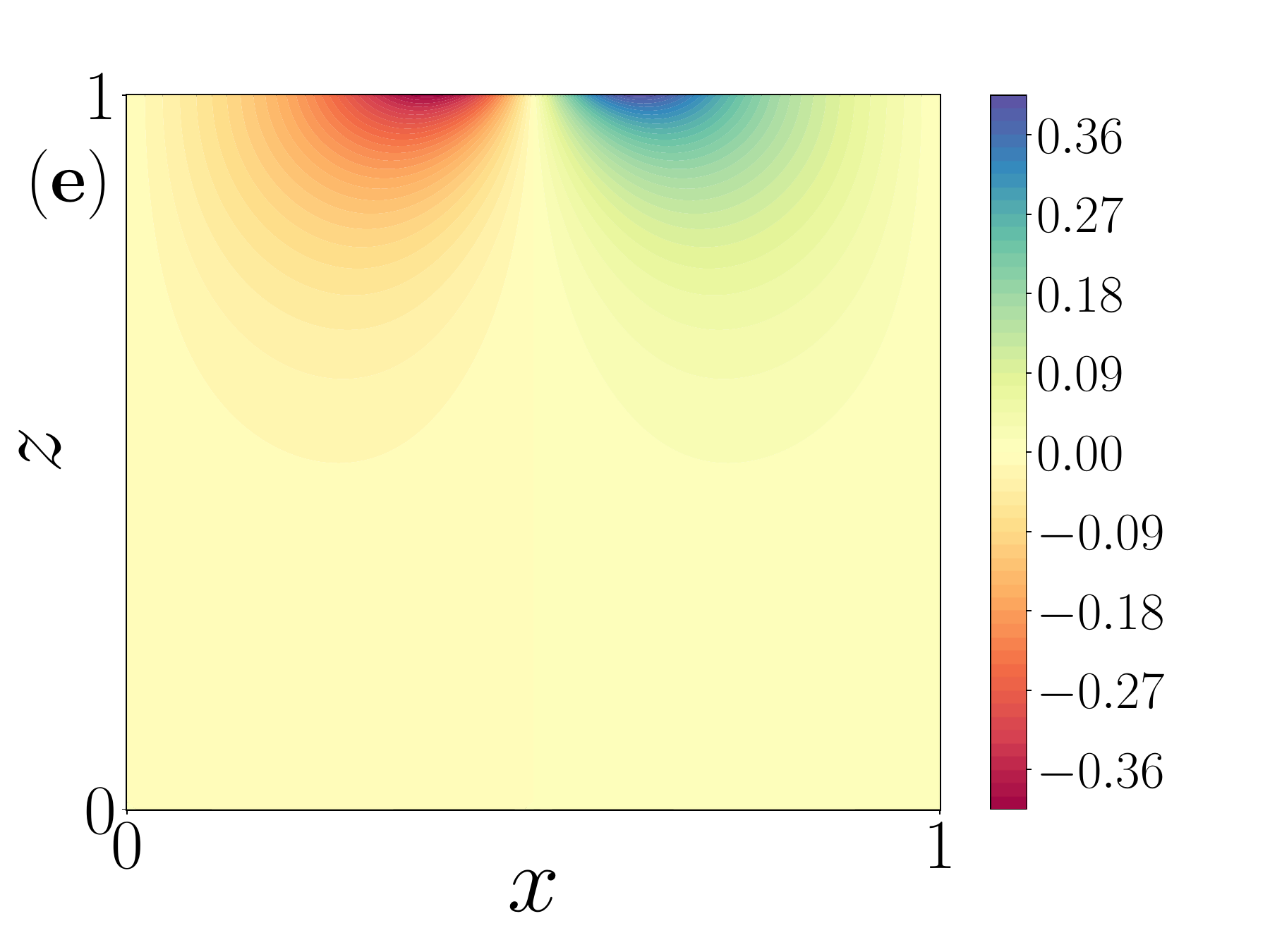}
\includegraphics[width=0.3\linewidth]{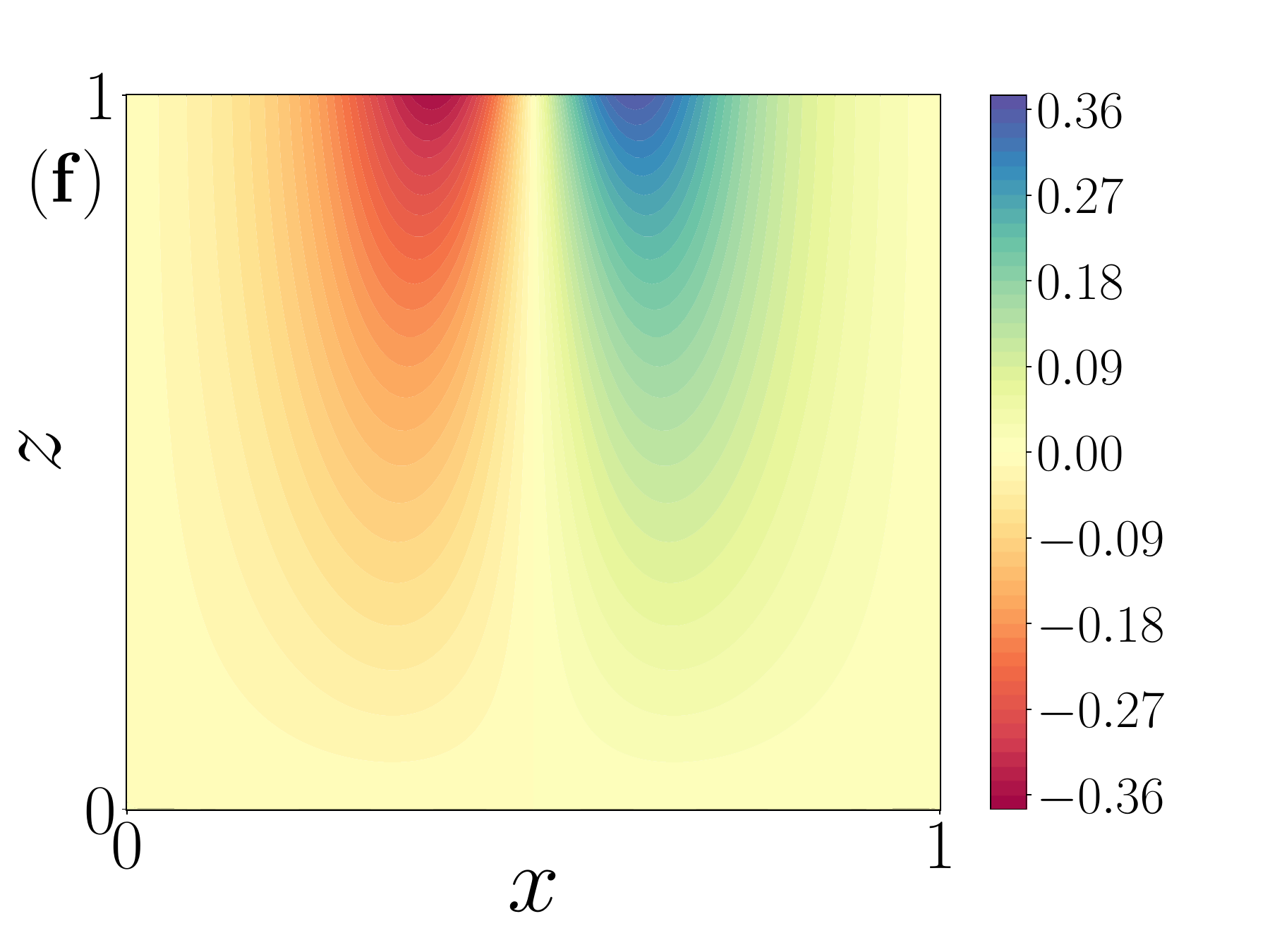}
\caption{The distribution of the spanwise velocity (given by Eq. \ref{eqn_w_measure}) across the channels for (a,b,c) the spanwise varying and (d,e,f) the extruded channel. The analysed aspect ratio are (a,d) $a = 0.2$, (b,e) $a = 1$, and (c,f) $a = 5$. Note that $d = 0.2$. }
\label{fig_w_contour}
\end{figure}

 We focus on these two competing mechanisms for the two different channel configurations. We adopt a measure of the spanwise velocity connected with the spanwise pressure gradient (independent of the film height)
\begin{equation}
\label{eqn_w_measure}
w_p = \frac{a}{2} \bigg( h^2 \frac{\partial p}{\partial z} \bigg),
\end{equation}
which is depicted across the channel in Figs. \ref{fig_w_contour}(a)-(c) for the spanwise varying and Figs. \ref{fig_w_contour}(d)-(f) for the extruded channel using three different values of the channel aspect ratio.
We observe for the spanwise varying, an increase close to the bulk region ($z=0$) which declines versus $a$ and a strong leakage rate at the boundary ($z=1$). For the extruded case, shown in Figs. \ref{fig_w_contour}(d)-(f), the bulk spanwise velocity is negligible, and rises as the width decreases. 

Hence, we define two characteristic spanwise velocity scales; (i) the velocity in the bulk and (ii) velocity at the boundary, which generate and diminish, respectively the effect of the viscoelasticity on the pressure (and then on the load).
For the spanwise varying case, the surface gradients along $z$ persist for a length $s$ (defined as the span of the Gaussian surface, see section \ref{sec_channelGeo}) and diminish beyond this limit. For small $a$, the spanwise velocity is effectively zero for $z > s$. Therefore, the reference spanwise velocity in this region is taken as $W_s = \max |w_p(x,z=s)|$. On the other hand, the velocity contributing to leakage along the spanwise boundary $W_a = \max |w_p(x,z=1)|$ continues to increase versus $a$, draining the channel of useful stretched polymers. For the extruded channel, the $xy$-cross-section is constant, covering the entirety of the width, leading to negligible velocity close to the channel bulk ($z = 0$).  For this case, only the leakage rate is considered significant. 

\begin{figure}
\centering
\includegraphics[width=0.45\linewidth]{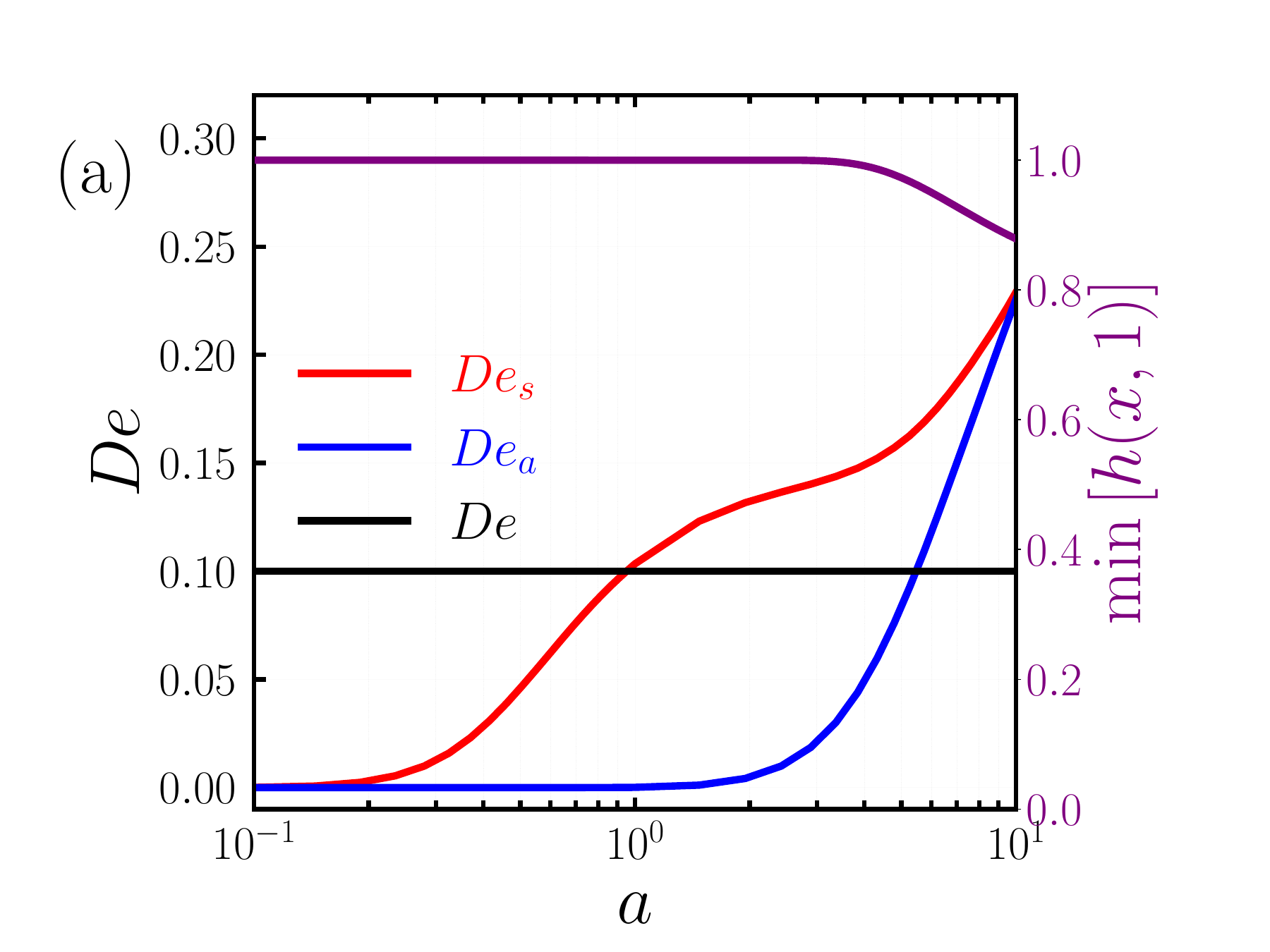}
\includegraphics[width=0.45\linewidth]{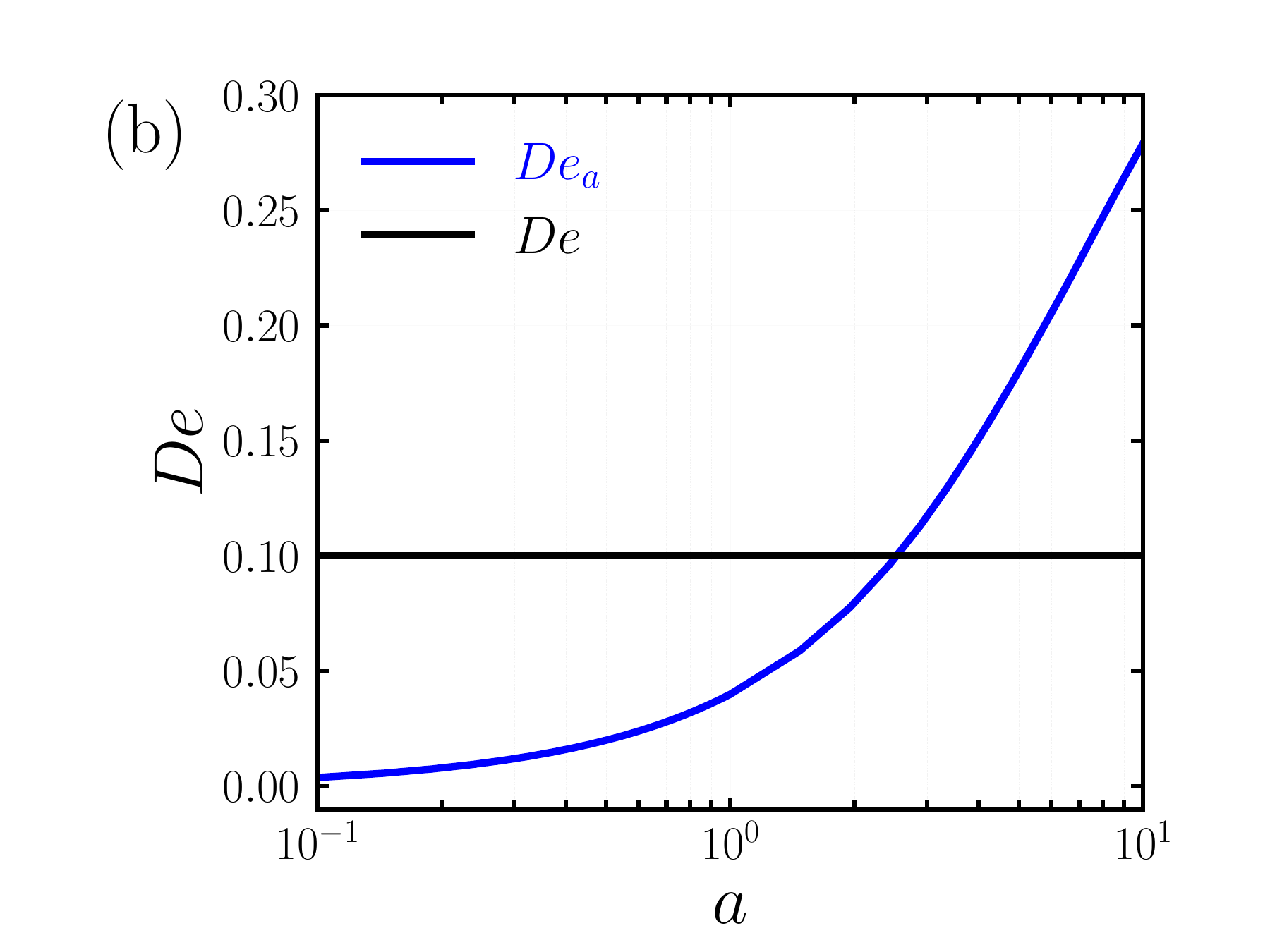}
\caption{The spanwise effective $De_s$, and the leakage $De_a$ versus the channel aspect ratio for the (a) spanwise varying and the (b) extruded case, using $d = 0.2$, and $De = 0.1$. }
\label{fig_De_spanwise}
\end{figure}

Owing to these two different and competing flow mechanisms, we have two different effective Deborah numbers along the width (i) the spanwise Deborah \begin{equation}
De_s = \frac{a}{s}W_s De,
\end{equation} and (ii) the leakage Deborah 
\begin{equation}
De_a = a W_a De,    
\end{equation} depicted in Fig. \ref{fig_De_spanwise} versus the channel aspect ratio for (a) the spanwise varying and (b) the extruded channels. 
For the spanwise varying case, the spanwise Deborah number tends to rise as the channel width decreases raising the polymeric contribution to the load, 
while $De_a$, acting against it, rises exponentially when the width is sufficiently small. A second increase in $De_s$ is observed around $a = 5$ because the Gaussian protrusion now touches the spanwise boundary and the bulk flow merges with the leakage region. At this point, the useful stretch is lost along the open boundary. Once $De_s$ is commensurable with $De$ (as $a\sim O(1)$) we reach the maximum load carrying capacity, see Fig. \ref{fig_fl_Wi100_150}(a).
For the extruded case given in Fig. \ref{fig_De_spanwise}(b), we only obtain a continual loss of polymers due to leakage, depicted by a rise in $De_a$, which eventually overtakes the streamwise $De$ number. This leads to a continual decline in load as observed in Fig. \ref{fig_fl_Wi100_150}(b). 

\subsection{Boundary conditions}
\label{sec_boundaryCondition}
\par The boundary conditions for the pressure have a strong influence on the net load carrying capacity within the channel. As pointed out in literature, the pressure is not properly defined in the viscoelastic case \citep{tichy_said_spherical}, and setting it equal to the ambient pressure, or any value, is not appropriate \citep{tichy_sawer_SOF_1998}. 
In a similar manner, excluding the stress altogether may also be equally inappropriate. Balancing the force at the open boundaries, we require the expressions for total force
\begin{equation}
    \boldsymbol{F}^* = \int_{s} \boldsymbol{\sigma}^* d\boldsymbol{s}^*
\end{equation}
to vanish at the streamwise ($d\boldsymbol{s}^* = dy^* \,dz^* \boldsymbol{e}_x$) and spanwise boundaries ($d\boldsymbol{s}^* = dy^* \,dx^* \boldsymbol{e}_z$). Non-dimensionalizing using Eqs. \ref{eqn_geo_scaling} and Eqs. \ref{eqn_nonDimSys1_stress}, and noting $\boldsymbol{\sigma}^* = \boldsymbol{\tau}^* - p^*\mathsfbi{I} + 2\eta_s\mathsfbi{D}^*$, we obtain the conditions, up to O($\epsilon^2$),
\refstepcounter{equation}
$$
p=\overline{\tau_{xx}}(0,z), \quad 
p=\overline{\tau_{xx}}(1,z), \quad 
p=\overline{\tau_{zz}}(x,1).
\eqno{(\theequation{\mathit{a},\mathit{b},\mathit{c}})}\label{eqns_bc_p}
$$
Note that the channels modeled in this work are symmetric about the spanwise direction. Hence, the pressure and stress condition along the axis of symmetry ($z = 0$) satisfies $\frac{\partial p}{\partial z} = \frac{\partial\tau_{ij}}{\partial z}=0$.
\par
\begin{figure}
\centering
\includegraphics[scale = 0.2]{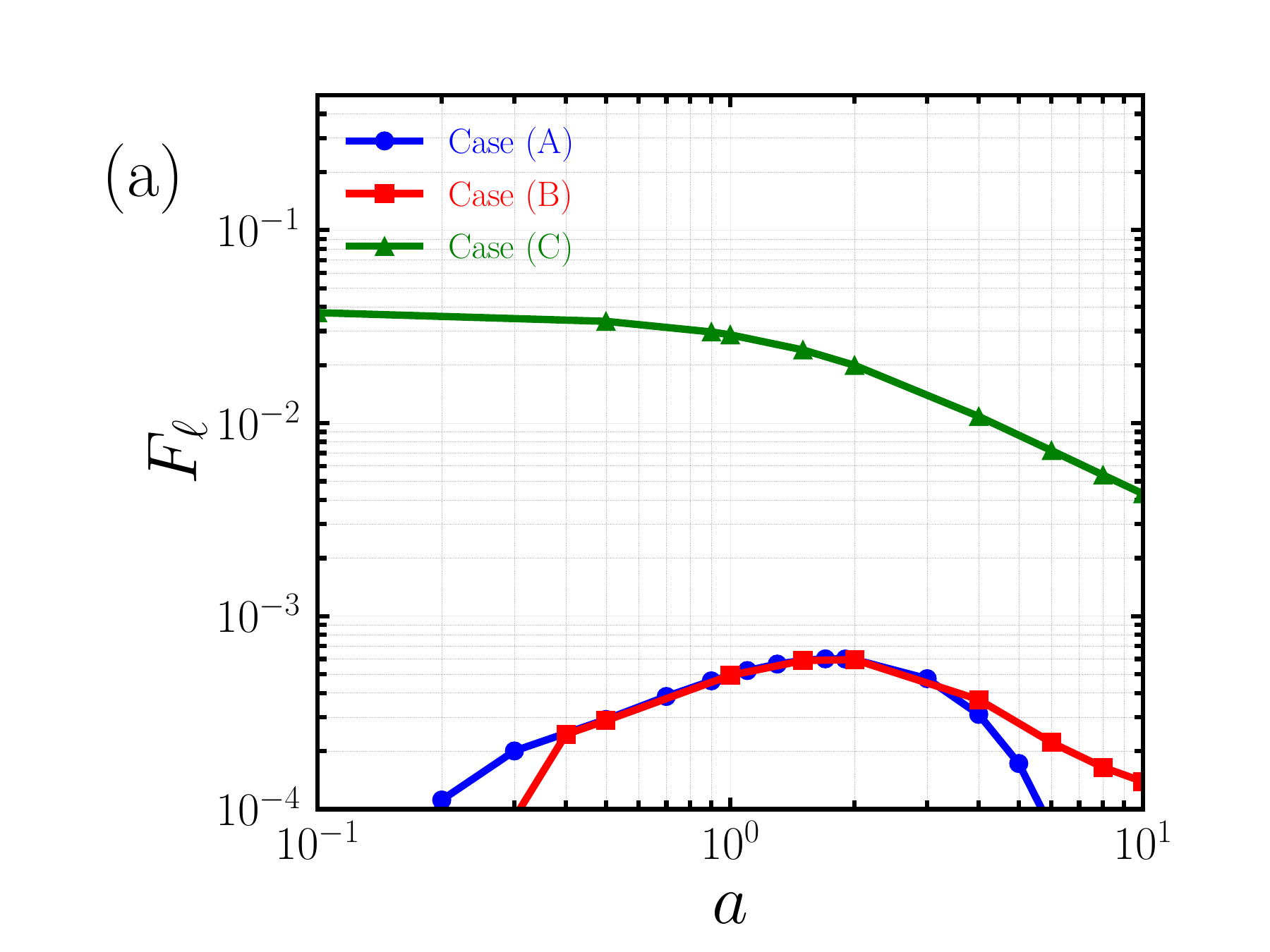}
\includegraphics[scale = 0.2]{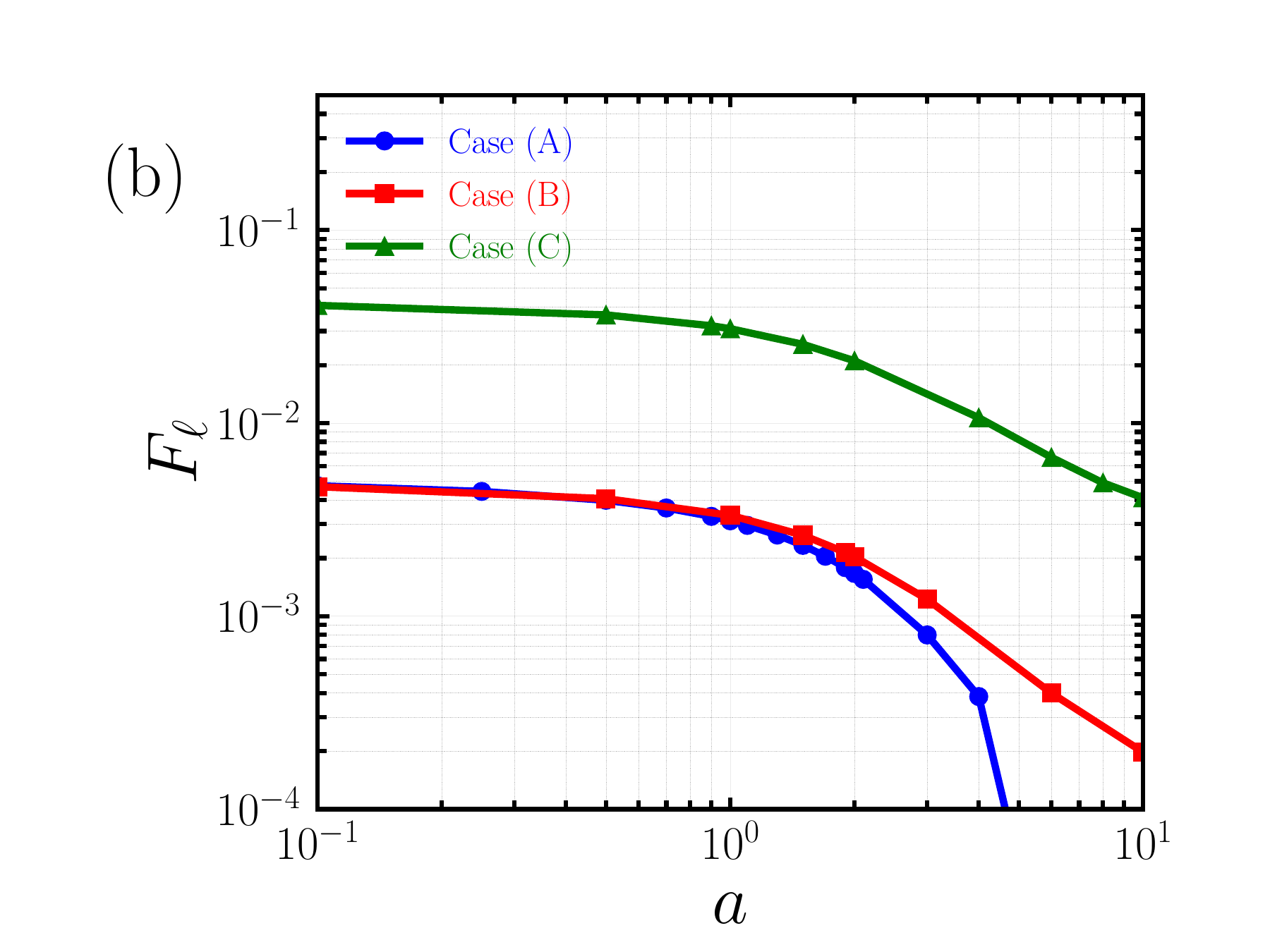}
\caption{The load carrying capacity for (a) the spanwise-varying and (b) extruded channel, considering three different cases of boundary conditions: (A) Newtonian pressure condition along all open boundaries, (B) Newtonian pressure condition only along the streamwise open boundaries, and (C) the pressure balanced by the average normal stress along all open boundaries (fully viscoelastic), using $De = 0.1$, and $d = 0.2$.}
\label{fig_bc}
\end{figure}

 Using Eqs. \ref{eqns_bc_p}, we compare the load carrying capacity versus the channel aspect ratio, as shown in Fig. \ref{fig_bc}, for three sets of boundary conditions: (i) the Newtonian condition $p=0$ (case A), (ii) Newtonian condition along the streamwise open boundaries, i.e., $p(0, z) = p(1, z) = 0$, and normal stress conditions along the spanwise boundaries (case B), and (iii) normal stress conditions along all open boundaries using Eqs. \ref{eqns_bc_p} (case C). 
\par {It is evident from Fig. \ref{fig_bc} that a finite increase in pressure, and consequently the load, occurs at the boundary owing to the extra stress of the polymer}. For both surface configurations, when comparing the three cases, the addition of the spanwise normal stress has a weak influence on the load (due to the absence of spanwise surface motion). As the channel width reduces further ($a > 1$), a net increase is observed, delaying an immediate drop to zero observed for the Newtonian conditions.
\par When all the open boundaries are influenced by the polymer normal stress components (i.e., Case C), a large net increase in the force is obtained, dominated by the streamwise normal stress component. This additional load persists over the entire range of $a$ examined, diminishing only when the channel width becomes small in comparison to its length ($a > 1$). At this point, the net shear strain rate that stretches the polymer chains (both streamwise and spanwise) is reducing owing to a decreasing pressure gradient. 
\par This additional gain is nearly one-order of magnitude greater than the Newtonian conditions. Despite having two distinct surface profiles, for small values of $a$, we obtain the same values of the load and almost identical trends. As $a$ decreases below one, the presence of any surface feature, no matter how sharp, is not felt by the fluid, since the terms involving spanwise variation scale with $a^2$. Thus, for small $a$, we have an effectively flat channel for which the extra stress is easily predicted via the linear model. This net increase in load was construed as spurious by \cite{tanner_1970_spurious}, since classical lubrication theory for Newtonian fluids plainly predicts a zero pressure field for  boundary driven flows. Experiments may be necessary to shed light on the true conditions prevailing at the open boundaries under strongly sliding conditions. In addition to this, we observe for $a > 1$, the curved portion of the surface intersects with the open boundary, leading to strong re-entry of the fluid at $z = 1$. In this case, assuming that the fluid beyond $z=1$ is at a fixed pressure condition, independent of the total stress (the blue curves in Fig. \ref{fig_bc}), may be inaccurate. 

\section{Conclusions}
\par The flow of a thin viscoelastic lubricating film along a channel with boundary motion was studied numerically for a 3D channel with a flat lower surface and an upper curved surface modeled as a Gaussian protrusion. Two special cases are considered for the curved surface; (i) $h = h(x,z)$ and (ii) an extrusion of the $xy$-cross-section, $h=h(x)$. For these two cases, we employ the viscoelastic Reynolds approach as the numerical procedure for solving the governing equations and, also, compare the models performance with an arguably simpler and efficient $De$-order perturbed model. The main observation is the possibility of enhancing the film's load bearing capability by exploiting the polymer's elastic properties (as evidenced via experiments).
\par An interesting feature of 3D channels is the ratio between the length and the width (aspect ratio) which strongly influences the forces developed in the channel. By varying the channel aspect ratio but keeping the apparent Deborah number constant (defined identically to a 2D configuration), we can modify the strength of viscoelastic effects. The load versus $a$ trend similarly to the two-dimensional case, i.e., monotonically increasing for small $De$ and then an eventual saturation for large $De$. This was the case for both surface configurations.
\par A prediction of the load was also made via a $De$-order perturbation model which served to validate the VR approach and demonstrate the viability of low-order models. Results compared favorably for low values of $De$ but exhibited a discrepancy with the VR for large values.
The reduced model in the three-dimensional case must be solved numerically, since a partial differential equation emerges for the leading-order and the first-order and, by extension, will also hold for the higher-order systems. While the $De$-order can still be tackled via classical finite difference methods (or any other discretization method), the appearance of mixed and higher order derivatives requires careful treatment. In fact, we infer that for the $De^2$-order, the stencil requirements for the FDM will become quite cumbersome and make the numerical treatment difficult.
\par The load enhancement was traced to the growth in the streamwise normal stress in the channel because it aligns with the sliding direction. For the spanwise varying channel, the increase is due to a rise in the bulk spanwise velocity, finite only in the vicinity of the protrusion and zero everywhere else. Defining as a spanwise Deborah number $De_s$, corresponding to this local spanwise velocity, we observe that $De_s$ rises as $a$ increases becoming commensurable to $De$ for $a\sim O(1)$ (i.e., when the load is maximum). However, increasing $a$ also narrows the width and exposes the useful stretched polymers to the open boundary. As the flow exits this boundary, the stretched polymers are lost and the load begins to drop. This leakage rate is measured by a competing leakage Deborah number $De_a$, which increases as well with $a$. Eventually $De_a$ overpowers $De_s$ and diminishes the load entirely. Note that for $a > 5$, the curved region covers the entirety of the channel, but the close proximity of the open boundaries causes the bulk flow to merge with the leakage, thereby nullifying the beneficial rise in the local effective $De$. 
For the extruded channel, the bulk spanwise velocity remains zero, and as a consequence the spanwise $De$ is zero. On the other hand, the leakage Deborah number rises (similar to the spanwise varying case), causing the load to always decline as $a$ increases.
\par Finally, we examine the influence of including the polymer stress in the boundary conditions. The presence of surface sliding (Couette flow) leads to finitely large normal stresses even when the surface is flat. For both channel configurations, including the normal stress at the streamwise and spanwise boundaries leads to a dramatic enhancement of the film load carrying capacity. To clarify whether this additional increase is indeed spurious will require an experimental effort that measures the excess pressure at the boundaries as they undergo sliding motion. 
\par This work focused explicitly on channel surface configurations modeled neatly via a Cartiesian coordinate system. However, many cases, e.g. tear film lubrication, synovial joint lubrication, to name a few, conform to a spherical coordinate system (if reduced-order modeling is desired). In such cases, the aspect ratio may be fixed (close to unity) and the problem prescribed entirely in three-dimensions, especially in the absence of any geometrical symmetry. In these instances, the VR approach offers a means to predict the useful quantities such as surface shear stresses and pressure over the useful range of $De$.

\appendix

\section{Viscoelastic Reynolds equation in three dimensions}\label{app_VR3D}
The viscoelastic Reynolds Eq. for a 2D channel can be extended to 3D channels (in Cartesian configuration) following the procedure in Ahmed and Biancofiore (2021),
\begin{subeqnarray}
\label{eqn_VRCrt1}
&\frac{\partial}{\partial x} \big( \frac{h^3}{12} \frac{\partial p}{\partial x} \big) + a^2 \frac{\partial}{\partial z} \big( \frac{h^3}{12} \frac{\partial p}{\partial z} \big) =  \frac{\partial \mathcal{F}}{\partial x} + \frac{\partial \mathcal{G}}{\partial z}  ,
\\
\label{eqn_VRCrt2}
& \mathcal{F} = \frac{\beta}{2}h + \mathcal{A} \big(\frac{\partial \tau_{xx}}{\partial x}\big) + \mathcal{B}(\tau_{xy}) + a\mathcal{A} \big( \frac{\partial \tau_{xz}}{\partial z} \big),
\\
\label{eqn_VRCrt3}
& \mathcal{G} = a^2 \mathcal{A} \big(\frac{\partial \tau_{zz}}{\partial z}\big) + a\mathcal{A} \big( \frac{\partial \tau_{xz}}{\partial x}\big) + a \mathcal{B}(\tau_{yz}),
\end{subeqnarray}
where,
\begin{subeqnarray}
\label{eqn_VRCrt_A}
& \mathcal{A}(\cdot) = \int_0^h \frac{y}{h} \int_{0}^{h} \int_{0}^{y^{'}} (\cdot) \, dy^{'}dydy - \int_0^h \int_{0}^{y^{''}} \int_{0}^{y^{'}} (\cdot) \, dy^{'}dy^{''}dy,
\\
& \mathcal{B}(\cdot) = \int_0^h \frac{y}{h} \int_{0}^{h} (\cdot) \, dy^{'}dy - \int_0^h \int_{0}^{y^{'}} (\cdot) \, dy^{'}dy,
\end{subeqnarray}
are integral operators. These operators can be further reduced by (i) applying integration by parts and (ii) switching to the computational configuration; $h(x,z) = H(X, Z)$, $y = YH(X)$,
\begin{subeqnarray}
& \int_0^h \int_{0}^{y^{'}} q \, dy^{'}dy = H^2\int_0^1 (Y-1)q dY = H^2\overline{[(Y-1)q]},
\\
& \int_0^h \frac{y}{h}\int_{0}^{h} q \, dy^{'}dy = \frac{H^2}{2}\overline{[q]},
\\
& \int_0^h \int_{0}^{y^{'}}\int_{0}^{y^{''}} q \, dy^{''}dy^{'}dy = H^3\int_0^1 \int_{0}^{Y^{'}}\int_{0}^{Y^{''}} q \, dY^{''}dY^{'}dY = H^3\overline{[(\frac{Y}{2}-Y+\frac{1}{2})q]},
\\
& \int_0^h \frac{y}{h}\int_{0}^{y^{'}}\int_{0}^{y^{''}} q \, dy^{''}dy^{'}dy = H^3\overline{[(-\frac{Y}{2}+\frac{1}{2})q]},
\end{subeqnarray}
for the quantity $q$, yielding (upon substitution into Eqs. \ref{eqn_VRCrt1} and Eqs. \ref{eqn_VRCrt_A})
\begin{subeqnarray}
\label{chap_PF_eqn_VRCrt1}
&\frac{\partial}{\partial X} \big( \frac{H^3}{12} \frac{\partial p}{\partial X} \big) + a^2 \frac{\partial}{\partial Z} \big( \frac{H^3}{12} \frac{\partial p}{\partial Z} \big) =  \frac{\partial \mathcal{F}}{\partial X} + \frac{\partial \mathcal{G}}{\partial Z}  ,
\\
& \mathcal{F} = \frac{\beta}{2}H + \frac{H^3}{2}\overline{[(Y-Y^2)\frac{\partial \tau_{xx}}{\partial x}]} + \frac{H^2}{2}\overline{[(2Y-1)\tau_{xy}]} + a\frac{H^3}{2}\overline{[(Y-Y^2)\frac{\partial \tau_{xz}}{\partial z}]},
\\
& \mathcal{G} = a^2\frac{H^3}{2}\overline{[(Y-Y^2)\frac{\partial \tau_{zz}}{\partial z}]} + \frac{H^2}{2}\overline{[(2Y-1)\tau_{zy}]} + a\frac{H^3}{2}\overline{[(Y-Y^2)\frac{\partial \tau_{xz}}{\partial x}]}.
\label{chap_PF_eqn_VRCrt2}
\end{subeqnarray}

\section{Three-dimensional curvilinear domain}\label{app_curvilinear}
\par In this appendix, following the work of \cite{boyko2024flow}, we provide the curvilinear coordinates used in the numerical treatment of the governing system of equations. The coordinates are transformed through
\refstepcounter{equation}
$$
  x^* = X^* + \epsilon Q^*(X^*, Y^*), \quad
  y^* = \frac{Y^*}{H_0 H^*(X^*, Z^*)} , \quad
  z^* = Z^* + \epsilon R^*(Z^*, Y^*), \eqno{(\theequation{\mathit{a},\mathit{b},\mathit{c}})}\label{eqn_coordTransform}
$$
where, the $^*$ denotes dimensional quantities, $Q^*$ and $R^*$ are the functions accounting for the channel variation along the streamwise and the spanwise directions. Using Eq. \ref{eqn_geo_scaling}, and following the procedure leading to an orthagonal coordinate system (see Appendix A in \cite{boyko2024flow}), we have for the curvilinear basis vectors
\begin{subeqnarray}
\label{eqn_curvilinearbasis1}
& \boldsymbol{e}_X = \boldsymbol{e}_x + \epsilon Y\frac{\partial H}{\partial X}\boldsymbol{e}_y,
\\
\label{eqn_curvilinearbasis2}
& \boldsymbol{e}_Y = -\epsilon Y\frac{\partial H}{\partial X}\boldsymbol{e}_x + \boldsymbol{e}_y - a\epsilon Y\frac{\partial H}{\partial Z}\boldsymbol{e}_z,
\\
\label{eqn_curvilinearbasis3}
& \boldsymbol{e}_Z = a\epsilon Y\frac{\partial H}{\partial Z}\boldsymbol{e}_y +\boldsymbol{e}_z,
\end{subeqnarray}
and the components of the velocity related through
\begin{subeqnarray}
&\boldsymbol{u}^* = \mathsfbi{M} \, \boldsymbol{U}^*,
\\
&\mathsfbi{M} = [\boldsymbol{e}_X \, \boldsymbol{e}_Y \, \boldsymbol{e}_Z],
\end{subeqnarray}
that are non-dimensionalized using Eqs. \ref{eqn_geo_scaling},
\refstepcounter{equation}
$$
U = u + O(\epsilon^2), \quad
V = v - Y\frac{\partial H}{\partial X} u - a Y \frac{\partial H}{\partial Z} w,\quad  W=w+O(\epsilon^2).\eqno{(\theequation{\mathit{a},\mathit{b},\mathit{c}})}\label{eqns_vel_comps}
$$
Eqs. \ref{eqns_vel_comps} are used in the numerical procedure, specifically in the flux conserving discretization of the stress advection.

\section{Implicit discretization}\label{app_num_meth}
\par The system of equations presented in this work are solved via the classical finite difference method applied to the curvilinear (or computational) domain $\boldsymbol{X}=(X, Y, Z)$ in which the grid spacing along each axis is constant. 
The numerical solution of the constitutive relation, a system of coupled hyperbolic equations, is achieved by first converting the advective component of the material derivative operator $(\boldsymbol{u} \cdot \boldsymbol{\nabla})\boldsymbol{\tau}$ into $ \boldsymbol{\nabla} \cdot (\boldsymbol{u}\otimes \boldsymbol{\tau})$, where, $\boldsymbol{u}\otimes \boldsymbol{\tau}$ is a third-order tensor. Note that by utilizing the continuity equation, we can show that the two expressions are identical. However, from a numerical perspective, the former expression causes the FDM to suffer from numerical dissipation when the solution ceases to be smooth. In these cases, the attempt to increase the accuracy by refining the mesh tends to have little effect if higher-order schemes are used.
\par We pursue a fully-implicit numerical approach which requires a sparse-matrix operator for each of the terms appearing in the constitutive relation (for each component of the extra stress tensor). First we convert the expression $ \boldsymbol{\nabla} \cdot (\boldsymbol{u}\otimes \boldsymbol{\tau})$, written explicitly as
\begin{equation}
\boldsymbol{\nabla} \cdot (\boldsymbol{u}\otimes \boldsymbol{\tau}) = \bigg( u\frac{\partial }{\partial x} + v\frac{\partial }{\partial y} + aw\frac{\partial }{\partial z} \bigg) \boldsymbol{\tau},
\end{equation}
into the computational domain via Eqs. \ref{eqns_vel_comps}($\mathit a, b, c$). Carrying out the simplification and noting that 
\begin{equation}
\frac{\partial (HU)}{\partial X} + \frac{\partial V}{\partial Y} + a\frac{\partial (HW)}{\partial Z}=0,
\end{equation}
we get,
\begin{equation}
\label{eqn_mat_der_comp}
\boldsymbol{\nabla} \cdot (\boldsymbol{u}\otimes \boldsymbol{\tau}) = \frac{1}{H}\bigg( \frac{\partial (HU\boldsymbol{\tau})}{\partial X} + \frac{\partial (V\boldsymbol{\tau})}{\partial Y} + a\frac{\partial (HW\boldsymbol{\tau})}{\partial Z} \bigg).
\end{equation}
\par Discretizing Eq. \ref{eqn_mat_der_comp} via the classical FDM will result in a matrix-vector product $[\boldsymbol{\nabla} \cdot (\boldsymbol{u}\otimes \boldsymbol{\tau})] = \mathsfbi{A}\boldsymbol{q}$, where, $\mathsfbi{A}$ is the sparse coefficient matrix, $\boldsymbol{q}$ is the unknown vector containing the extra stress components $\tau_{ij}$. The sparse-matrix operators needed for the implicit treatment are then obtained,
\begin{equation}
\label{eqn_mat_der_comp_mats}
\mathsfbi{A} = \mathsfbi{D_{Hi}} (\mathsfbi{D_{\partial X}} \mathsfbi{D_{HU}} + \mathsfbi{D_{\partial Y}} \mathsfbi{D_{V}} + a\mathsfbi{D_{\partial Z}} \mathsfbi{D_{HW}})
\end{equation}
where, $\mathsfbi{D_{Hi}} = \textit{diag}(H^{-1})$, $\mathsfbi{D_{HU}} = \textit{diag}(HU)$, $\mathsfbi{D_V} = \textit{diag}(V)$, $\mathsfbi{D_{HW}} = \textit{diag}(HW)$ are sparse-diagonal matrices, $\mathsfbi{D_{\partial X}}$, $\mathsfbi{D_{\partial Y}}$, $\mathsfbi{D_{\partial Z}}$ are the sparse-difference matrix operators.
\par In this setting, a fourth-order scheme can be used for the operators along $Y$ and $Z$, and based on $De$, a second-order flux-conserving scheme for $X$. The latter being more sensitive at high $De$ and leading to poor convergence (requiring a large number of iterations) if higher-order schemes are used. It was observed that increasing the nodes along $X$, and using a first-order hybrid scheme that is unconditionally numerically stable was far more effective in terms of convergence and accuracy at higher $De$. Despite these measures, no noticeable improvement in the HWNP was observed since it persists within the polynomial-based finite-difference, volume and element discretization measures.

\section{Coordinate transformation}\label{app_coordinate_transformation}
\par Here, we describe the relationship between the derivatives appearing in the governing equations when moving from the rectilinear $\boldsymbol{x} = (x, y, z)$ to the curvilinear $\boldsymbol{X} = (X, Y, Z)$ domain. When converting the first-order scalar derivatives, a simple application of the total derivative of a quantity and the chain rule will yield, 
\begin{subeqnarray}
\label{eqn_Transform1stDer1}
& \frac{\partial }{\partial X} = \frac{\partial x}{\partial X}\frac{\partial }{\partial x} + \frac{\partial y}{\partial X}\frac{\partial }{\partial y} + \frac{\partial z}{\partial X}\frac{\partial }{\partial z}, 
\\
& \frac{\partial }{\partial Y} = \frac{\partial x}{\partial Y}\frac{\partial }{\partial x} + \frac{\partial y}{\partial Y}\frac{\partial }{\partial y} + \frac{\partial z}{\partial Y}\frac{\partial }{\partial z}, 
\\
& \frac{\partial }{\partial Z} = \frac{\partial x}{\partial Z}\frac{\partial }{\partial x} + \frac{\partial y}{\partial Z}\frac{\partial }{\partial y} + \frac{\partial z}{\partial Z}\frac{\partial }{\partial z}.
\label{eqn_Transform1stDer2}
\end{subeqnarray}
The $De$-order linear model (Eq. \ref{chap_PF_eqn_lin3d}) contains higher-order derivatives. Therefore, we extend Eqs. \ref{eqn_Transform1stDer1} to the second-order (including the mixed partial derivatives), 
\begin{subeqnarray}
\label{eqn_Transform2ndDer1}
\nonumber
& \frac{\partial^2 }{\partial X^2} = \frac{\partial^2 x}{\partial X^2}\frac{\partial }{\partial x} + \frac{\partial^2 y}{\partial X^2}\frac{\partial }{\partial y} + \frac{\partial^2 z}{\partial X^2}\frac{\partial }{\partial z} + \big(\frac{\partial x}{\partial X}\big)^2\frac{\partial^2}{\partial x^2} + \big(\frac{\partial y}{\partial X}\big)^2\frac{\partial^2}{\partial y^2} + \big(\frac{\partial z}{\partial X}\big)^2\frac{\partial^2}{\partial z^2}
\\ & + 2 \big( \frac{\partial x}{\partial X} \frac{\partial y}{\partial X} \big) \frac{\partial^2}{\partial x \partial y} + 2 \big( \frac{\partial x}{\partial X} \frac{\partial z}{\partial X} \big) \frac{\partial^2}{\partial x \partial z} + 2 \big( \frac{\partial z}{\partial X} \frac{\partial y}{\partial X} \big) \frac{\partial^2}{\partial z \partial y}, 
\\
\nonumber
& \frac{\partial^2 }{\partial Y^2} = \frac{\partial^2 x}{\partial Y^2}\frac{\partial }{\partial x} + \frac{\partial^2 y}{\partial Y^2}\frac{\partial }{\partial y} + \frac{\partial^2 z}{\partial Y^2}\frac{\partial }{\partial z} + \big(\frac{\partial x}{\partial Y}\big)^2\frac{\partial^2}{\partial x^2} + \big(\frac{\partial y}{\partial Y}\big)^2\frac{\partial^2}{\partial y^2} + \big(\frac{\partial z}{\partial Y}\big)^2\frac{\partial^2}{\partial z^2}
\\ & + 2 \big( \frac{\partial x}{\partial Y} \frac{\partial y}{\partial Y} \big) \frac{\partial^2}{\partial x \partial y} + 2 \big( \frac{\partial x}{\partial Y} \frac{\partial z}{\partial Y} \big) \frac{\partial^2}{\partial x \partial z} + 2 \big( \frac{\partial z}{\partial Y} \frac{\partial y}{\partial Y} \big) \frac{\partial^2}{\partial z \partial y}, 
\\
\nonumber
& \frac{\partial^2 }{\partial Z^2} = \frac{\partial^2 x}{\partial Z^2}\frac{\partial }{\partial x} + \frac{\partial^2 y}{\partial Z^2}\frac{\partial }{\partial y} + \frac{\partial^2 z}{\partial Z^2}\frac{\partial }{\partial z} + \big(\frac{\partial x}{\partial Z}\big)^2\frac{\partial^2}{\partial x^2} + \big(\frac{\partial y}{\partial Z}\big)^2\frac{\partial^2}{\partial y^2} + \big(\frac{\partial z}{\partial Z}\big)^2\frac{\partial^2}{\partial z^2}
\\ & + 2 \big( \frac{\partial x}{\partial Z} \frac{\partial y}{\partial Z} \big) \frac{\partial^2}{\partial x \partial y} + 2 \big( \frac{\partial x}{\partial Z} \frac{\partial z}{\partial Z} \big) \frac{\partial^2}{\partial x \partial z} + 2 \big( \frac{\partial z}{\partial Z} \frac{\partial y}{\partial Z} \big) \frac{\partial^2}{\partial z \partial y}, 
\\
\nonumber
& \frac{\partial^2 }{\partial X\partial Y} = \frac{\partial^2 x}{\partial X\partial Y}\frac{\partial }{\partial x} + \frac{\partial^2 y}{\partial X\partial Y}\frac{\partial }{\partial y} + \frac{\partial^2 z}{\partial X\partial Y}\frac{\partial }{\partial z} + \big(\frac{\partial x}{\partial Y}\frac{\partial x}{\partial X}\big)\frac{\partial^2}{\partial x^2} + \big(\frac{\partial y}{\partial Y}\frac{\partial y}{\partial X}\big)\frac{\partial^2}{\partial y^2} + \big(\frac{\partial z}{\partial Y}\frac{\partial z}{\partial X}\big)\frac{\partial^2}{\partial z^2}
\\ & + \big( \frac{\partial x}{\partial X} \frac{\partial y}{\partial Y} + \frac{\partial x}{\partial Y} \frac{\partial y}{\partial X} \big) \frac{\partial^2}{\partial x \partial y} + \big( \frac{\partial x}{\partial X} \frac{\partial z}{\partial Y} + \frac{\partial x}{\partial Y} \frac{\partial z}{\partial X} \big) \frac{\partial^2}{\partial x \partial z} + \big( \frac{\partial z}{\partial X} \frac{\partial y}{\partial Y} + \frac{\partial z}{\partial Y} \frac{\partial y}{\partial X} \big) \frac{\partial^2}{\partial z \partial y}, 
\\
\nonumber
& \frac{\partial^2 }{\partial X\partial Z} = \frac{\partial^2 x}{\partial X\partial Z}\frac{\partial }{\partial x} + \frac{\partial^2 y}{\partial X\partial Z}\frac{\partial }{\partial y} + \frac{\partial^2 z}{\partial X\partial Z}\frac{\partial }{\partial z} + \big(\frac{\partial x}{\partial X}\frac{\partial x}{\partial Z}\big)\frac{\partial^2}{\partial x^2} + \big(\frac{\partial y}{\partial X}\frac{\partial y}{\partial Z}\big)\frac{\partial^2}{\partial y^2} + \big(\frac{\partial z}{\partial X}\frac{\partial z}{\partial Z}\big)\frac{\partial^2}{\partial z^2}
\\ & + \big( \frac{\partial x}{\partial X} \frac{\partial y}{\partial Z} + \frac{\partial x}{\partial X} \frac{\partial y}{\partial Z} \big) \frac{\partial^2}{\partial x \partial y} + \big( \frac{\partial x}{\partial X} \frac{\partial z}{\partial Z} + \frac{\partial x}{\partial Z} \frac{\partial z}{\partial X} \big) \frac{\partial^2}{\partial x \partial z} + \big( \frac{\partial z}{\partial X} \frac{\partial y}{\partial Z} + \frac{\partial z}{\partial Z} \frac{\partial y}{\partial X} \big) \frac{\partial^2}{\partial z \partial y}, 
\\
\nonumber
& \frac{\partial^2 }{\partial Y\partial Z} = \frac{\partial^2 x}{\partial Y\partial Z}\frac{\partial }{\partial x} + \frac{\partial^2 y}{\partial Y\partial Z}\frac{\partial }{\partial y} + \frac{\partial^2 z}{\partial Y\partial Z}\frac{\partial }{\partial z} + \big(\frac{\partial x}{\partial Y}\frac{\partial x}{\partial Z}\big)\frac{\partial^2}{\partial x^2} + \big(\frac{\partial y}{\partial Y}\frac{\partial y}{\partial Z}\big)\frac{\partial^2}{\partial y^2} + \big(\frac{\partial z}{\partial Y}\frac{\partial z}{\partial Z}\big)\frac{\partial^2}{\partial z^2}
\\ & + \big( \frac{\partial x}{\partial Y} \frac{\partial y}{\partial Z} + \frac{\partial x}{\partial Z} \frac{\partial y}{\partial Y} \big) \frac{\partial^2}{\partial x \partial y} + \big( \frac{\partial x}{\partial Y} \frac{\partial z}{\partial Z} + \frac{\partial x}{\partial Z} \frac{\partial z}{\partial Y} \big) \frac{\partial^2}{\partial x \partial z} + \big( \frac{\partial z}{\partial Y} \frac{\partial y}{\partial Z} + \frac{\partial z}{\partial Z} \frac{\partial y}{\partial Y} \big) \frac{\partial^2}{\partial z \partial y}.
\label{eqn_Transform2ndDer2}
\end{subeqnarray}
\par Eqs. \ref{eqn_Transform2ndDer1} relate the derivative operators in the rectilinear coordinates ($x,y,z$) to the curvilinear coordinates ($X, Y, Z$). In order to use the classical FDM in which the stencil width is fixed, we require expressions for the derivatives in terms of the physical coordinates for which we express the system of Eqs. \ref{eqn_Transform2ndDer1} as a single linear algebraic equation
\begin{equation}
\label{eqnTransfromMatVec}
\boldsymbol{\partial X}
=
\mathsfbi{A}
\boldsymbol{\partial x},
\end{equation}
where,
\begin{subeqnarray}
& \boldsymbol{\partial X}^T = \big[ \frac{\partial }{\partial X} , \frac{\partial }{\partial Y} , \frac{\partial }{\partial Z} , \frac{\partial^2 }{\partial X^2} , \frac{\partial^2 }{\partial Y^2} , \frac{\partial^2 }{\partial Z^2} , \frac{\partial^2 }{\partial X \partial Y} , \frac{\partial^2 }{\partial X \partial Z} , \frac{\partial^2 }{\partial Y \partial Z} \big],
\\
& \boldsymbol{\partial x}^T = \big[ \frac{\partial }{\partial x} , \frac{\partial }{\partial y} , \frac{\partial }{\partial z} , \frac{\partial^2 }{\partial x^2} , \frac{\partial^2 }{\partial y^2} , \frac{\partial^2 }{\partial z^2} , \frac{\partial^2 }{\partial x \partial y} , \frac{\partial^2 }{\partial x \partial z} , \frac{\partial^2 }{\partial y \partial z} \big],
\end{subeqnarray}
and $\mathsfbi{A}$ is the Jacobian of the transformation. The detailed contents of $\textbf{A}$ are omitted for the sake of brevity but can be obtained from Eqs. \ref{eqn_coordTransform}. 
\par In this work, we only pursue the transformations given by Eqs. \ref{eqn_Transform1stDer1} and Eqs. \ref{eqn_Transform2ndDer2} to gain a computational advantage, namely, that the operators $\boldsymbol{\partial X}$ can be constructed independently of $H$ and they enable the discretization via the classical finite difference schemes which are easily available. Furthermore, the difference operators are then used to construct an implicit system of algebraic equations that results in a numerically efficient solution procedure, see \cite{ahmed2023modeling} for more details.

\section*{Acknowledgements}
The authors would like to acknowledge the Turkish National Research
Agency (T\"UB$\mathrm{\dot{I}}$TAK) for supporting this work under the project
221N576.

\section*{Declaration of interest}

The authors declare that they have no known competing financial
interests or personal relationships that could have appeared to
influence the work reported in this paper.

\bibliographystyle{jfm}
\bibliography{ref}

\begin{thebibliography}{54}
\expandafter\ifx\csname natexlab\endcsname\relax\def\natexlab#1{#1}\fi
\def\au#1{#1} \def\ed#1{#1} \def\yr#1{#1}\def\at#1{#1}\def\jt#1{\textit{#1}} \def\bt#1{#1}\def\bvol#1{\textbf{#1}} \def\vol#1{#1} \def\pg#1{#1} \def\publ#1{#1}\def\arxiv#1{#1}\def\org#1{#1}\def\st#1{\textit{#1}}

\bibitem[Abbaspur {\em et~al.\/}(2023)Abbaspur, Norouzi, Akbarzadeh, Vaziri, Sharghi, Kim \& Kim]{abbaspur2023analytical}
{\sc \au{Abbaspur, A.}, \au{Norouzi, M.}, \au{Akbarzadeh, P.}, \au{Vaziri, S.~A.}, \au{Sharghi, M.~M.}, \au{Kim, K.~C.} \& \au{Kim, M.}} \yr{2023}  \at{An analytical study on nonlinear viscoelastic lubrication in journal bearings}.  \jt{Scientific Reports}  \bvol{13}~(1),  \pg{16836}.

\bibitem[Ahmed \& Biancofiore(2021)]{ahmed2021new}
{\sc \au{Ahmed, H.} \& \au{Biancofiore, L.}} \yr{2021}  \at{A new approach for modeling viscoelastic thin film lubrication}.  \jt{Journal of Non-Newtonian Fluid Mechanics}  \bvol{292},  \pg{104524}.

\bibitem[Ahmed \& Biancofiore(2022)]{ahmed2022modified}
{\sc \au{Ahmed, H.} \& \au{Biancofiore, L.}} \yr{2022}  \at{A modified viscosity approach for shear thinning lubricants}.  \jt{Physics of Fluids}  \bvol{34}~(10).

\bibitem[Ahmed \& Biancofiore(2023)]{ahmed2023modeling}
{\sc \au{Ahmed, H.} \& \au{Biancofiore, L.}} \yr{2023}  \at{Modeling polymeric lubricants with non-linear stress constitutive relations}.  \jt{Journal of Non-Newtonian Fluid Mechanics}  \bvol{321},  \pg{105123}.

\bibitem[Akyildiz \& Bellout(2004)]{akyildiz2004viscoelastic}
{\sc \au{Akyildiz, F.~T.} \& \au{Bellout, H.}} \yr{2004}  \at{Viscoelastic lubrication with phan-thein-tanner fluid (ptt)}.  \jt{J. Trib.}  \bvol{126}~(2),  \pg{288--291}.

\bibitem[Alves {\em et~al.\/}(2021)Alves, Oliveira \& Pinho]{alves2021numerical}
{\sc \au{Alves, M.~A.}, \au{Oliveira, P.~J.} \& \au{Pinho, F.~T.}} \yr{2021}  \at{Numerical methods for viscoelastic fluid flows}.  \jt{Annual Review of Fluid Mechanics}  \bvol{53}~(1),  \pg{509--541}.

\bibitem[Beris {\em et~al.\/}(1986)Beris, Armstrong \& Brown]{beris1986finite}
{\sc \au{Beris, A.~N.}, \au{Armstrong, R.~C.} \& \au{Brown, R.~A.}} \yr{1986}  \at{Finite element calculation of viscoelastic flow in a journal bearing: Ii. moderate eccentricity}.  \jt{Journal of non-newtonian fluid mechanics}  \bvol{19}~(3),  \pg{323--347}.

\bibitem[Bertocchi {\em et~al.\/}(2013)Bertocchi, Dini, Giacopini, Fowell \& Baldini]{bertocchi2013fluid}
{\sc \au{Bertocchi, Luca}, \au{Dini, Daniele}, \au{Giacopini, Matteo}, \au{Fowell, Mark~T} \& \au{Baldini, Andrea}} \yr{2013}  \at{Fluid film lubrication in the presence of cavitation: a mass-conserving two-dimensional formulation for compressible, piezoviscous and non-newtonian fluids}.  \jt{Tribology International}  \bvol{67},  \pg{61--71}.

\bibitem[Bird {\em et~al.\/}(1987)Bird, Armstrong \& Hassager]{bird1987dynamics}
{\sc \au{Bird, R.~B.}, \au{Armstrong, Robert~C.} \& \au{Hassager, O.}} \yr{1987}  \at{Dynamics of polymeric liquids. vol. 1: Fluid mechanics} .

\bibitem[Boyko \& Christov(2023)]{boyko2023non}
{\sc \au{Boyko, E.} \& \au{Christov, I.~C.}} \yr{2023}  \at{Non-newtonian fluid--structure interaction: Flow of a viscoelastic oldroyd-b fluid in a deformable channel}.  \jt{Journal of Non-Newtonian Fluid Mechanics}  \bvol{313},  \pg{104990}.

\bibitem[Boyko {\em et~al.\/}(2024)Boyko, Hinch \& Stone]{boyko2024flow}
{\sc \au{Boyko, E.}, \au{Hinch, J.} \& \au{Stone, H.~A.}} \yr{2024}  \at{Flow of an oldroyd-b fluid in a slowly varying contraction: theoretical results for arbitrary values of deborah number in the ultra-dilute limit}.  \jt{Journal of Fluid Mechanics}  \bvol{988},  \pg{A10}.

\bibitem[Boyko \& Stone(2021)]{boyko2021reciprocal}
{\sc \au{Boyko, E.} \& \au{Stone, H.~A.}} \yr{2021}  \at{Reciprocal theorem for calculating the flow rate--pressure drop relation for complex fluids in narrow geometries}.  \jt{Physical Review Fluids}  \bvol{6}~(8),  \pg{L081301}.

\bibitem[Boyko \& Stone(2022)]{boyko2022pressure}
{\sc \au{Boyko, E.} \& \au{Stone, H.~A.}} \yr{2022}  \at{Pressure-driven flow of the viscoelastic oldroyd-b fluid in narrow non-uniform geometries: analytical results and comparison with simulations}.  \jt{Journal of Fluid Mechanics}  \bvol{936},  \pg{A23}.

\bibitem[{\c{C}}am {\em et~al.\/}(2023){\c{C}}am, Giacopini, Dini \& Biancofiore]{ccam2023numerical}
{\sc \au{{\c{C}}am, Mert~Yusuf}, \au{Giacopini, Matteo}, \au{Dini, Daniele} \& \au{Biancofiore, Luca}} \yr{2023}  \at{A numerical algorithm to model wall slip and cavitation in two-dimensional hydrodynamically lubricated contacts}.  \jt{Tribology International}  \bvol{184},  \pg{108444}.

\bibitem[Chakraborty {\em et~al.\/}(2010)Chakraborty, Bajaj, Yeo, Friend, Pasquali \& Prakash]{chakraborty2010viscoelastic}
{\sc \au{Chakraborty, D.}, \au{Bajaj, M.}, \au{Yeo, L.}, \au{Friend, J.}, \au{Pasquali, M.} \& \au{Prakash, J.~R.}} \yr{2010}  \at{Viscoelastic flow in a two-dimensional collapsible channel}.  \jt{Journal of non-newtonian fluid mechanics}  \bvol{165}~(19-20),  \pg{1204--1218}.

\bibitem[Dowson \& Taylor(1979)]{dowson1979cavitation}
{\sc \au{Dowson, D} \& \au{Taylor, CM}} \yr{1979}  \at{Cavitation in bearings}.  \jt{Annual Review of Fluid Mechanics}  \bvol{11}~(1),  \pg{35--65}.

\bibitem[Dunn {\em et~al.\/}(2013)Dunn, Tichy, Urue{\~n}a \& Sawyer]{dunn2013lubrication}
{\sc \au{Dunn, A.~C.}, \au{Tichy, J.~A.}, \au{Urue{\~n}a, J.~M.} \& \au{Sawyer, W.~G.}} \yr{2013}  \at{Lubrication regimes in contact lens wear during a blink}.  \jt{Tribology International}  \bvol{63},  \pg{45--50}.

\bibitem[Fattal \& Kupferman(2004)]{fattal2004constitutive}
{\sc \au{Fattal, R.} \& \au{Kupferman, R.}} \yr{2004}  \at{Constitutive laws for the matrix-logarithm of the conformation tensor}.  \jt{Journal of Non-Newtonian Fluid Mechanics}  \bvol{123}~(2-3),  \pg{281--285}.

\bibitem[Fattal \& Kupferman(2005)]{fattal2005time}
{\sc \au{Fattal, R.} \& \au{Kupferman, R.}} \yr{2005}  \at{Time-dependent simulation of viscoelastic flows at high weissenberg number using the log-conformation representation}.  \jt{Journal of Non-Newtonian Fluid Mechanics}  \bvol{126}~(1),  \pg{23--37}.

\bibitem[Feng \& Jabbarzadeh(2024)]{feng2024rheological}
{\sc \au{Feng, Y.} \& \au{Jabbarzadeh, A.}} \yr{2024}  \at{Rheological properties of water-based amino acid ionic liquids}.  \jt{Physics of Fluids}  \bvol{36}~(1).

\bibitem[Fernandes {\em et~al.\/}(2017)Fernandes, Araujo, Ferr{\'a}s \& N{\'o}brega]{fernandes2017improved}
{\sc \au{Fernandes, C.}, \au{Araujo, M. S.~B.}, \au{Ferr{\'a}s, L.~L.} \& \au{N{\'o}brega, J.~M.}} \yr{2017}  \at{Improved both sides diffusion (ibsd): A new and straightforward stabilization approach for viscoelastic fluid flows}.  \jt{Journal of Non-Newtonian Fluid Mechanics}  \bvol{249},  \pg{63--78}.

\bibitem[Gamaniel {\em et~al.\/}(2021)Gamaniel, Dini \& Biancofiore]{gamaniel2021effect}
{\sc \au{Gamaniel, S.~S.}, \au{Dini, D.} \& \au{Biancofiore, L.}} \yr{2021}  \at{The effect of fluid viscoelasticity in lubricated contacts in the presence of cavitation}.  \jt{Tribology International}  \bvol{160},  \pg{107011}.

\bibitem[Hinch {\em et~al.\/}(2024)Hinch, Boyko \& Stone]{hinch2024fast}
{\sc \au{Hinch, J.}, \au{Boyko, E.} \& \au{Stone, H.~A.}} \yr{2024}  \at{Fast flow of an oldroyd-b model fluid through a narrow slowly varying contraction}.  \jt{Journal of Fluid Mechanics}  \bvol{988},  \pg{A11}.

\bibitem[Housiadas \& Beris(2023)]{housiadas2023lubrication}
{\sc \au{Housiadas, K.~D.} \& \au{Beris, A.~N.}} \yr{2023}  \at{Lubrication approximation of pressure-driven viscoelastic flow in a hyperbolic channel}.  \jt{Physics of Fluids}  \bvol{35}~(12).

\bibitem[Housiadas \& Beris(2024{\natexlab{{\em a\/}}})]{housiadas2024pressure}
{\sc \au{Housiadas, K.~D.} \& \au{Beris, A.~N.}} \yr{2024{\natexlab{{\em a\/}}}}  \at{Pressure-drop and trouton ratio for oldroyd-b fluids in hyperbolic converging channels}.  \jt{Physics of Fluids}  \bvol{36}~(2).

\bibitem[Housiadas \& Beris(2024{\natexlab{{\em b\/}}})]{housiadas2024viscoelastic}
{\sc \au{Housiadas, K.~D.} \& \au{Beris, A.~N.}} \yr{2024{\natexlab{{\em b\/}}}}  \at{Viscoelastic flow with slip in a hyperbolic channel}.  \jt{Journal of Rheology}  \bvol{68}~(3),  \pg{415--428}.

\bibitem[Jeng {\em et~al.\/}(1986)Jeng, Hamrock \& Brewe]{jeng1986piezoviscous}
{\sc \au{Jeng, Y.}, \au{Hamrock, B.~J.} \& \au{Brewe, D.~E.}} \yr{1986}  \at{Piezoviscous effects in nonconformal contacts lubricated hydrodynamically}.  \jt{ASLE transactions}  \bvol{30}~(4),  \pg{452--464}.

\bibitem[Johnson \& Tevaarwerk(1977)]{johnson1977shear}
{\sc \au{Johnson, K.~L.} \& \au{Tevaarwerk, J.~L.}} \yr{1977}  \at{Shear behaviour of elastohydrodynamic oil films}.  \jt{Proceedings of the Royal Society of London. A. Mathematical and Physical Sciences}  \bvol{356}~(1685),  \pg{215--236}.

\bibitem[Keunings(1986)]{keunings1986high}
{\sc \au{Keunings, Roland}} \yr{1986}  \at{On the high weissenberg number problem}.  \jt{Journal of Non-Newtonian Fluid Mechanics}  \bvol{20},  \pg{209--226}.

\bibitem[Kundu {\em et~al.\/}(2015)Kundu, Cohen \& Dowling]{kundu2015fluid}
{\sc \au{Kundu, P.~K.}, \au{Cohen, I.~M.} \& \au{Dowling, D.~R.}} \yr{2015} {\em Fluid mechanics\/}.  \publ{Academic press}.

\bibitem[Li(2014)]{li2014non}
{\sc \au{Li, X.}} \yr{2014}  \at{Non-newtonian lubrication with the phan-thien--tanner model}.  \jt{Journal of Engineering Mathematics}  \bvol{87},  \pg{1--17}.

\bibitem[Mortier {\em et~al.\/}(2010)Mortier, Orszulik \& Fox]{mortier2010chemistry}
{\sc \au{Mortier, R.~M.}, \au{Orszulik, S.~T.} \& \au{Fox, M.~F.}} \yr{2010} {\em Chemistry and technology of lubricants\/}, ,  \vol{vol. 107115}.  \publ{Springer}.

\bibitem[Oliveira(2002)]{oliveira2002exact}
{\sc \au{Oliveira, P.~J.}} \yr{2002}  \at{An exact solution for tube and slit flow of a fene-p fluid}.  \jt{Acta Mechanica}  \bvol{158}~(3),  \pg{157--167}.

\bibitem[Owens \& Phillips(2002)]{owens2002computational}
{\sc \au{Owens, R.~G.} \& \au{Phillips, T.~N.}} \yr{2002} {\em Computational rheology\/}.  \publ{World Scientific}.

\bibitem[Pandey {\em et~al.\/}(2016)Pandey, Karpitschka, Venner \& Snoeijer]{pandey2016lubrication}
{\sc \au{Pandey, A.}, \au{Karpitschka, S.}, \au{Venner, C.~H.} \& \au{Snoeijer, J.~H.}} \yr{2016}  \at{Lubrication of soft viscoelastic solids}.  \jt{Journal of fluid mechanics}  \bvol{799},  \pg{433--447}.

\bibitem[Phan-Thien {\em et~al.\/}(1985)Phan-Thien, Dudek, Boger \& Tirtaatmadja]{phan1985squeeze}
{\sc \au{Phan-Thien, N.}, \au{Dudek, J.}, \au{Boger, D.~V.} \& \au{Tirtaatmadja, V.}} \yr{1985}  \at{Squeeze film flow of ideal elastic liquids}.  \jt{Journal of non-newtonian fluid mechanics}  \bvol{18}~(3),  \pg{227--254}.

\bibitem[Phan-Thien \& Tanner(1983)]{phan1983viscoelastic}
{\sc \au{Phan-Thien, N.} \& \au{Tanner, R.~I.}} \yr{1983}  \at{Viscoelastic squeeze-film flows--maxwell fluids}.  \jt{Journal of Fluid Mechanics}  \bvol{129},  \pg{265--281}.

\bibitem[Phan-Thien \& Tanner(1984)]{phan1984lubrication}
{\sc \au{Phan-Thien, N.} \& \au{Tanner, R.~I.}} \yr{1984}  \at{Lubrication squeeze-film theory for the oldroyd-b fluid}.  \jt{Journal of non-newtonian fluid mechanics}  \bvol{14},  \pg{327--335}.

\bibitem[Rastogi \& Gupta(1991)]{rastogi1991accounting}
{\sc \au{Rastogi, A} \& \au{Gupta, RK}} \yr{1991}  \at{Accounting for lubricant shear thinning in the design of short journal bearings}.  \jt{Journal of Rheology}  \bvol{35}~(4),  \pg{589--603}.

\bibitem[Renardy \& Thomases(2021)]{renardy2021mathematician}
{\sc \au{Renardy, M.} \& \au{Thomases, B.}} \yr{2021}  \at{A mathematician’s perspective on the oldroyd b model: Progress and future challenges}.  \jt{Journal of Non-Newtonian Fluid Mechanics}  \bvol{293},  \pg{104573}.

\bibitem[Sari {\em et~al.\/}(2024)Sari, Putignano, Carbone \& Biancofiore]{sari2024effect}
{\sc \au{Sari, M.~H.}, \au{Putignano, C.}, \au{Carbone, G.} \& \au{Biancofiore, L.}} \yr{2024}  \at{The effect of fluid viscoelasticity in soft lubrication}.  \jt{Tribology International}  \bvol{195},  \pg{109578}.

\bibitem[Sawyer \& Tichy(1998)]{tichy_sawer_SOF_1998}
{\sc \au{Sawyer, W.~G.} \& \au{Tichy, J.~A.}} \yr{1998}  \at{{Non-Newtonian Lubrication With the Second-Order Fluid}}.  \jt{Journal of Tribology}  \bvol{120}~(3),  \pg{622--628}.

\bibitem[Schuh {\em et~al.\/}(2017)Schuh, Lee, Allison \& Ewoldt]{schuh2017design}
{\sc \au{Schuh, J.~K.}, \au{Lee, Y.}, \au{Allison, J.~T.} \& \au{Ewoldt, R.~H.}} \yr{2017}  \at{Design-driven modeling of surface-textured full-film lubricated sliding: validation and rationale of nonstandard thrust observations}.  \jt{Tribology Letters}  \bvol{65},  \pg{1--17}.

\bibitem[Szeri(2010)]{szeri2010fluid}
{\sc \au{Szeri, A.~Z.}} \yr{2010} {\em Fluid film lubrication\/}.  \publ{Cambridge university press}.

\bibitem[Tanner(1969)]{tanner_1970_spurious}
{\sc \au{Tanner, R.~I.}} \yr{1969}  \at{{Increase of Bearing Loads Due to Large Normal Stress Differences in Viscoelastic Lubricants}}.  \jt{Journal of Applied Mechanics}  \bvol{36}~(3),  \pg{634--635}.

\bibitem[Tanner(2000)]{tanner2000engineering}
{\sc \au{Tanner, R.~I.}} \yr{2000} {\em Engineering rheology\/}, ,  \vol{vol.~52}.  \publ{OUP Oxford}.

\bibitem[Tichy(1996)]{tichy_lin_1996}
{\sc \au{Tichy, J.~A.}} \yr{1996}  \at{{Non-Newtonian Lubrication With the Convected Maxwell Model}}.  \jt{Journal of Tribology}  \bvol{118}~(2),  \pg{344--348}.

\bibitem[Tichy \& Bou-Saïd(2008)]{tichy_said_spherical}
{\sc \au{Tichy, J.~A.} \& \au{Bou-Saïd, B.}} \yr{2008}  \at{{The Phan-Thien and Tanner Model Applied to Thin Film Spherical Coordinates: Applications for Lubrication of Hip Joint Replacement}}.  \jt{Journal of Biomechanical Engineering}  \bvol{130}~(2),  \pg{021012}.

\bibitem[Tichy \& Winer(1978)]{tichy_winer_squeeze}
{\sc \au{Tichy, J.~A.} \& \au{Winer, W.~O.}} \yr{1978}  \at{{An Investigation Into the Influence of Fluid Viscoelasticity in a Squeeze Film Bearing}}.  \jt{Journal of Lubrication Technology}  \bvol{100}~(1),  \pg{56--64}.

\bibitem[Venkatesh {\em et~al.\/}(2022)Venkatesh, Anand \& Narsimhan]{venkatesh2022peeling}
{\sc \au{Venkatesh, A.}, \au{Anand, V.} \& \au{Narsimhan, V.}} \yr{2022}  \at{Peeling of linearly elastic sheets using complex fluids at low reynolds numbers}.  \jt{Journal of Non-Newtonian Fluid Mechanics}  \bvol{309},  \pg{104916}.

\bibitem[Wolff \& Kubo(1996)]{wolf_kubo}
{\sc \au{Wolff, R.} \& \au{Kubo, A.}} \yr{1996}  \at{{A Generalized Non-Newtonian Fluid Model Incorporated Into Elastohydrodynamic Lubrication}}.  \jt{Journal of Tribology}  \bvol{118}~(1),  \pg{74--82}.

\bibitem[Yousfi {\em et~al.\/}(2013)Yousfi, Bou-Sa{\"\i}d \& Tichy]{yousfi2013analytical}
{\sc \au{Yousfi, M.}, \au{Bou-Sa{\"\i}d, B.} \& \au{Tichy, J.~A.}} \yr{2013}  \at{An analytical study of the squeezing flow of synovial fluid}.  \jt{Mechanics \& Industry}  \bvol{14}~(1),  \pg{59--69}.

\bibitem[Zhang {\em et~al.\/}(2023)Zhang, Zhang, Wang, Li, Li \& Li]{zhang2023role}
{\sc \au{Zhang, H.}, \au{Zhang, W.}, \au{Wang, X.}, \au{Li, Y.}, \au{Li, X.} \& \au{Li, F.}} \yr{2023}  \at{On the role of tensor interpolation in solving high-wi viscoelastic fluid flow}.  \jt{Physics of Fluids}  \bvol{35}~(3).

\bibitem[Zheng {\em et~al.\/}(2023)Zheng, Xie, Chen, Liu, Yang, Xu \& Huang]{zheng2023squeeze}
{\sc \au{Zheng, Z.}, \au{Xie, H.}, \au{Chen, X.}, \au{Liu, X.}, \au{Yang, W.}, \au{Xu, Y.} \& \au{Huang, W.}} \yr{2023}  \at{Squeeze flow of a maxwell fluid between two parallel disks or two spheres}.  \jt{Physics of Fluids}  \bvol{35}~(8).

\end{thebibliography}

\end{document}